\documentclass[12pt]{iopart}
\usepackage{graphicx} 
\usepackage{upgreek}
\usepackage[hidelinks]{hyperref}

\begin{document}

\title{Roadmap on Nanoscale Magnetic Resonance Imaging}
\author{Guest editors: Raffi Budakian, Amit Finkler \& Alex Eichler}
\date{December 2023}
\begin{abstract}
    The field of nanoscale magnetic resonance imaging (NanoMRI) was started 30 years ago. It was motivated by the desire to image single molecules and molecular assemblies, such as proteins and virus particles, with near-atomic spatial resolution and on a length scale of 100 nm. The realization of this goal requires the development of spin detection techniques that are many orders of magnitude more sensitive than conventional MRI, capable of detecting and controlling nanoscale ensembles of spins. Over the years, a number of different technical approaches to NanoMRI have emerged, each possessing a distinct set of capabilities for basic and applied areas of science. The goal of this roadmap article is to report the current state of the art in NanoMRI technologies, outline the areas where they are poised to have impact, identify the challenges that lie ahead, and propose methods to meet these challenges. This roadmap also shows how developments in NanoMRI techniques can lead to breakthroughs in emerging quantum science and technology applications.
\end{abstract}

\maketitle

\section*{Introduction}
Magnetic resonance imaging (MRI) has revolutionized medical diagnosis due to its unparalleled combination of elemental selectivity, noninvasive detection, and three-dimensional imaging capabilities. Translating these same advantages to the nanoscale holds immense potential for structural biology and quantum device design, but achieving this feat presents a formidable challenge: the signal strength in MRI typically decreases with the cube of the object's length, rendering conventional techniques ineffective at the nanoscale. To overcome this hurdle, novel experimental platforms and theoretical methods are being actively pursued. This roadmap provides an overview of the current state of the art and future directions in this exciting field at the intersection of physics, chemistry, and materials science.

The concept of NanoMRI can be traced back to John A. Sidles, a theoretical physicist working in the Department of Orthopedics at the University of Seattle in the early 1990s. Recognizing the need for a method to investigate the structure of biological molecules on the atomic scale, Sidles drew inspiration from the invention of the atomic force microscope a few years earlier. He proposed a technique based on force detection \cite{Sidles1991}. Experimental proof-of-principle studies soon followed in the group of Dan Rugar at IBM Almaden \cite{Rugar1992}, sparking interest beyond the nanomechanics community.

NanoMRI has evolved over the years to encompass a diverse array of experimental techniques, including magnetic probes based on nitrogen-vacancy color centers in diamond, magnetic scanning tunneling microscopy, and microscopic superconducting circuits. Each approach offers unique advantages and caters to specific applications. This diversity has fostered a vibrant and expanding NanoMRI community, whose original focus on structural biology has broadened to encompass new frontiers such as nanoscale nuclear magnetic resonance, quantum data storage, and atomic-scale surface investigations.

The roadmap article is organized as follows: section \ref{force} reviews magnetic resonance force microscopy and its offshoots, with perspectives to the future. Section \ref{NV} discusses color centers (the NV among them) as an emerging technique to explore magnetic resonance on these very small length scales. Section \ref{STM} presents scanning tunneling microscopy combined with electron spin resonance, a relatively new entrant to the field. In section \ref{SC}, a demonstration of the power of superconducting qubits is shown with an outlook to what’s next, and finally, section \ref{SEQ} looks at various control schemes and quantum-enhanced techniques that bring about significant improvements to spin detection at the nanoscale. Together, all five sections outline the current state-of-the-art, future challenges and opportunities in NanoMRI.
\newpage

\tableofcontents
\newpage

\section{Force Detection}
\label{force}
\subsection{Magnetic resonance force microscopy: from cantilevers to nanowires}

Martino Poggio\\
Department of Physics \& Swiss Nanoscience Institute, University of Basel, 4056 Basel, Switzerland\\
martino.poggio@unibas.ch

\subsubsection*{\textbf{Status}}
The invention of the scanning tunneling microscope (STM) in the early 1980s and its use in making the first images of individual surface atoms cleared the way for a new type of microscopy, in which a sharp tip is scanned over a sample surface. By using a variety of probes to measure different quantities, nanometer-scale scanning probe microscopy was then extended to a wide variety of contrasts and surfaces, beyond measuring tunneling current on conducting materials. Atomic force microscopy (AFM) is perhaps the most widely-used of these techniques, with applications spanning across physics, chemistry, biology, and materials science. Functionalization of AFM probes led to further refinements, including the possibility to measure electric and magnetic fields via, for example, electrostatic or magnetic force microscopy (MFM). 

The application of these concepts to magnetic resonance imaging (MRI) is known as magnetic resonance force microscopy (MRFM) \cite{Poggio2018}. Proposed by Sidles \cite{Sidles1991, Sidles1992} in the early 1990s as a means to image individual molecules atom-by-atom, MRFM relies on the measurement of the minute force between a magnetic scanning probe and resonantly flipping sample spins. Spatial maps of this force can be transformed into 3D images of spin density with spatial resolution orders of magnitude better than conventional inductively detected MRI, albeit not yet at the atomic scale.  The reason for this advantage – and the key to further improvement – is the exquisite sensitivity of the force transducers that are used. 

Today, these force transducers are the direct descendants of the metal foils and Si cantilevers used in early AFM. They are designed to optimize sensitivity to small forces by minimizing thermomechanical noise, which, at a given temperature $T$, depends on mechanical dissipation $\Gamma$. Since smaller mechanical transducers tend to have less dissipation \cite{Braakman2019}, there has been a notable trend toward ever tinier transducers, ranging from ultra-thin cantilevers to nanowires (NWs) and even nanotubes.  Smaller transducers also tend to have higher mechanical resonance frequencies \cite{Braakman2019}, decoupling them from common sources of noise, such as external vibrations or the force noise caused by electronic fluctuators as a probe approaches a surface (non-contact friction).  

Remarkable progress has been made in improving force sensitivity, which in the classical limit is set by the fluctuation-dissipation theorem and given by:
$$ 
    F_\mathrm{min} = \sqrt{4k_BT \Gamma},
$$
where $k_B$ is the Boltzmann constant \cite{Mamin2001}. Starting in 2004, ultra-thin Si and diamond cantilevers designed for MRFM achieved around $F_\mathrm{min} = 1$ aN/(Hz)$^{1/2}$ sensitivity \cite{Chui}, followed by ‘nanoladder’ levers of the same materials with under 200 zN/(Hz)$^{1/2}$ \cite{Heritier2018}, and doubly clamped carbon nanotube (CNT) resonators with 1 zN/(Hz)$^{1/2}$ \cite{Moser2014}. Doubly clamped transducers like CNTs, strings, and membranes, however, are not easily amenable to the protruding tip geometry of SPM. For this reason, the demonstration of NW cantilevers with 1 aN/Hz$^{1/2}$ sensitivity has proven particularly promising \cite{Nichol2012}. When arranged in the pendulum geometry – that is with their long axes perpendicular to the sample surface – NW cantilevers are ideal scanning probes that avoid snapping into contact when close to a surface.  By virtue of their high frequencies and small size, NWs experience orders of magnitude lower non-contact friction than larger cantilevers with lower resonance frequencies, making their near-surface force sensitivities much better \cite{Nichol2012}. It is no accident that since the 2009 demonstration of MRI with sub-10-nm spatial resolution via cantilever MRFM \cite{Degen2009}, MRI with similar resolution has only been achieved in 2013 using a NW transducer \cite{Nichol2013}. 

\begin{figure}
    \includegraphics[width = \textwidth]{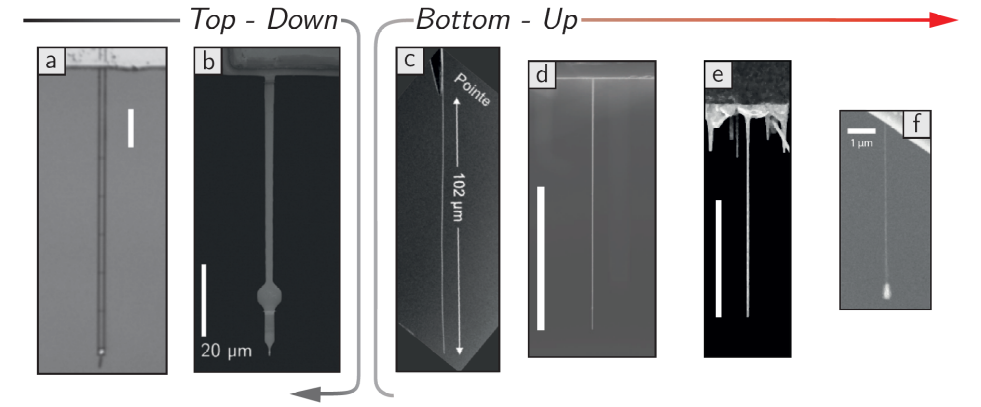}
    \caption{Reducing transducer size. Mechanical sensors made by `top-down' fabrication methods include (a) diamond `nanoladder' cantilevers \cite{Heritier2018} (scale bar 20\,$\upmu$m) and (b) undoped Si cantilevers \cite{Chui} Smaller transducers are typically made by `bottom-up' or self-assembly techniques, such as (c) SiC NWs \cite{Gloppe2014}, (d) Si NWs \cite{Nichol2012, Nichol2013}, (scale bar 10\,$\upmu$m), (e) GaAs NWs (scale bar 10\,$\upmu$m) \cite{Rossi2019}, and (f) CNTs \cite{Tavernarakis2018}.}
\end{figure}

\subsubsection*{\textbf{Current and Future Challenges}}
Since the first demonstration of MRFM in the early 1990s, sensitivity has improved from the equivalent of $10^9$ to about $10^2$ $^1$H magnetic moments. In order to detect individual nuclear spins and to image molecules with atomic resolution, at least another 2 orders of magnitude are required. While such a boost could come from refinements to a number of components, including the magnetic tip or the measurement protocol, improvement of the mechanical transducer, i.e., a reduction of mechanical dissipation $\Gamma$ is likely to be the most impactful approach.

The central challenge will be to develop ever tinier cantilever transducers, using NWs, CNTs, or similar structures, with high mechanical quality, whose displacement can be measured and controlled. This poses a number of technical obstacles. First, while displacement detection of conventional AFM cantilevers and micron-scale transducers can be achieved via the deflection of a reflected laser beam, such schemes are poorly suited to transducers with sub-wavelength dimensions. Without a scheme sensitive enough to resolve a transducer’s thermal motion, forces cannot be detected down to its ultimate sensitivity limits. Second, as a transducer shrinks in size, its linear dynamic range decreases. Strategies for dealing with this reduction, for maintaining easy control of the transducer’s motion, or for exploiting nonlinear effects in force measurements have therefore to be developed. Third, any remaining sources of mechanical dissipation have to be addressed and reduced as much as possible. The most prominent sources include dissipation related to the clamping of cantilever transducer, i.e., the path through which energy is lost to the support structure, and dissipation related to the transducer’s surface. As a mechanical resonator is shrunk down, its surface-to-volume ratio increases, resulting in a large contribution of surface defects or contamination to overall mechanical dissipation. In fact, for a 100-nm-thick Si cantilever, it was shown that 85\% of the mechanical dissipation can be traced to the 2-nm thick native oxide layer on its surface \cite{Tao2015}. 

Further challenges are related to the functionalization of increasingly small cantilever transducers. As recently as 2004, the magnetic tip for the MRFM experiment demonstrating the detection of a single electron spin was broken off of a macroscopic magnet and glued to the cantilever using micromanipulators under an optical microscope \cite{Rugar2004}. The tip was subsequently milled using a focused ion beam. Such techniques are not likely to scale for the placement of magnets or samples on NW and CNT devices, requiring new strategies.

\subsubsection*{\textbf{Advances in Science and Technology to Meet Challenges}}
A number of advances have to be made to meet the aforementioned challenges. For the detection of mechanical displacement, a pair of optical techniques based on interferometry or scattering have been shown to work with NWs and to be compatible with SPM applications \cite{Rossi2016, Lepinay2016}. Despite the sub-wavelength diameter of the NWs, in both cases, sufficient light is scattered to sensitively detect displacement. Whether such schemes will continue to work with smaller NW and CNT cantilevers, remains an open question. A technique based on detecting the inelastically scattered electrons from a focused electron beam has also been demonstrated to work on a CNT transducer in an SPM setup \cite{Siria2017}. Another very promising technique, combining a conventional AFM with a NW or CNT transducer such that their modes hybridize, allows the mechanical displacement to be detected with the ease of an AFM and sensitivities to approach those only possible with much smaller transducers \cite{Reiche2015}.

Despite the potential for force amplification, nonlinear motion has so far not been exploited in experiments on force sensing with NW or CNT cantilevers. Work in this area and on making use of the various mechanical modes of NW and CNT cantilevers, such as in vectorial force detection \cite{Gloppe2014}, could yield important advances. Excitation, feedback, and control of cantilever displacement has improved considerably in recent years with the proliferation of all-optical mechanical driving via photothermal actuation. Compared to excitation via piezoelectric actuators, this technique allows for the isolation of the transducer’s mechanical modes without interference from spurious modes in the microscope. This, in turn, allows for the application of feedback and other control required to suppress unwanted nonlinearities. 

To reduce mechanical dissipation in NW or CNT cantilevers, losses related to be clamping should be addressed. Improvements may come from implementing phononic band-gap supports. These schemes have already improved clamping losses in doubly clamped strings and membranes, but have yet to be implemented on cantilever scanning probes. Surface-related losses could be mitigated by the removal of amorphous layers and/or surface passivation, as was already demonstrated in larger Si cantilevers. A complete quenching of surface losses may only be possible by operating transducers under ultra-high vacuum conditions, as in STM where even atomically thin contaminant layers can be removed.

Strategies for the functionalization of the cantilever tip with a nanometer-scale magnet are also being developed. In one case, a NW transducer grown by molecular beam epitaxy was terminated, during growth, with a magnetic MnAs tip \cite{Rossi2019}. In subsequent NW MFM experiments, this tip proved to have excellent properties. Focused ion or electron beam induced deposition has also been used to deposit nanometer-scale Co directly on the tips of NWs. The magnetic quality of these tips is yet to be determined, but initial results appear promising \cite{Mattiat2020}.

\subsubsection*{\textbf{Concluding Remarks}}
In addition to potentially improving the resolution of mechanically detected MRI, the development of NW cantilevers designed for ultra-sensitive force detection may have impact in other applications. Unlike conventional AFM cantilevers that are designed to match the high stiffness required for atomic-resolution AFM, these exceptionally compliant probes allow for sensitive mapping of tip-sample force fields and energy losses. Since energy dissipation plays a central role in the breakdown of topological protection, it may provide important contrast in spatial studies of strongly correlated states in 2D vdW materials. In fact, dissipation contrast using cantilevers in the pendulum geometry has already been used to observe superconducting \cite{Kisiel2011} and bulk structural phase transitions \cite{Kisiel2015}, as well as the local density of states. In addition, the combination of high force sensitivity, high spatial resolution, and low invasiveness of magnetic NW probes also has the potential to expand the applicability of MFM to much more subtle contrast than is conventionally accessible. Sensitive MFM could be particularly useful for measuring the spatial localization of flowing currents, as in edge states, and for the determination of length scales such as magnetic domain sizes and coherence lengths \cite{Marchiori2021}. In particular, NW transducers should allow MFM imaging of the tiny current densities and magnetization associated with many emerging 2D materials and their heterostructures.
  
\subsubsection*{Acknowledgements}
MP acknowledges support from the Canton Aargau.
\label{Poggio}
\newpage

\subsection{Force-detected NanoMRI with high-Q resonators}

Alexander Eichler and Christian L. Degen\\
Institute for Solid State Physics, ETH Zurich, Otto-Stern-Weg 1, 8093 Zurich\\
eichlera@ethz.ch, degenc@ethz.ch

\subsubsection*{\textbf{Status}}
Over the past decade, we witnessed a leap in the ability to design and fabricate nanomechanical systems from silicon nitride with quality factors beyond $Q=10^8$. The main innovation on this path was a technique commonly referred to as “dissipation dilution”: applying tensile strain to a mechanical resonator increases its resonance frequency $f_0$ without affecting the dissipation rate $\Gamma=2\pi f_0/Q$, hence increasing $Q$ \cite{Unterreithmeier2010, Schmid2011}. Later, the implementation of mode shape engineering led to further significant improvements in $\Gamma$ and $Q$ \cite{Tsaturyan2017, Ghadimi2018, Gisler2022, Bereyhi2022} (see \cite{Eichler2022} for a more comprehensive list of references). These developments were primarily fuelled by the need for quantum devices with long thermal coherence times and high $Qf_0$ products, but the resulting devices also open up exciting perspectives in force sensing.

The sensitivity of force sensors is typically limited by their thermomechanical force noise power spectral density $S_f = 4k_BTm\Gamma$, where $k_B$ is Boltzmann’s constant, $T$ is the temperature, and $m$ is the effective resonator mass. In most cases, $S_f$ is reduced by designing resonators with low masses and resonance frequencies, resulting in long and floppy devices (see section \ref{Poggio}). The new generation of strained silicon nitride force sensors has the potential to reach similar or better performances while operating at significantly higher frequencies. This change comes with several advantages: first, it is expected that detrimental $1/f$ fluctuation effects, such as non-contact friction, have a smaller impact at higher frequencies, preserving the resonator performance in the presence of nearby surfaces \cite{Heritier2021}. Second, static bending will be reduced due to the higher spring constant $k=4m\pi^2f_0^2$, making the system more stable towards bending and mechanical instability \cite{Krass2022}. Third, silicon nitride resonators can be integrated into cavity optomechanical systems to achieve quantum-limited displacement readout and mode cooling \cite{Aspelmeyer2014}. In total, optomechanical force sensors with high quality factors could therefore become a valuable resource for ultrasensitive scanning force microscopy, and in particular for nanoscale MRI \cite{Sidles1991, Poggio2010}.

The design of nanomechanical resonators remains an active research area. The community is only just beginning to study the behavior of silicon nitride optomechanical resonators at temperatures below 100 mK, and to discover the opportunities offered by on-chip integrated devices. Further improvements of the figures of merit and more sophisticated measurement protocols will support the development of spin detection and other state-of-the-art scanning force applications.

\begin{figure}[h]
    \includegraphics[width = \textwidth]{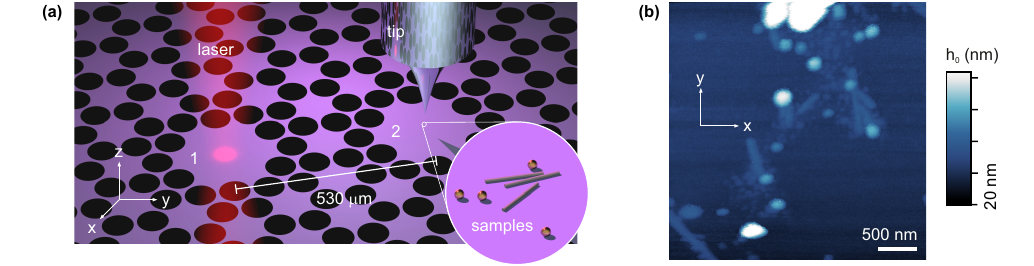}
    \caption{Demonstration of scanning force microscopy with a high-Q optomechanical membrane system. (a) Artistic rendering of a scanning force microscopy setup using a membrane resonator as force sensor. Black holes in the membrane surface establish a phononic bandgap that bestows high quality factors upon two drum resonators (1 and 2). Samples are placed on one of the resonators and interact with a sharp, static scanning tip. The resonators form normal modes through evanescent coupling, such that the sample-tip interactions taking place on resonator 2 can be monitored with a laser interferometer on resonator 1. (b) Topography map of gold nanoparticles and tobacco mosaic viruses on the membrane surface measured with the scanning force microscope. Reprinted graphs with permission from \cite{Haelg2021}. Copyright 2022 by the American Physical Society.}
    \label{Eichler:fig1}
\end{figure}

\subsubsection*{\textbf{Current and Future Challenges}}
While strained silicon nitride resonators exhibit excellent nominal force sensitivities, their use in real applications is not straightforward. Some of the issues that we foresee are connected to the requirement to work inside a dilution refrigerator or a different cryogenic apparatus to reduce $S_f$ as much as possible.  The most pressing experimental challenges are:

(i) in order to enforce strain along the axis of a beam or in the plane of a membrane, the device must be clamped on opposite ends, making it impossible to employ the typical singly-clamped cantilever geometry that is found in almost all scanning force instruments. This leads to the question of how the scanning tip, the sample, and the readout mechanism can be arranged inside a scanning force microscope. Such geometrical incompatibilities become more pronounced when placing the resonator inside an optical cavity, such as is nowadays routinely done for membranes or trampoline devices to enhance the readout efficiency.

(ii) It remains to be shown how far the force noise of a sensor in a realistic application coincides with the values found for bare, isolated resonators. In particular, non-contact friction and heating due to absorption of light will at some point become noticeable and limit the sensitivity. It is likely that the final performance that can be achieved in sensing instruments will result from a compromise between optical heating, detector noise, and quantum backaction. Further research is necessary to evaluate e.g. the optical absorption and the heat conductivity of silicon nitride at dilution refrigerator temperatures.

(iii) For the special case of nanoscale MRI, there arises also a challenge related to spin sensing protocols. Strained silicon nitride resonators with resonance frequencies in the MHz regime cannot be coupled to nuclear spins in the same way as is typically done with kHz cantilevers. Specifically, established nuclear spin inversion protocols cannot be applied with MHz repetition rates due to the limited amplitude of the AC magnetic fields available in dilution refrigerators. Note that for electron spins, the issue is easier to overcome due to the much higher gyromagnetic ratio, which allows for MHz Rabi oscillations with relatively small AC field amplitudes.

\subsubsection*{\textbf{Advances in Science and Technology to Meet Challenges}}
Significant progress towards nanoscale MRI with silicon nitride resonators was accomplished in the past few years. First measurements of macroscopic electron spin ensembles with membrane and trampoline resonators were published in 2016 \cite{Scozzaro2016} and 2019 \cite{Fischer2019}, motivating further work in this new direction (see also section \ref{Polzik} for related experiments). In 2021, an important step towards nanoscale applications was reached with the first demonstration of a scanning force microscope using a silicon nitride membrane resonator as a sensor \cite{Haelg2021}. In order to overcome the geometry issue arising from the clamping on all sides, the microscope was defined in an inverted fashion relative to a typical atomic force microscope: the membrane served at the same time as the sample plate and as the vibrating sensor, see Fig.\,\ref{Eichler:fig1}(a). The samples then interacted with a sharp, but static, scanning tip. This offers a general solution for challenge (i), as the ``inverted scanning approach'' can be translated to trampoline and string resonators as well. The experiment also showed that the noncontact friction at tip-surface distances of a few nanometers is smaller than with traditional cantilever devices, providing a partial answer to the challenge (ii).

On-chip integration of the optical readout \cite{Gavartin2012} is expected to increase the optomechanical coupling strength, thereby reducing the optical power required for sensitive readout. However, it remains to be demonstrated to what level the absorptive heating can be lowered relative to interferometer-based approaches \cite{Gisler2022}. Challenge (ii) is thus likely to remain a topic of research for some years to come.

In order to overcome challenge (iii), various protocols have been proposed and demonstrated to allow spin-mechanics coupling at MHz frequencies. One technique that was invented in the context of nanoscale MRI with bottom-up grown nanowires is based on a modulation of the magnetic gradient fields \cite{Nichol2012} (see section \ref{Budakian} for details). Other ideas include parametric coupling between various modes \cite{Dougherty1996, Kosata2020} and resonant coupling between the mechanical resonator and spin ensembles \cite{Sidles1992a}. Which approach proves most useful in the future will depend on their respective practical requirements and technical limitations.

\subsubsection*{\textbf{Concluding Remarks}}
The field of nanoscale MRI with force sensors enjoys a rejuvenation with the advent of strained silicon nitride resonators with ultrahigh quality factors. The evolution from traditional cantilever devices in the kHz range to these new sensors in the MHz range offer several advantages but also leads to challenges in the setup geometry and the measurement protocols. These challenges are being tackled with promising approaches at the interface between the fields of optomechanics and nanomechanical spin sensing.

\subsubsection*{Acknowledgements}
We would like to acknowledge the Swiss National Science Foundation (SNSF) through the National Center of Competence in Research in Quantum Science and Technology (NCCR QSIT), the Sinergia grant (CRSII5 177198/1), and the project Grant 200020-17886, as well as the ETH Zurich through the Research Grant (ETH-03 16-1).
\newpage

\subsection{Magnetic resonance force microscopy using current-generated gradients}
\label{Budakian}
Sahand Tabatabaei and Raffi Budakian\\
Department of Physics and Astronomy, University of Waterloo, Canada\\
Institute for Quantum Computing, University of Waterloo, Canada\\
ss4tabat@uwaterloo.ca, rbudakian@uwaterloo.ca

\subsubsection*{\textbf{Status}}
Magnetic resonance force microscopy (MRFM) was originally proposed as a technique for 3D imaging of macromolecules utilizing the chemical specificity, and sub-surface imaging capabilities of magnetic resonance imaging (MRI) \cite{Sidles1992, Sidles1995}. A unique feature of MRFM is the ability to detect spins in volumes on the (100 nm)$^3$ scale with high sensitivity. This capability is particularly important for studying a large class of biological systems, such as protein complexes and virus particles, that possess 3D structures on length scales below 100 nm. Since the introduction of the initial concept, MRFM has witnessed significant technical advances, that have given rise to such notable achievements as the detection of single electron spins of defect centers in SiO$_2$ \cite{Rugar2004}, 3D imaging of proton spins in tobacco mosaic virus
particles with 4 nm resolution \cite{Degen2009}, and sub-angstrom precision nuclear magnetic resonance (NMR) diffraction measurements of $^{31}$P spin ensembles in InP \cite{Haas2022}.

In MRFM, the spins within the detection volume are placed in a magnetic field gradient, which produces a force that is detected using an ultrasensitive nanomechanical oscillator. The interaction with the spin is typically mediated using a series of radio-frequency (RF) pulses which modulate the longitudinal spin component to create a force at the resonance frequency of the oscillator. In most MRFM experiments, a small ferromagnetic particle placed near the sample produces the magnetic field gradient required for spin detection, while also confining the interaction with the sample to a thin ``resonance slice'' that can be swept through the detection volume to yield a volumetric data set of the spin density.

\begin{figure}
    \includegraphics[width = \textwidth]{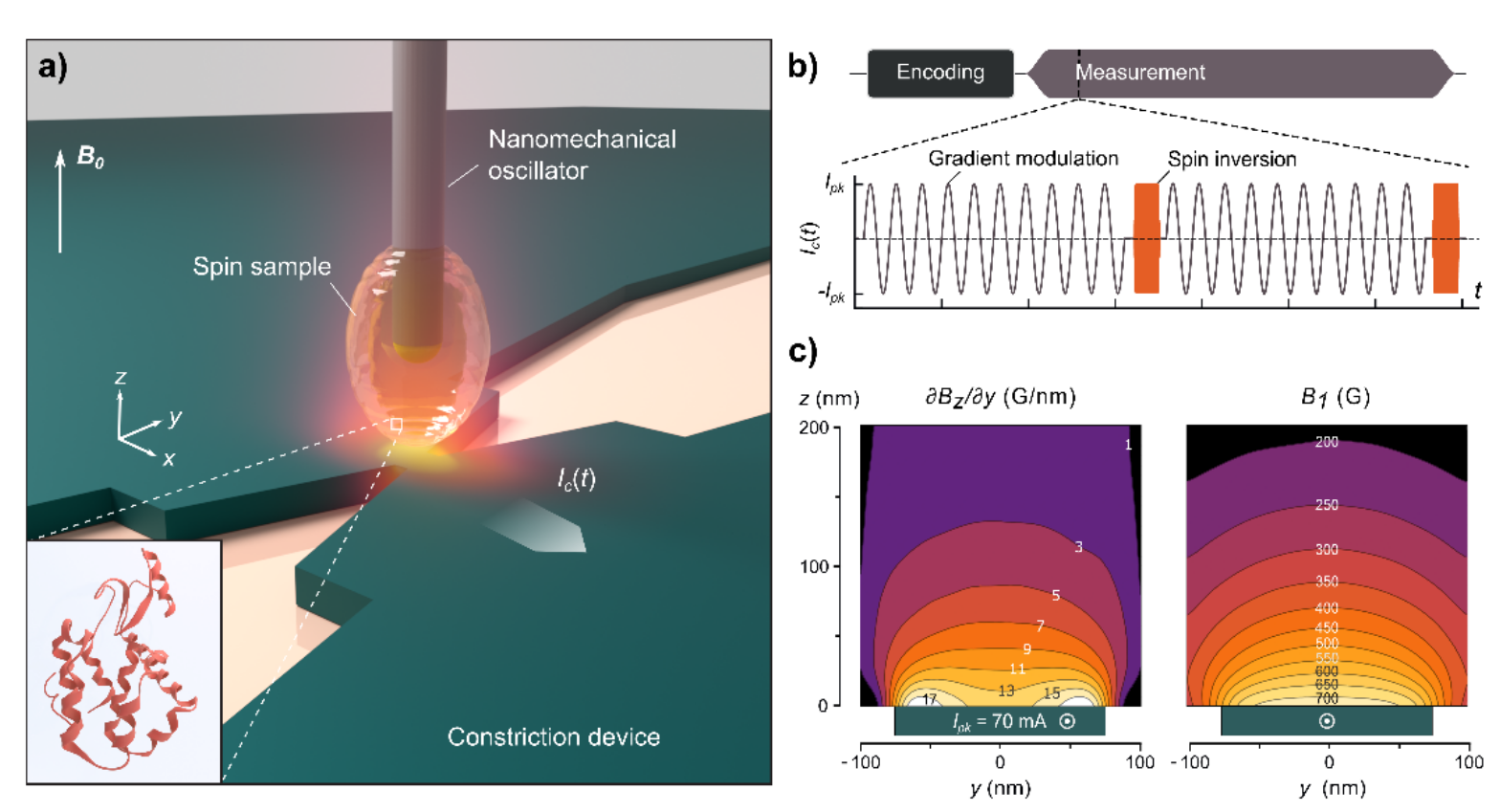}
    \caption{(a) Experimental setup for current-driven MRFM. The sample depicted here as protein molecules in a cryoprotectant – is attached
to the tip of the nanomechanical oscillator, and positioned above a narrow metallic constriction. The intense field gradients generated by flowing current through the constriction couple the spins to the oscillator, and the resulting motion is detected using optical interferometry. (b) Typical experimental protocol in current-driven MRFM, consisting of an encoding sequence – used, e.g., for Fourier encoding – and a measurement sequence in which the longitudinal component of the spins is coupled to the mechanical oscillator. (c) Contour plot of the readout gradient and $B_1$ field along the cross-section of a representative device, with the Rabi field $B_1$ being the circularly-polarized component of the RF field used for spin control. The plots correspond to a peak current of $I_{pk} = 70$ mA flowing through the constriction.}
    \label{Budakian:fig1}
\end{figure}

An alternative approach utilizes electrical currents that flow through a narrow metallic constriction to produce both the time-dependent magnetic fields at the Larmor frequency required for spin manipulation, and the magnetic field gradients for MRFM detection \cite{Nichol2012, Nichol2013} (Fig.\,\ref{Budakian:fig1}). In this approach, a particular MRFM experiment is divided into an encoding sequence, during which the detection gradient is turned off and the desired spin control sequence is applied, followed by the measurement sequence during which the longitudinal spin component is detected. The measurement sequence
consists of the application of a time-dependent magnetic field gradient at the resonance frequency of the oscillator, the phase of which is periodically inverted to cancel any electrical forces on resonance (Fig.\,\ref{Budakian:fig1}) \cite{Rose2018}. The spin orientation is also inverted with each phase flip using an RF pulse – typically an adiabatic full passage – to generate a spin force on resonance. The associated displacement of the oscillator is detected using optical interferometry. The ability to separate the spin control and detection sequences enables a great deal of flexibility in the kinds of quantum control sequences that can be incorporated into MRFM measurements. In the following sections, we outline the current challenges and advancements in current-driven MRFM.

\subsubsection*{\textbf{Current and Future Challenges}}
The key technical challenge to achieving high resolution molecular-scale imaging with MRFM is improving the detection signal-to-noise ratio (SNR). In an MRFM experiment, the SNR is primarily determined by the 1) spin polarization in the sample volume, 2) thermal noise of the oscillator, 3)
magnitude of the readout field gradient and 4) force noise caused by electric field fluctuations near the surface. As discussed in the next section, significant advancements have been made in each area, however integrating these approaches into a single MRFM setup capable of angstrom-resolution imaging of $\sim$ (100 nm)$^3$ volumes remains a challenge.

The total measurement time required for constructing an image depends on the measurement protocol. Reciprocal space techniques that utilize data multiplexing, such as Hadamard \cite{Eberhardt2007} and Fourier encoding \cite{Nichol2013} are routinely used in conventional MRI, providing an efficient means of imaging. In current-driven MRFM, spatial encoding of spins is achieved by varying the static and RF field gradients, in the same way as conventional MRI. While 1D and 2D imaging have been demonstrated using current-driven MRFM \cite{Nichol2013, Rose2018}, the highly non-uniform field gradients generated by the constriction complicates extraction of the real-space spin density. Furthermore,  generating the encoding gradients for 3D Fourier imaging is not possible using a single constriction. To overcome these challenges, in the following section, we present a new design for a gradient source capable of applying spatially uniform field gradients in 3D in a $\sim$(100 nm)$^3$ volume.

Another important consideration for phase encoding techniques is the spin coherence time, which determines the maximum encoding wavevector that can be realized. Although the dephasing time for most organic solids tends to be quite short–typically of order $T_2 \sim 20\,\mathrm{\mu s}$ – dynamical decoupling (DD) sequences can be used to extend the coherence time, thereby increasing the image resolution that can be achieved. The highly inhomogeneous magnetic field generated by the constriction presents a challenge to applying global rotation pulses – the basic building blocks of these sequences – and potentially limits the set of NMR sequences, including DD sequences, that can be realized. Given the rich repertoire of spin control techniques in NMR, this also limits MRFM’s potential use as a tool for spectroscopy and studying spin dynamics in nanoscale materials. In the next section, we discuss the implementation of numerical pulse engineering methods that have been successfully used to realize high-fidelity coherent spin control, thereby overcoming the challenges posed by the field inhomogeneity.

\subsubsection*{\textbf{Advances in Science and Technology to Meet Challenges}}
Over the years, major developments have been made to improve detection sensitivity in MRFM. These include polarization enhancement using cross polarization \cite{Eberhardt2007a} and dynamic nuclear polarization \cite{Issac2016}, achieving readout field gradients as high as 10$^7$ T/m using disk write heads \cite{Tao2016} and  engineering various ultrasensitive nanomechanical force sensors, e.g., lithographically-made cantilevers \cite{Heritier2018}, nanowires \cite{Sahafi2019} and carbon nanotubes \cite{Moser2013}. Fluctuating electric fields near the surface can give rise to noncontact friction that can degrade detection sensitivity \cite{Stipe2001}. Approaches proposed to mitigate this include chemically treating the surfaces and utilizing sharper tips \cite{Tao2015a}.

The integration of MRFM and high-fidelity quantum control schemes is creating powerful new approaches in NanoMRI. Numerical pulse optimization is a well-established technique in quantum information science for designing control pulses in demanding experimental conditions, much like
those encountered in MRFM. The large field inhomogeneities of MRFM can be readily accounted for in existing numerical schemes, which utilize a gradient-descent algorithm to optimize the control waveform over a range of field values, while also constraining the optimization to adhere to the
amplitude and bandwidth limitations of the experimental setup \cite{Haas2019, Tabatabaei2021}. Furthermore, undesirable spin Hamiltonians such as chemical shifts and spin-spin interactions can be efficiently accounted for by minimizing corresponding perturbative corrections to the spin propagator. Utilizing such schemes has been essential for recent advancements in current-driven MRFM. 1D Fourier transform imaging of $^1$H spins with $\sim 2$ nm resolution was achieved using DD sequences that employ numerically optimized $\pi$/2 pulses \cite{Rose2018}. Recently, numerically engineered control pulses were used to dynamically decouple $^{31}$P spins in an InP nanocrystal and encode a z-axis magnetization grating with a $\sim 4$ \AA\, modulation wavelength. An NMR diffraction-based experiment using the encoded grating enabled the detection of an angstrom-scale displacement of the InP sample with a precision of 0.07 \AA\,\cite{Haas2022}.

\begin{figure}[h]
    \centering
    \includegraphics[width = \textwidth]{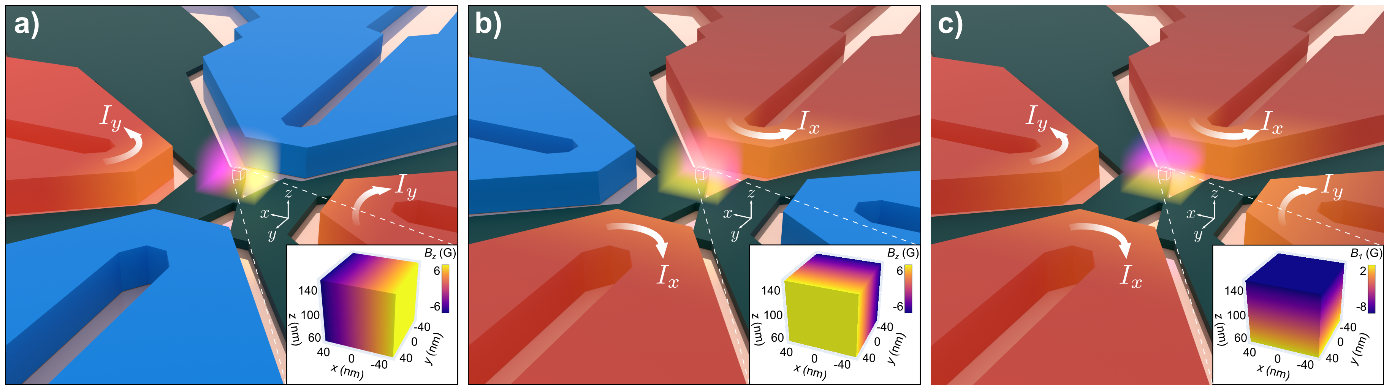}
    \caption{Next-generation gradient source with four additional current paths for generating uniform field gradients in three dimensions.
The subfigures show different current configurations for encoding along (a) $x$ (uniform $\partial B_z/\partial x$), (b) $y$ (uniform $\partial B_z/\partial y$), and (c) $z$
(uniform $\partial B_1/\partial z$). Each inset shows the calculated field distribution for 150 mA peak current in a (100 nm)$^3$ volume centered above the constriction, derived using finite-element simulations. The simulations indicate that over this volume, the relevant gradient in each
configuration deviates by less than 0.5\% relative to its average value, which was computed to be 1.35 $\times$ 10$^4$ T/m in all three cases.}
    \label{Budakian:fig2}
\end{figure}

Achieving 3D Fourier transform imaging in current-driven MRFM will require devices capable of generating uniform field gradients in all three dimensions. Upcoming experiments will utilize a new design (Fig.\,\ref{Budakian:fig2}) that incorporates four additional current paths around the central constriction for generating highly uniform encoding gradients over a $\sim$ (100 nm)$^3$ volume above the central constriction. Figs.\,\ref{Budakian:fig2}(a, b) show the static currents $I_y$ $(I_x)$ used to produce uniform resonance offset field gradients $\partial B_z/\partial x$ $(\partial B_z/\partial y)$ for phase encoding in the $x$ ($y$) directions. To encode in the $z$ direction, an RF pulse at the Larmor frequency is applied to both gradient wire pairs with a $\pi$/2 phase shift between $I_x$ and $I_y$, (Fig.\,\ref{Budakian:fig2}(c)), which results in a circularly polarized field with a uniform Rabi field gradient $\partial B_1/\partial z$ in the rotating frame.

\subsubsection*{\textbf{Concluding Remarks}}
Over the three decades since its inception, MRFM has proven to be a prominent method for imaging and characterizing materials on the nanometer scale. However, significant work is still needed to achieve the sensitivity required for imaging relevant biological systems with atomic resolution. Among the various MRFM approaches, the current-driven modality has been shown to be particularly adept at incorporating coherent spin control by utilizing numerical pulse engineering methods. These methods have not only been essential in extending the coherence time and hence resolution of Fourier imaging in this modality, but have also created great potential for bringing NMR’s 70-year-long history of spin control techniques to the nanometer scale. As it stands, the missing ingredient in current driven MRFM is the ability to do full 3D phase encoding with uniform gradients. With next-generation gradient source devices demonstrating such capabilities, the combination of large uniform field
gradients with numerical spin control is expected to enable 3D Fourier imaging, as well as NMR diffraction on periodic spin systems such as protein nanocrystals. This could additionally open up new avenues for conducting spatially-resolved spectroscopy and studying spin transport on the atomic length scale.

\subsubsection*{Acknowledgements}
This work was undertaken due in part to funding from the US Army Research Office through Grant W911NF1610199, the Canada First Research Excellence Fund (CFREF), and the Natural Sciences and Engineering Research Council of Canada. The University of Waterloo’s Quantum-Nano Fabrication and Characterization Facility (QNFCF) was used for this work. This infrastructure would not be possible without the significant contributions of Transformative Quantum Technologies (CFREF-TQT), Canada Foundation for Innovation (CFI), Industry Canada, the Ontario Ministry of Research and Innovation, and Mike and Ophelia Lazaridis. Their support is gratefully acknowledged.
\newpage

\subsection{Nanoscale ferromagnetic resonance imaging using high-sensitivity mechanical force detection}
Inhee Lee and P. Chris Hammel \\
Department of Physics, The Ohio State University, Columbus, OH 43210, USA\\
lee.2338@osu.edu, hammel@physics.osu.edu

\subsubsection*{\textbf{Status}}
\begin{figure}[h]
    \centering
    \includegraphics[width = 0.5\textwidth]{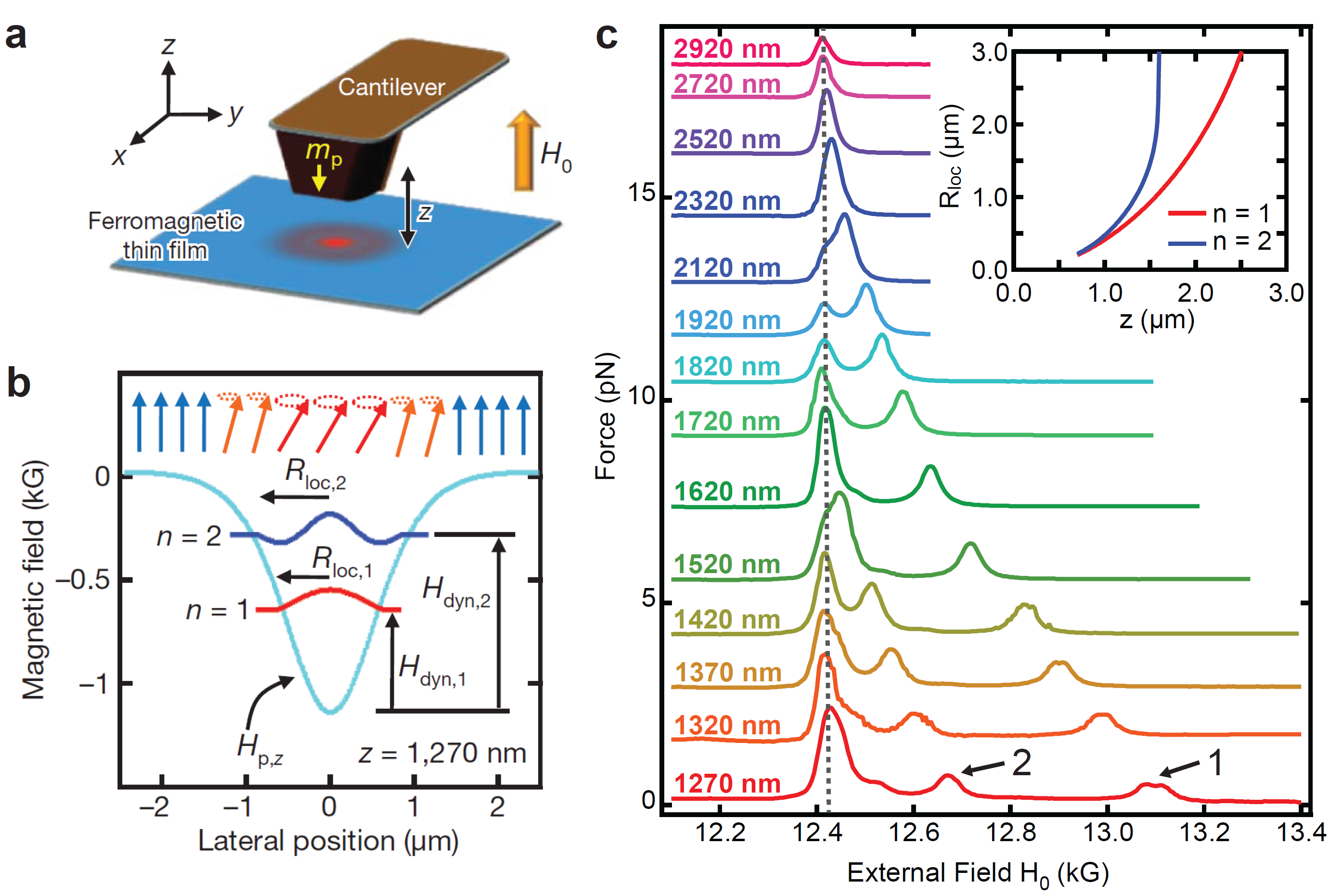}
    \caption{(a) FMRFM schematic. The moment $m_p$ of the micromagnetic probe is antiparallel to the external magnetic field $H_0$ to create localized FMR modes and is separated from the film by a distance $z$ which is the distance from the sample surface to the center of the probe magnet. (b) The micromagnetic probe creates a well of magnetic field, shown by the solid light-blue line, that localizes spin-wave excitations, indicated by the solid red and blue lines for the first two modes. The probe field $H_p$, the dynamic field $H_\mathrm{dyn}$ and the mode amplitudes are calculated for $z$ = 1,270 nm. (c) FMRFM spectra of a continuous thin film for the indicated values of $z$. The vertical dotted line shows the resonance field for the uniform FMR mode which is independent of $z$. The first and second confined modes indicated by the arrows shift to the higher field as $z$ decreases. Inset: Calculated local-mode radius $R_\mathrm{loc,n}$ versus $z$ for the first two magnetostatic modes obtained using the variational method. Adapted with permission from \cite{Lee2010} Copyright 2010 Nature Publishing Group.}
    \label{Hammel:fig1}
\end{figure}

Ferromagnetic resonance (FMR) measures the interaction of a collection of ordered spins with its environment, allowing accurate measurement of internal fields, anisotropies and dipolar fields within a material or arising from interfaces. Ferromagnetic resonance force microscopy (FMRFM) is a spectroscopic imaging technique based on mechanical force detection with sensitivity of 1 – 100 attonewtons. FMRFM detects the spin excitation modes in FMR of a magnetic sample through cantilever oscillations driven by the dipole force between the micromagnetic particle on the cantilever and the suppressed longitudinal magnetization resulting from magnetization dynamics excited by FMR. This enables scanned, spectroscopic imaging capable of quantitatively studying microscopic magnetic properties of inhomogeneous thin films and complex magnetic devices, including buried structures in multi-component materials, at the nanoscale. 
The FMR modes observed in the early FMRFM reflect the global magnetic properties in which the collective modes of the entire sample are excited due to strong spin exchange interactions, so local magnetic properties are not spatially resolved \cite{Zhang1996, Wago1998, Urban2006, Mewes2006, Loubens2007, Obukhov2008, Klein2008}. The first FMR mode localization was demonstrated using a strong dipole field generated from a micromagnetic tip \cite{Lee2010, Lee2011} (Fig.\,\ref{Hammel:fig1}a and \ref{Hammel:fig1}b). Fig.\,\ref{Hammel:fig1}c shows FMR spectra measured at various probe-sample separations $z$ on a 40 nm thick permalloy film. Two localized FMR modes are spectroscopically separated, occurring at higher fields than the $z$-independent uniform FMR mode. A localized mode (LM) is confined to nanoscale area by the field well created by strong field of micromagnetic probe at small $z$ (see inset, Fig.\,\ref{Hammel:fig1}c). These LMs were used to image the spatial variation of the internal field inside a permalloy (Py) thin film and the inhomogeneous demagnetizing field of individual Py microdisks with lateral spatial resolution down to 200 nm. 
Since this demonstration, LM-FMRFM has been applied to image the internal field or damping changes of various heterostructures: interfaces separating two regions with different saturation magnetizations of a Py film modified by region-selective ion irradiation \cite{Du2015}, interfaces between two regions with different uniaxial magnetic anisotropy in a Y$_3$Fe$_5$O$_{12}$ (YIG) thin film induced by interfacial interactions coupling the YIG and a gold overlayer \cite{Wu2020}, perpendicular magnetic anisotropy and Gilbert damping enhancement regions in a YIG film due to the influence of few-layer WTe$_2$, a high spin-orbit coupling transition metal dichalcogenide, overlying YIG \cite{Wu2022}. Spin angular momentum transport from the LM across its well boundary was observed through the inverse proportionality of the Gilbert damping constant $\alpha$ of the LM on its radius $R_\mathrm{loc}$ (see $R_\mathrm{loc}$ in Fig.\,\ref{Hammel:fig1}(b)) \cite{Adur2014}.

\begin{figure}[h]
    \centering
    \includegraphics[width = 0.5\textwidth]{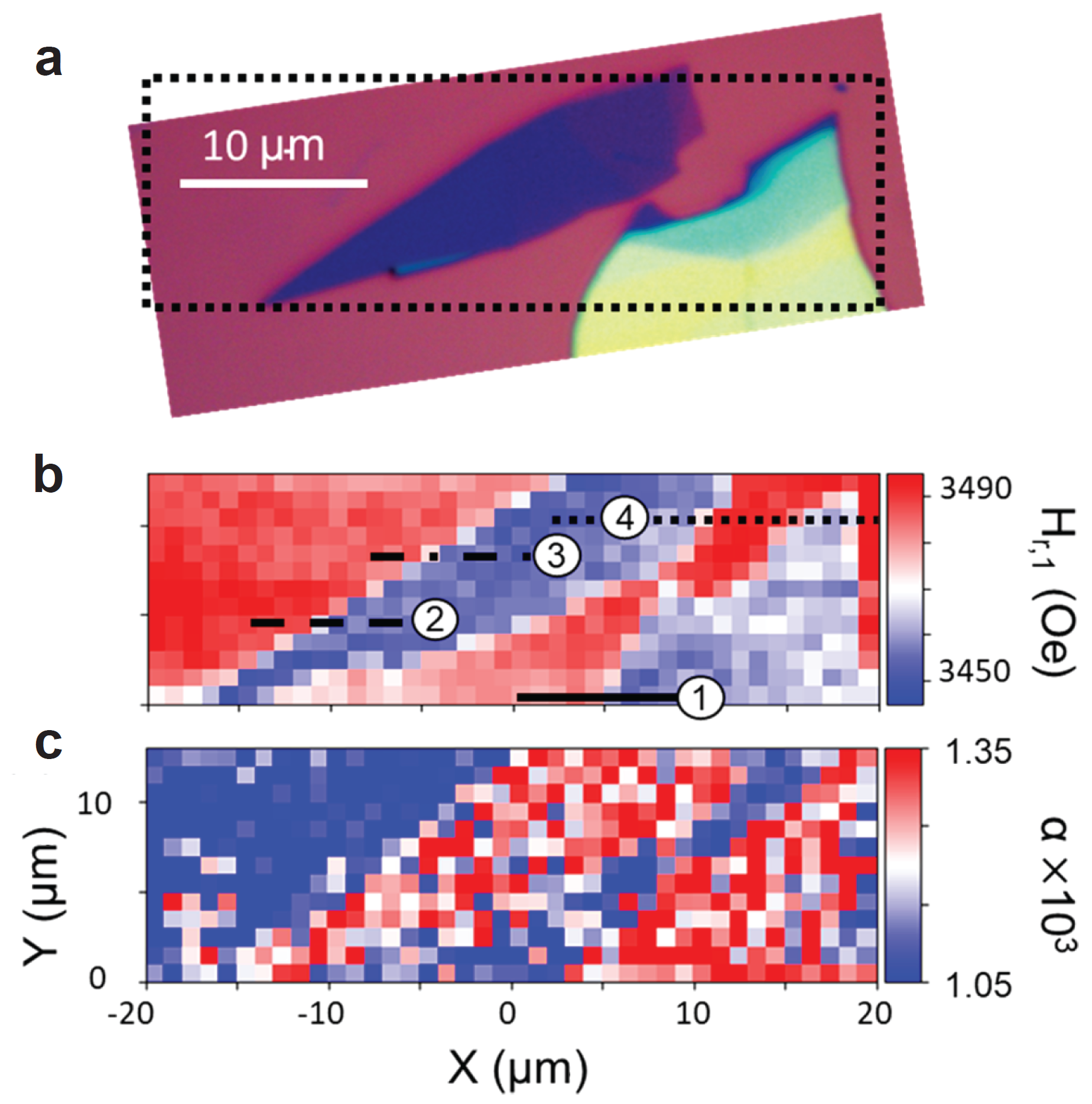}
    \caption{Two-dimensional FMRFM scan resolving the spatial variation of magnetic anisotropy and Gilbert damping of YIG in the presence of WTe$_2$ flakes. (a) Optical micrograph showing the color contrast of different thickness WTe$_2$ flakes (ranging from 4.7 to 44.8 nm). The black dashed box outlines the FMRFM scanned area for 2D mapping. (b) 2D map of the $n = 1$ localized mode resonance field. (c) 2D mapping of the Gilbert damping extracted from the $n = 1$ localized mode resonance peak amplitude. Reprinted with permission from \cite{Wu2022} Copyright 2022 American Chemical Society.}
    \label{Hammel:fig2}
\end{figure}
\subsubsection*{\textbf{Current and Future Challenges}}
Improving LM-FMRFM spatial resolution requires the minimization of the LM size and improvement of detection sensitivity through reduction of spurious forces on the cantilever.\\
\emph{Spatial resolution}: The ultimate spatial resolution is limited by the exchange length of the ferromagnet. Exchange imposes an energy cost for reducing $R_\mathrm{loc}$ which is balanced by the negative tip field and the positive dynamic field. Decreasing $z$ reduces $R_\mathrm{loc}$ but exchange sets a minimum LM wavelength and hence the smallest achievable $R_\mathrm{loc}$ allowed for localized spin wave excitation. \\
\emph{Spectroscopic precision vs. sensitivity}: A larger cantilever oscillation amplitude improves the accuracy of frequency detection measurements of FMRFM, but this degrades the spectroscopic precision. Large amplitude cantilever oscillations using a probe tip with high field gradient cause FMR peak broadening as the sample spins experience a wider range of static fields during oscillation.\\ 
\emph{Tip fabrication}: LMs require a tip moment $m_p$ antiparallel $H_0$ sufficient to saturate the film (Fig.\,\ref{Hammel:fig1}(a)) \cite{Lee2010, Lee2011}. Higher spatial resolution requires both a smaller tip and a larger field. Micron-scale high-coercivity ($> 2$ T) SmCo$_5$ probe tips were fabricated using focused ion beam (FIB). In principle, the highest coercivity is expected with very small domains. However, reducing tip dimensions to the nanoscale using FIB machining often degrades coercivity because the tip surface magnetization is impaired and the local pinning and domain nucleation that maintain high coercivity are reduced. Also, very small nanosized tips may be subject to quantum fluctuations of magnetic moments or energy states, which can affect the force detection sensitivity and spin relaxation in FMR.\\   
\emph{Tip characterization}: Accurate tip characterization is essential to know the LM spatial profile, which is necessary to convert spectroscopic maps into images of local magnetic properties. Modelling the tip as a dipole whose moment is obtained from cantilever magnetometry \cite{Banerjee2010} works well for large $z$, but accurate tip characterization at small $z$ is confounded by inadequate knowledge of the magnetization distribution, including the dead layer. Also, if the change in the sample’s internal field is large enough to be comparable to the intensity of the local tip field, it deforms the LM spatial profile, limiting its suitability for imaging \cite{Du2015, Wu2020}.\\ 
\emph{Reducing spurious coupling and surface noise}: To synchronize GHz-frequency magnetization dynamics with kHz-frequency cantilever oscillation, the microwave power amplitude is modulated at the cantilever frequency. The direct interaction of the microwaves with the cantilever produces a spurious force on the cantilever. Coulomb interactions between the bound charge on cantilever and sample surface can damp the cantilever oscillation, lowering the Q-factor and hence detection sensitivity.

\subsubsection*{\textbf{Advances in Science and Technology to Meet Challenges}}
Many efforts have been made to improve FMRFM performance through fabrication and application of various cantilevers, magnetic tips, microwires etc., to minimize spurious coupling and surface noise. Non-conductive materials are chosen to reduce spurious couplings, and deionization reduces surface charge. Diamond cantilevers exhibit smaller spurious coupling than commonly used silicon cantilevers \cite{Adur2014}. An iron filled carbon nanotube (FeCNT), a 10 – 40 nm diameter ferromagnetic nanowire enclosed in a protective carbon tube, has been studied for use as a FMRFM probe with high spatial resolution, high coercivity, and small surface noise. \cite{Banerjee2010}. Indeed, an FeCNT was successfully used as a scanning magnetic monopole probe at close $z$ with images of vortex and multi-domain structures in magnetic force microscopy (MFM) measurements \cite{Wolny2011}. The integrated cantilever and tip lithographically fabricated for custom use in the desired size, shape, and configuration is another good option for FMRFM probe. For example, an integrated magnetic tip consisting of a narrow magnetic nanorod suspended from the leading edge of a cantilever was lithographically fabricated and used in magnetic resonance force microscopy, which allows for large tip field gradient while minimizing surface noise \cite{Hickman2010}. Fabrication of the nanowire capable of generating a highly localized microwave field can also significantly reduce spurious coupling \cite{Haas2022}.     
In addition to the out-of-plane field geometry that requires a high coercive magnetic tip, FMR mode localization is also possible in the in-plane field geometry \cite{Chia2012}. In this configuration, the tip field and sample magnetization are aligned parallel to the applied magnetic field, but the local tip field experienced by the sample is antiparallel to the magnetization, which is possible even with a soft magnetic tip. Because of its ease, the in-plane field mode localization was later applied to the spin-torque FMR \cite{Zhang2017} and auto-oscillation \cite{Zhang2021} of LM, although the LM spatial profiles are not isotropic like those of out-of-plane field geometry. 
The development of micromagnetic simulations has greatly improved understanding of complex magnetic dynamics of FMR localized modes. Micromagnetic simulations show how spatial mode profiles are deformed by abrupt spatial variations in the internal field across a boundary in a magnetic heterostructure \cite{Du2015, Wu2020}. 

\subsubsection*{\textbf{Concluding Remarks}}
Localized mode FMRFM is a powerful spectroscopic tool for imaging ferromagnets with high spatial resolution. This technique measures the internal field by locally exciting the FM’s spin dynamics, which is clearly distinct from other magnetic scanning microscopies that measure the stray field from the magnetic samples, such as MFM, scanning SQUID microscopy and NV optically detected magnetic resonance imaging. Currently, the best lateral spatial resolution of 100 nm has been achieved on YIG films with this method, but this can be addressed through improvements in the nanofabrication of cantilevers and magnetic tips. LM-FMRFM can provide the microscopic details quantitatively needed for the characterization of ferromagnetic materials used in fields ranging from spintronics to biomagnetism.  This method is applicable to buried and surface magnets, and as it is a resonance spectroscopy technique, it measures local internal fields and other magnetic properties and spin interactions at interfaces.

\subsubsection*{Acknowledgements}
This work was primarily supported by the Center for Emergent Materials, an NSF MRSEC, under award number DMR-2011876 and by the NanoSystems Laboratory at OSU.
\newpage

\subsection{Quantum sensing in the negative mass reference frame}
Eugene S. Polzik\\
Niels Bohr Institute, University of Copenhagen, Blegdamsvej 17, Copenhagen, 2100 Denmark\\
polzik@nbi.dk
\label{Polzik}

\subsubsection*{\textbf{Status}} 
\begin{figure}[h]
    \centering
    \includegraphics[width = 0.75\textwidth]{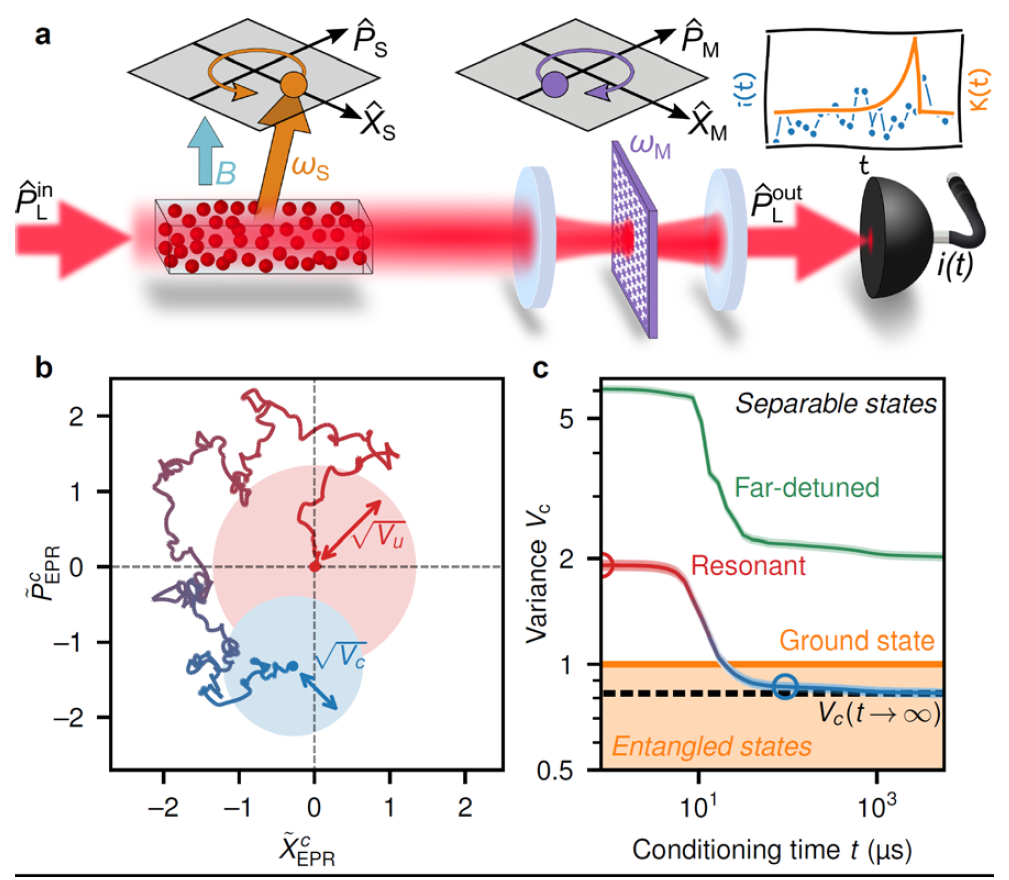}
    \caption{Generation of an entangled state between a mechanical oscillator and a spin oscillator which acts as an effective negative mass reference frame (NMRF). Panel a: detection of light propagating through the spin system and the mechanical oscillator placed inside an optical cavity generates entanglement between the two systems. Panel b: trajectory of the mechanical oscillator referenced to the NMRF (blue) and the standard quantum limit (red area). Panel c: the joint state of the mechanics and the spin systems becomes entangled as the measurement progresses in time. Reproduced with permission from \cite{Thomas2020}.}
    \label{Polzik:fig1}
\end{figure}

Over the past two decades, a new approach to measurements of fields, motion and forces has emerged \cite{Julsgaard2001, Polzik2014, Hammerer2009}. Based on the measurement in the so-called negative mass reference frame (NMRF), this new approach quantum sensing provides, in principle, unbound sensitivity, not limited by the Heisenberg uncertainty bound (also referred to as the standard quantum limit (SQL)). The origin of the SQL is in the impossibility of simultaneous measurement of non-commuting variables, $X$ and $P$ with arbitrary accuracy. The negative mass reference frame is a quantum system \cite{Polzik2014, Hammerer2009, Tsang2012, Woolley2013}, which was first realized by an optically pumped atomic spin oscillator \cite{Julsgaard2001}, and has been later extended to optomechanical \cite{Woolley2013}, and ultra-cold atomic systems \cite{Zhang2013}. Sensing beyond SQL in the negative mass reference frame is realized by generation of an Einstein-Podosky-Rosen (EPR) entangled state between the sensor variables $X$, $P$ and the negative mass reference frame variables $X_0$ ,$P_0$. Such an entangled state has been experimentally demonstrated for spin oscillators \cite{Julsgaard2001}, two mechanical oscillators \cite{OckeloenKorppi2018, Riedinger2018}, and composite mechanical-spin oscillators \cite{Moeller2017, Thomas2020, Karg2020}. For spin oscillators, $X$ and $P$ are realized by collective spin variables $J_x = X\sqrt{J}$ and $J_y = P\sqrt{J}$ which are the quantum projections of the collective spin orthogonal to the mean collective spin $J$. After an entangled state is generated between the sensor and the reference system, the external force acting on the sensor (but not on the reference system) can be applied. The subsequent collective measurement on both the sensor and the reference system then reveals perturbations due to the external force exerted on both $X$ and $P$ of the sensor. Fig.\,\ref{Polzik:fig1} illustrates the layout of entanglement generation between the sensor – a mechanical oscillator ad the spin oscillator – the NMRF. Magnetic field sensing beyond SQL using the negative mass reference frame has been already demonstrated \cite{Wasilewski2010} and proposals to extend such sensing onto other systems with nearly arbitrary spectral properties have been developed \cite{Khalili2018, Zeuthen2019, Zeuthen2022}. Application of the NMRF to gravitational wave interferometry beyond SQL has been proposed \cite{Zeuthen2019}.

\subsubsection*{\textbf{Current and Future Challenges}}
With the proof-of-principle demonstrations of sensing beyond SQL achieved, the challenge is to improve the technology to the point where a sizable improvement is sensitivity in practical applications can be achieved. A figure of merit for strong coupling is quantum cooperativity \cite{Hammerer2009, Moeller2017, Thomas2020}. For an optomechanical interaction $C_{qM} = 4g^2/\Gamma_M$ where $g = \omega/L\sqrt{\hbar n/m\Omega}$ with $\omega$ – optical frequency, $m$ and $\Omega$ mass and frequency of the mechanical system, $L$ and $n$ – the length of the optical resonator and the photon number. $\Gamma_M$ is the decoherence rate. In principle, high $C_{qM}$ can be achieved by using advanced mechanical systems with quality factors reaching $10^9$ and above \cite{Rossi2018} for membrane resonators combined with high intra-cavity photon number $n$. In practice, however, $n$ is often limited by photo-thermal effects. High cooperativity regime is easier to reach at millikelvin temperatures in a dilution refrigerator \cite{OckeloenKorppi2018, Riedinger2018}, although it has been also achieved with rather modest cryogenic cooling \cite{Moeller2017, Thomas2020}.
High cooperativity for the spin system $C_{qS} = \kappa^2\Gamma_S$ can be achieved by using systems with a high optical depth $\sim\kappa^2$ combined with long spin coherence time. Macroscopically sized room temperature spin ensembles contained in cells with walls covered with coatings preventing collisional spin relaxation have been a key element in a number of experiments [1,7,10,12]. Those conditions can also be achieved by using cold and trapped atoms and cavity enhanced interactions \cite{Karg2020}.
As with many other quantum sensing protocols based on detection of continuous variable entanglement, sensing using NMRF requires low optical losses. For mechanical oscillators it means low scattering and absorption losses. That is why silicon nitride membranes characterized by negligible losses in the near infrared spectral range and super-polished cavity mirrors are often the elements of choice. For room temperature atomic ensembles topical losses are typically in a few percent range limited mainly by the optical losses on the cell windows.

\subsubsection*{\textbf{Advances in Science and Technology to Meet Challenges}}
\begin{figure}[ht!]
    \centering
    \includegraphics[width = 0.5\textwidth]{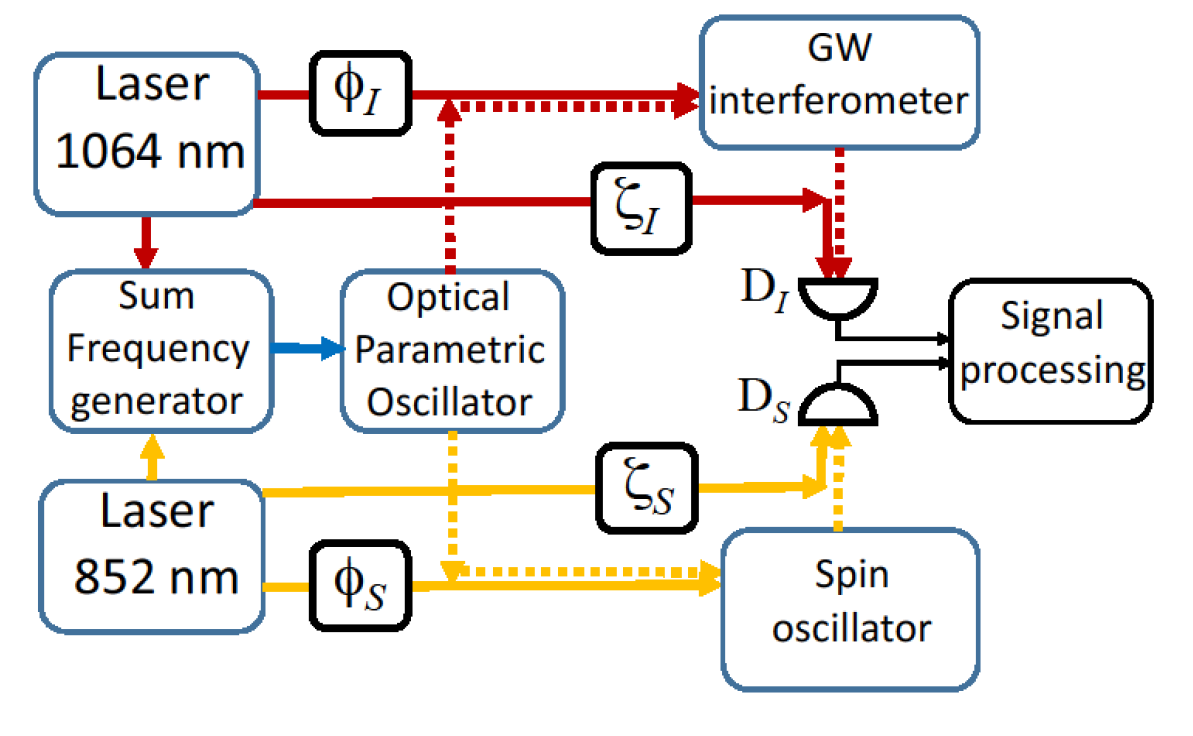}
    \caption{A layout for sensing of motion beyond SQL using the negative mass reference frame (NMRF) of the spin oscillator. Here the sensor is a gravitational wave interferometer, but the scheme can be utilized for a wide range of interferometric measurements. The sensor and the spin (NMRF) are addressed by two entangled light fields, which allow to have the sensor response function to be centred at a wavelength different from the resonant wavelength of the NMRF. In addition, the choice of phases $\zeta_I$ and $\zeta_S$ allow the frequency spectral response of the NMRF to be tuned to the response of the sensor (interferometer) using the oscillator virtual rigidity. Reproduced with permission from \cite{Zeuthen2019}.}
    \label{Polzik:fig2}
\end{figure}

Development of high Q, low loss mechanical oscillators is one of the key challenges on the way towards sensing of motion beyond the SQL. $Q\Omega$, the product of the quality factor and resonance frequency is a figure of merit which allow to calculate the number of periods over which an oscillator will stay in a given quantum state before thermal decoherence becomes an issue. With this product beyond $10^{13}$ an oscillator will stay coherent over a single period of oscillations at room temperature. $Q\Omega$ values exceeding this threshold by several orders of magnitude have been reported for surface acoustic wave resonators \cite{Shao2019} and nano-acoustic resonators \cite{MacCabe2020}. Those systems appear to offer exciting options as future platforms for sensing using NMRF, if optical losses can be minimized.
Spin oscillators with sufficiently high cooperativity and low losses can be implemented by further developments of room temperature systems, such as vapour cell with antireflection coatings described above, solid state systems, such as collective spins of nitrogen vacancies in diamond \cite{Wood2022} and more exotic systems \cite{Hahn2021}.
NMRF systems can be used for sensing and metrology for sensors with nearly arbitrary spectrum which can be very different from the NMRF system’s spectrum \cite{Zeuthen2019, Zeuthen2022}. In this case, the sensor and the NMRF system cannot be probed with the same optical (or microwave) field. In order to generate entanglement and achieve sensitivity beyond the SQL in this case, the two systems have to be probed with entangled light fields. The two entangled modes are tuned to the corresponding resonances of the two systems. An example of such system is the implementation of the proposal \cite{Zeuthen2019} for gravitational wave interferometry beyond the SQL. Here, the two entangled modes of light probe the gravitational wave detector and the NMRF system implemented by a spin ensemble with an effective negative mass, respectively (Fig.\,\ref{Polzik:fig2}).

\subsubsection*{\textbf{Concluding Remarks}}
The outlook for sensing beyond the SQL is promising. Applications of such sensing range from gravitational wave detection to medical diagnostics to communication. Measurements of magnetic and electrical fields are ubiquitous. Until recently, quantum limits of measurements of those fields have been achieved only in a limited number of applications. With the attention presently focused on quantum sensing and quantum technologies, technical improvements will inevitably bring more sensing applications closer to quantum limited sensitivity. Overcoming those limits using the NMRF methodology offers exciting new capabilities.

\subsubsection*{Acknowledgements}
The author’s research reviewed in this article has been supported by the European Research Council (ERC) under the European Union’s Horizon 2020 research and innovation programme (grant agreement No 787520) and by VILLUM FONDEN under a Villum Investigator Grant (grant no. 25880).

\newpage
\section{NV}
\label{NV}
\subsection{Towards imaging complex spin systems and molecules with spin-based quantum sensors}
Tim H. Taminiau\\
QuTech and Kavli Institute of Nanoscience, Delft University of Technology\\
Amit Finkler\\
Department of Chemical and Biological Physics, Weizmann Institute of Science\\
t.h.taminiau@tudelft.nl, amit.finkler@weizmann.ac.il	

\subsubsection*{\textbf{Status}}
\underline{Brief history and status}: In recent years, quantum sensors based on optically accessible defect spins have emerged as a promising platform for imaging complex spin systems, such as molecules and quantum devices. Whereas traditional NMR and MRI methods require averaging over ensembles, spin-based quantum sensors might enable the magnetic imaging of individual systems and single-molecules, with {\aa}ngstr\"om spatial resolution and sensitivity down to single nuclear spins. Such nano- and atomic-scale MRI might open new opportunities to determine the structure of single molecules like proteins, to investigate qubit systems for quantum information tasks, and to characterize spin-based quantum simulators. 
A variety of theoretical proposals has been put forward  \cite{Ajoy2015, Kost2015, Wang2016, Perunicic2016, Perunicic2021}. Pioneering experiments have reported the magnetic resonance signal of single molecules \cite{Shi2015, Lovchinsky2016}, and the reconstruction of the positions of individual electron and nuclear spins with atomic-scale resolution, including for systems with up to approximately 30 spins \cite{Abobeih2019, Yudilevich2022, Cujia2022}. To be able to image more spatially extended systems, arrays of sensors or scanning probe based sensors are explored (See Sec.\,\ref{Jayich}).
\begin{figure}
    \centering
    \includegraphics[width = 0.5\textwidth]{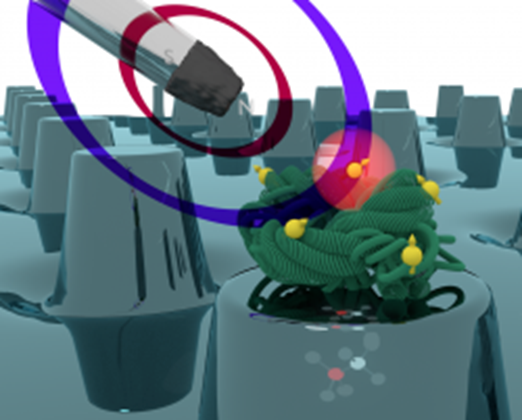}
    \caption{An artist rendition of NanoMRI for molecular structure. $\copyright$ 2017, Philipp Scheiger. The figure depicts an array of diamond nanopillars, each hosting an NV center. The molecule of interest is placed on the diamond surface, and a magnetic field gradient enables an improved NanoMRI spatial resolution.}
\end{figure}
\underline{Why it is still important}: Molecular structure is part of a triumvirate, the other members being function and dynamics. NanoMRI combines the non-invasiveness and chemical specificity of magnetic resonance with the advantages of single molecule spectroscopy. While extremely challenging, observing structure and dynamics on the single molecule level (NanoMRI) might lead to the capability to observe the different stages in time of the same molecule. It can reveal outliers in the statistical distribution and might eventually allow for the use of far smaller quantities of material for the initial (and very expensive) stages of development of materials, for example new drugs.

In the field of quantum materials and technologies, the capability to image spins and currents on the atomic- and nano-scale can open new opportunities to investigate quantum materials (see Sec.\,\ref{Jayich}), to investigate various quantum bits and magnetic noise environment, and to characterize spin-based quantum simulators both inside and outside of the host material \cite{Cai2013, Randall2021}.  

\underline{What will be gained}: The different platforms are now in an advanced stage, in both our understanding of their underlying physical principles and also the technological readiness level. Further advances now need to tailor material science, chemical, electrical and mechanical engineering with these platforms (mostly developed by physicists). This integration promises to enable a transition from method development and demonstration to applied studies of molecules of interest, ranging from relatively simple amino-acids and eventually all the way to biology’s building blocks, proteins. While the latter is a very ambitious goal, there are several interim stages, such as the use of small spin clusters for quantum simulation, which on their own already constitute a formidable achievement and are worth the entire endeavor.

\subsubsection*{\textbf{Current and Future Challenges}}
The general concept of imaging spin systems with defect-spin-based quantum sensors is the following. The quantum sensor consists of a single electron spin, or an array of electron spins, that can be detected and read out optically or electrically, and controlled through microwave pulses. Advanced examples of such quantum sensors are the nitrogen-vacancy (NV) center and other defects in diamond and defects in silicon-carbide, as well as donor spins in silicon \cite{Perunicic2021}.  

This electron-spin sensor is brought in proximity to the sample so that it interacts with the sample spins. Various pulse sequences that manipulate the sensor, as well as the sample spins, can then be used to detect the sample spins and obtain spectroscopic information on dynamics and characteristic precession frequencies. Spatial structure is encoded in the spectral information through a combination of an external field gradient, the field gradient generated by the electron-spin sensor itself and/or interactions and chemical shifts internal to the sample spins. In that sense, nano-scale magnetic imaging of individual systems blends together the principles of MRI and NMR.     

We divide our treatment of the challenges ahead into three categories: (1) high-quality sensors in close proximity to the sample, (2) sample preparation and fixation, and (3) methods for efficient extraction of structure. Meeting these challenges will require advances in theoretical understanding and computation methods, as well as improvements in materials and hardware. 

\begin{enumerate}
    \item \emph{High-quality near-surface sensors} – A key challenge is to bring the electron-spin sensor in close proximity to the sample, while maintaining good coherence properties. Because the spin-spin interaction signal typically falls off as $1/r^3$, a small sensor-sample distance is generally essential to obtain a significant S/N. Additionally, when using the field gradient from the sensor electron spin to localize the sample spins, close proximity tends to translate in high-spatial resolution. However, the required small distance from the sensor defect to the host material surface introduces electromagnetic noise from surface defects and dangling bonds, and causes charge state instabilities \cite{Janitz2022}.    
    \item \emph{Sample preparation} – A second challenge is how to place interesting samples in close proximity of the sensor material surface. For this functionalization and fabrication methods must be developed. An interesting general solution is to use defect spin sensors embedded in scanning probes so that they can be positioned over samples, but the small desired sensor-sample distances (for example less than 5 nm) remain a challenge.  Additionally, a trade-off exists between fixation in a solid matrix to enable long interrogation times and allowing dynamics. While ideally, one would like to study for example biologically relevant molecules at room temperature in liquids, diffusion can limit interrogation times and linewidths.
    \item \emph{Efficient spectroscopy and imaging} – A final challenge is to develop methods for high resolution spectroscopy that enable imaging complex spin systems with high spatial resolution and manageable acquisition times. Current proof-of-principle demonstrations \cite{Abobeih2019, Cujia2022} require long measurements, and improved signal-to-noise, as well as more efficient data collection and data analysis are required to enable imaging more complex systems. 
\end{enumerate}

\subsubsection*{\textbf{Advances in Science and Technology to Meet Challenges}}
For applications aiming to magnetically detect and image individual molecules or nanoscale devices it will be essential to  apply surface science and chemical engineering to realize reliable \& reproducible templates for fixating molecules, surface functionalization \cite{Abendroth2022} and modification of the environment \cite{Bayliss2020} (so far most experiments were not done on external molecules).
Finally, our reliance on surface functionalization will play a decisive role in any technique which aims at measuring the magnetic properties of single molecules. Here, the constant advances in chemical engineering, understanding surface termination \cite{Xie2022}, controlling the spin bath environment and modifying specific ligands to have stronger and coherent bonds is pivotal for an efficient and useful sample preparation process.
The availability of the so-called electronic- or quantum-grade materials must become abundant and accessible for the community as a whole to benefit. This is already happening in quantum foundries being set-up around the globe. 

As the application of nanoscale MRI to molecules makes use of the spin degree-of-freedom of both electrons and nuclei, the concept of hyperpolarization comes about almost immediately. Ideas from the NMR community have diffused and propagated over the last decade (Sec.\,\ref{Ajoy}), but when dealing with systems of a very small cluster of spins, an all-out polarization is yet to be demonstrated.
To reduce acquisition time and in general to improve readout fidelity, a combination of pulse engineering (e.g. \cite{Wang2022}, and see also Sec.\,\ref{Cappellaro}), adaptive algorithms (Sec.\,\ref{Bonato}), machine-learning and yet even better materials, e.g., commercially available $^{12}$C diamonds, are needed. We have ample experience from the past two decades, and even a richer library of possible improvements from the NMR community in the past half-century, both of which provide the resources for ingenious tailoring of sequences for manipulating spins. 
Regardless of the specific method for data gathering, algorithms for streamlining the post-processing (or real-time processing) of the data for the purpose of reconstructing the positions of the various spins is critical. We expect some deviations from the ensemble-based techniques and argue that such algorithms, already in their abstract form, will provide a substantially high launch point for a full molecular structure reconstruction in the limit of a few tens of spins \cite{Abobeih2019}.
With all NanoMRI techniques, the issue of knowing the precise position of the sensor will be critical. For example, in the case of color centers, more deterministic positioning of sensors \cite{Luehmann2019} will enable a reliable and reproducible method of acquiring data. Naturally, some techniques have this capability by definition, but nevertheless this will bring about, so to speak, a closure to the sample-sensor fabrication process, including the nanometer-positioning of RF and MW antennae relative to the sensor and the molecule under investigation \cite{Nguyen2019}. Antennae that enact magnetic field gradients \cite{Rose2018} will in-effect be a force multiplier, making it much easier to create a distinct spectral separation between neighboring spins.

\subsubsection*{\textbf{Concluding Remarks}}
The nanoscale approach to determining molecular structure by means of magnetic resonance methods is finally gaining momentum and in our eyes is set to soon reach the critical point of technical readiness for applications. Looking at the different techniques we have in the NanoMRI community, they all seem to support each other for an overall toolkit, which is surprisingly powerful and versatile, and can complement existing methods in time, frequency and space domains. In the coming years, we expect the maturation of technological tools to advance the field beyond the threshold of proof-of-concept towards practical single-molecule imaging using magnetic resonance.

\subsubsection*{Acknowledgements}
THT is a group leader at QuTech and the Kavli Institute of Nanoscience and acknowledges support by the Netherlands Organization for Scientific Research (NWO) through a Vidi grant and the European Research Council (ERC) under the European Union’s Horizon 2020 research and innovation programme (grant agreement No. 852410). AF is the incumbent of the Elaine Blond Career Development Chair and acknowledges support from Israel Science Foundation (grant agreement No. 418/20) and the Minerva Stiftung as well as the Abramson Family Center for Young Scientists and the Willner Family Leadership Institute for the Weizmann Institute of Science.

\newpage

\subsection{Micron-scale NV-NMR}
Ronald L. Walsworth\\
University of Maryland (2218 Kim Engineering Building, College Park, MD 20742, USA\\
walsworth@umd.edu    

\subsubsection*{\textbf{Status}}
Nuclear magnetic resonance (NMR) spectroscopy is an important chemical analysis tool in many areas of basic and applied life science, due to its unparalleled ability to determine molecular identity and structure under ambient conditions. However, conventional NMR requires macroscopic sample volumes because it relies on detecting the weak magnetization of thermally-polarized nuclear spins via inductive detection. The poor sensitivity of conventional NMR has restricted its utility in many important areas where sample volume, molecular concentrations, and/or total molecular number are intrinsically limited, e.g., single-cell metabolomics, chemical analysis of mass-limited samples used for screening of drugs and catalysts, and label-free sensing of biomarkers.

\begin{figure}[h]
    \centering
    \includegraphics[width = 0.325\textwidth]{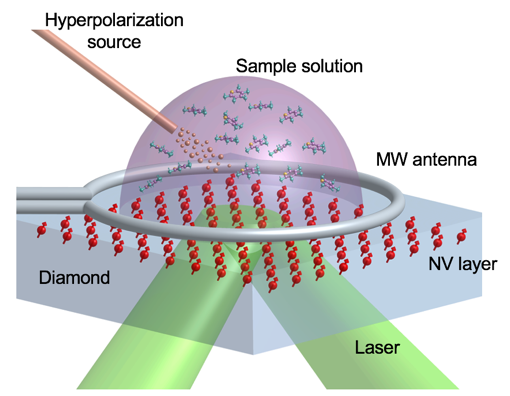}
    \includegraphics[width = 0.625\textwidth]{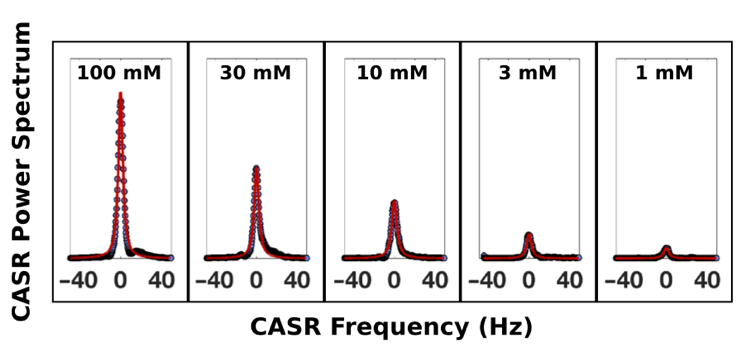}
    \includegraphics[width = \textwidth]{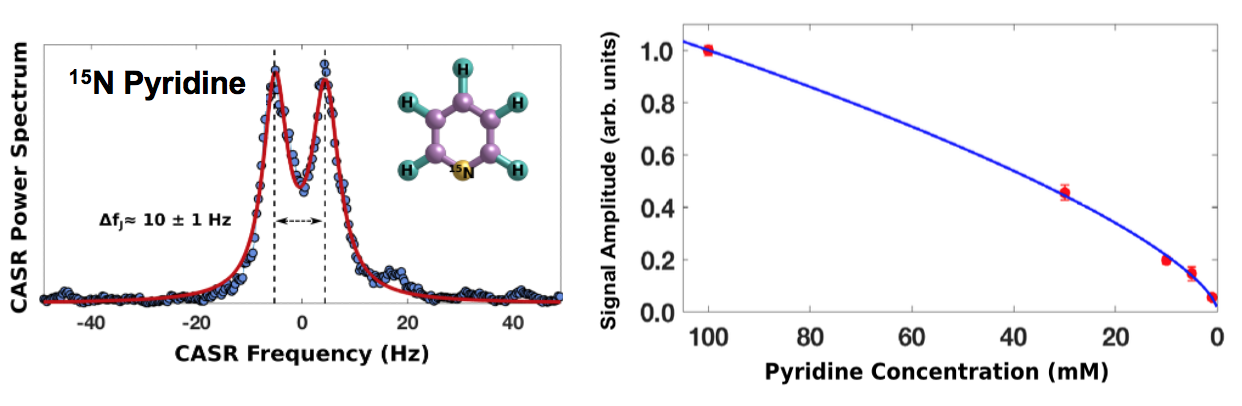}
    \caption{Example micron-scale NV-NMR results using SABRE hyperpolarization \cite{Arunkumar2021}. (Upper left) Experimental schematic.  (Lower left) NV-NMR spectrum of $^{15}$N-labeled pyridine at 100 mM concentration in methanol, with an effective sample volume of about 10 picoliters. Few-Hz splitting of the NMR spectrum due to J-coupling between the $^{15}$N nucleus and protons in pyridine is easily observable. (Upper and lower right) Micron-scale NV-NMR signal spectra, here for unlabeled pyridine at various concentrations in methanol, have observable peak amplitudes down to $\sim$1 mM molecular concentration, equivalent to $\sim$10 femtomole molecular number sensitivity.}
    \label{Walsworth:fig1}
\end{figure}

Recent advances using optically-addressable nitrogen vacancy (NV) color centers in diamond hold great promise to overcome this sensitivity limitation, thereby enabling high-resolution NMR spectroscopy and imaging (MRI), under ambient conditions, at the scale of single biological cells — i.e., the micron or picoliter scale. Recent work demonstrated NV-NMR spectroscopy with chemical resolution in very small liquid sample volumes ($\sim$10 picoliters): i.e., detection of both J-couplings and chemical shifts in small molecules in solution, by coherent averaging of repeated NV-NMR measurements synchronized to an external clock \cite{Glenn2018}. This micron-scale NV-NMR capability was then integrated with hyperpolarization and pre-polarization techniques \cite{Smits2019, Bucher2020, Arunkumar2021} to boost the molecule concentration sensitivity for micron-scale samples to as low as $\sim$1 millimolar (mM), thereby reaching $\sim$10 femtomole absolute molecular number sensitivity; see Fig.\,\ref{Walsworth:fig1}. No other NMR technology can provide this combination of picoliter volume applicability, molecular sensitivity, and chemical specificity for samples under ambient conditions.

Some of the potentially transformative applications of micron-scale NV-NMR are the following:

\begin{itemize}
    \item Single-cell metabolomics: NV-NMR may provide high-resolution NMR spectroscopy of small molecules and proteins within single-cells, enabling physiologically relevant single-cell metabolomics. Compared to the genome or transcriptome, the metabolome of a single cell or organelle is much more difficult to measure because metabolites exhibit much larger structural diversity and dynamic range and are difficult to amplify. Nevertheless, the metabolome provides the most immediate and dynamic picture of a cellular phenotype. As such, a reliable NMR method to perform single-cell metabolomic measurements is expected to have a large impact in a wide area in biology and medicine, thus providing a high value target for our new technology development.
    \item Chemical analysis of mass-limited samples: High-throughput chemical analysis is essential for screening of drugs, catalysts, and many other applications. Today, economically efficient chemical synthesis is often performed on microfluidic chips with picomole or smaller samples in a highly parallelized way. The corresponding chemical analysis is thus restricted to small-mass-sensitive methods like fluorescence spectroscopy or mass spectrometry. Importantly, conventional NMR — which gives superior molecular structural information compared to these methods — cannot be used for such applications due to its inherently low sensitivity and lack of parallelization. NV-NMR may overcome this limitation, allowing high-throughput, chip-based NMR spectroscopy applicable to femtomole-quantity, picoliter-volume samples. NV fluorescence readout onto a camera could also allow interrogation of many NMR samples in parallel, thereby enabling practical high-throughput NMR chemical analysis of mass-limited samples.
\end{itemize}

\subsubsection*{\textbf{Current and Future Challenges}}
As outlined above, recent work has shown that micron-scale NV-NMR can spectrally resolve small molecule J-couplings and chemical shifts in liquid samples under ambient conditions, with absolute NMR frequency resolution $\sim$1 Hz and molecular concentration sensitivity $\sim$1 mM (currently, when using hyperpolarization). No other NMR technology can approach this combination of small sample volume, high spectral resolution, and good sensitivity. However, as chemical shifts scale linearly with $B_0$, the applied static (bias) magnetic field, a micron-scale NV-NMR system with $B_0> 1$ T will be needed for identification of molecular species and associated kinetics and dynamics when using standard NMR chemical analysis techniques. In addition, a tesla-scale bias field will increase the sample nuclear spin polarization and hence the concentration sensitivity of a micron-scale NV-NMR system by $\sim$10-100$\times$ when not using hyperpolarization.

Thus, a current goal of micron-scale NV-NMR technology development is to realize instruments that operate at tesla-scale bias field \cite{Aslam2017, Fortman2021, Ren2023} while maintaining or improving millimolar concentration sensitivity, spectral resolution able to resolve J-couplings and chemical shifts, and efficient handling of small-volume samples. A core associated challenge will be realization of efficient, high-power microwave signal sources and antennas for coherent manipulation of NV electronic spins at the large microwave frequencies associated with a tesla-scale bias field: e.g., $\approx$ 40 GHz central frequency and $\approx$ 120 MHz Rabi frequency at $B_0\approx$ 1.5 T. Other key challenges include the development and integration of several sub-systems and capabilities:
\begin{itemize}
    \item A magnet providing a tesla-scale bias field with suitable homogeneity, stability, and pole spacing for operation of a micron-scale NV-NMR measurement module providing high spectral resolution.
    \item A microwave integrated GaN transmitter chip operating at $\sim$ 40-GHz, including a power amplifier and a planar coil for fast (high power) ESR excitation of the NV electronic spins.
    \item Microfluidic or other methods to localize picoliter volume samples with the ensemble NV sensor, or arrays of samples on the diamond surface for parallel NV-NMR spectroscopy systems, while not broadening NMR spectral lines via magnetic susceptibility effects.
    \item A compact NV-NMR measurement module, including efficient delivery and handling of optical pumping light and collection of NV fluorescence; an RF spectrometer and delivery antenna to manipulate sample nuclear spins with any prescribed pulse sequence at the appropriate resonant frequency (e.g., 60 MHz for proton spins at $B_0\approx$ 1.5 T); integrated delivery of (optional) hyperpolarization sources to the liquid sample; and mechanical supports that do not induce significant magnetic susceptibility broadening of NMR spectral lines.
    \item Improved micron-scale NV-NMR sensor performance to allow micromolar/millimolar concentration sensitivity with/without use of hyperpolarization.
\end{itemize}

\subsubsection*{\textbf{Advances in Science and Technology to Meet Challenges}}
As one approach to the above challenges, a collaboration at Harvard and the University of Maryland (UMD) is developing an integrated micron-scale NV-NMR instrument operating at bias field $B_0\approx$ 1.5 T and 60 MHz proton NMR frequency. The associated coherent excitation of NV electron spins at $\approx$ 40 GHz microwave central frequencies and $\approx$ 120 MHz Rabi frequency will be provided by a GaN microwave transmitter chip; and a silicon (CMOS) RF spectrometer chip will be used to manipulate sample nuclear spins with any prescribed pulse sequence \cite{Kruger2022}. The GaN microwave transmitter chip will include an on-chip coil and a power amplifier with large amplitude tuning; and the CMOS RF chip will include an arbitrary pulse sequencer for nuclear spin manipulation and a power amplifier; both chips will have a small (mm$^2$) footprint. The 1.5 T micron-scale NV-NMR instrument will also include microfluidic manipulation of samples integrated with delivery of hyperpolarization sources; a compact NV-NMR measurement module; and advanced spin control and measurement protocols. The system has a design goal of $\sim$1 Hz absolute spectral resolution and small molecule concentration sensitivity $\sim$10 $\upmu$M/10 mM with/without hyperpolarization in sensing volumes down to $\sim$ 10 picoliters, thereby enabling diverse applications from single-cell biology to chemical analysis of mass-limited samples.

Meeting this target NV-NMR sample molecular concentration sensitivity in micron-scale samples will be aided by the increased bias field, together with a combination of other improvements and techniques, including: high-density NV spin ensembles in optimized diamond material (with minimized strain and paramagnetic impurities, and stabilized NV charge states); robust quantum control techniques that allow simultaneous suppression of the undesired effects associated with NV spin-spin interactions, disorder, and control and measurement protocol imperfections (recently shown to yield a 5$\times$ enhancement in NV coherence time \cite{Zhou2020, Choi2020}); and use of quantum logic enhanced (QLE) sensing for NV ensembles. In QLE sensing, the NV electronic spin population difference carrying the measured AC magnetic field information is mapped onto the $^{15}$N nuclear spin degree of freedom on-board each NV in the ensemble. The NV nuclear and electronic spin states are then entangled into a pseudo-spin-singlet state. Next, the electronic spin state information is repeatedly read out optically and re-correlated with the (unperturbed) nuclear spin state in a quantum non-demolition (QND) measurement. As shown in \cite{Arunkumar2023}, the QLE technique can provide $>30\times$ enhanced SNR for ensemble NV AC magnetometry for MHz-scale signal fields; and improved signal magnitude sensitivity of about 10$\times$, depending on the measurement protocol.

\subsubsection*{\textbf{Concluding Remarks}}
Recent years have seen exciting progress in the development of micron-scale NMR with good spectral resolution and sensitivity, using dense ensembles of nitrogen vacancy (NV) color centers in diamond as precision NMR sensors. Ongoing efforts seek to extend this capability to few tesla bias magnetic fields; and to integrate the NV-NMR instrument with compact sub-systems for NV and sample spin manipulation, efficient manipulation of picoliter volume samples (with optional hyperpolarization of samples spins), high-efficiency optical measurement of NV fluorescence, and quantum control and measurement techniques. Once fully developed, few-tesla micron-scale NV-NMR can be expected to have wide-ranging, high-impact applications — from label-free sensing of biomarkers to chemical analysis of mass-limited samples that are expensive or difficult to synthesize, e.g., high-throughput screening in drug and catalyst discovery. It may also open up new lines of scientific inquiry, including single-cell metabolomics for quantitative cell biology and cell-based drug screening; and functional and structural probing of biological tissues and organisms with cellular resolution and chemical specificity.

\subsubsection*{Acknowledgements}
The Harvard-UMD collaboration on micron-scale NV-NMR includes major contributions from the research groups of Profs.\,Donhee Ham, Mikhail Lukin, and Hongkun Park. Ongoing support for this work is provided by the Moore Foundation. Additional support for specific aspects of this collaboration are listed in the related scientific articles referenced below.

\newpage

\subsection{Nanoscale imaging with NV centers}
\label{Jayich}
Paz London and Ania Bleszynski Jayich\\
UC Santa Barbara, Department of Physics, University of California, Santa Barbara California 93106, USA\\
ania@physics.ucsb.edu, pazlondon@ucsb.edu

\subsubsection*{\textbf{Status}}
\begin{figure}[h]
    \centering
    \includegraphics[width = 0.6\textwidth]{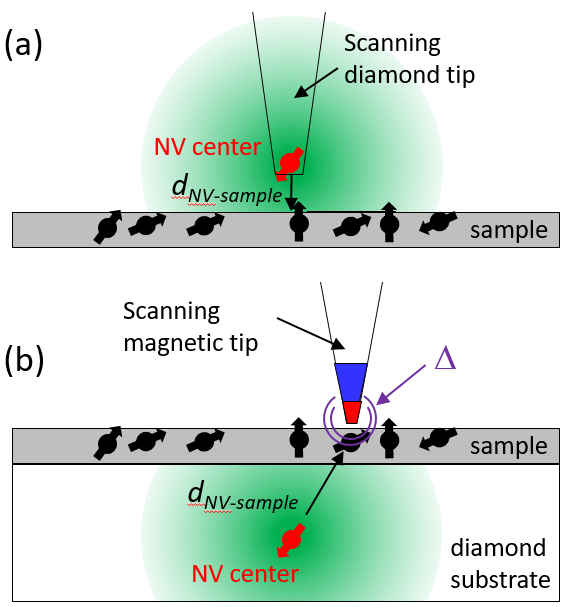}
    \caption{Scanning probe microscopy with a single NV center. (a) Scanning NV sensor configuration, where the tip-sample distance ($d_\mathrm{NV-sample}$) determines both spatial resolution and signal strength. (b) Scanning magnetic tip configuration. Here, the spatial resolution is given by $\Delta = 1/\gamma\tau\nabla B_\mathrm{tip}$, where $\gamma$ is the target spin's gyromagnetic ratio, $\tau$ is the interrogation time, and $\nabla_\mathrm{tip}$ is the gradient of the tip's magnetic field at the target spin's position. Purple lines bound this ``resonant slice'' of width $\Delta$ in which spins precess at a frequency to which the NV center is selectively sensitive. The fading spheres in (a,b) illustrate the volume at which the NV has a single electron sensitivity.}
    \label{Jayich:fig1}
\end{figure}

A powerful approach for nanoscale imaging is a scanning nitrogen vacancy (NV) center sensor. The NV center in diamond is a point-like defect in diamond whose spin degree of freedom has high sensitivity to magnetic and electric fields, as well as temperature. Moreover, the spin can be initialized, manipulated, and read out even at ambient conditions, making the NV center an easy-to-use, versatile, non-invasive, nanoscale sensor. To date, NV-based imaging has largely focused on magnetic phenomena.
Scanning NV center magnetometry provides high spatial resolution magnetic images, where the resolution can greatly exceed the optical diffraction limit and approach the nanometer-scale. An NV-scanning probe microscope (NV SPM) apparatus can be realized in two conceptual configurations (Fig.\,\ref{Jayich:fig1}): In a scanning sensor configuration, a single NV center is placed at the apex of an atomic force microscope (AFM) tip and is scanned above the surface of the sample (Fig.\,\ref{Jayich:fig1}a) \cite{Balasubramanian2008, Maletinsky2012}. In a scanning magnetic tip geometry, the NV center is placed in the vicinity of a sample and the tip’s magnetic gradient provides spatial discrimination of different regions of the sample as in MRI (Fig.\,\ref{Jayich:fig1}b) \cite{Balasubramanian2008, Grinolds2014, Bian2021}; this technique is suitable for samples that exhibit an electron or nuclear spin resonance with sufficiently narrow linewidth. While the former allows a larger scanned area (not limited by the NV center’s sensing volume), the latter offers spatial resolution that is limited by the achievable gradient only, and not by the sensor-target distance. We note that in an alternative imaging approach that does not involve a scanning tip one can apply variable magnetic gradients to encode the spatial coordinate via Fourier imaging \cite{Arai2015}.
In the last decade, NV SPM was used to image several condensed matter (CM) systems and phenomena: single electron spins \cite{Grinolds2014, Grinolds2013}, the dynamics of magnetic domains and domain walls at room temperature and cryogenic temperatures \cite{Tetienne2014, Thiel2019, Sun2021}, superconductivity \cite{Pelliccione2016, Thiel2016}, and hydrodynamic current flow in interacting electron systems \cite{Jenkins2022}. For a wider list of condensed matter phenomena to which NV SPM has been applied, see \cite{Casola2018} and references within. NV SPM can also be applied to biological systems \cite{Wang2019}, though progress on this front is slower due to the variability of biological environments, non-planar morphologies inherent to biological systems, and sensitivity to laser and microwave excitations used for NV measurements. The versatility of the NV SPM is also manifest in the several sensing modalities that are possible, giving access to DC, AC, and incoherently fluctuating magnetic \cite{Ariyaratne2018} and electric fields \cite{Bian2021}.

\subsubsection*{\textbf{Current and Future Challenges}}
The major challenge to achieving truly nanoscale ($< 10\,\mathrm{nm}$) spatial resolution while maintaining high sensitivity with NV SPM is the deleterious effect of the proximal surface on the NV center’s charge state stability \cite{Bluvstein2019} and spin coherence \cite{Myers2014}, effects that have been characterized extensively but whose microscopic origins remain unclear, resulting in a lack of mitigation strategies. Surface-related charge traps, paramagnetic impurities, and Fermi level modifications are all suspected culprits that shorten coherence time and NV charge state stability, hence limiting sensitivity and even prohibiting a measurement (Fig.\,\ref{Jayich:fig2}a). Currently, scanning experiments typically utilize NV centers that are $> \sim 10\,\mathrm{nm}$ deep to avoid the surface effects, which are exacerbated in a scanning experiment as the surface can change due to unintentional tip-sample contact.
The high throughput production of robust NV probes is another outstanding challenge that is rooted in several reasons: quantum-grade, single-crystal diamond is expensive and is limited to small (mm-scale chips); thin-film diamond is difficult to form because heteroepitaxial growth of high-quality, single-crystal diamond has not been achieved; and diamond is notoriously difficult to polish and etch. In addition, the best route towards deterministic creation of well positioned NV centers in nanostructures is still an open question. Even deterministic methods such as attaching a pre-characterized nanodiamond to the end of a probe tip \cite{Balasubramanian2008, Tetienne2014}, or laser activation \cite{Chen2019} require time-consuming, probe-by-probe assembly.
Pixel-by-pixel imaging is inherently slow, compounding the two challenges detailed above. For NV imaging, acquisition rates are largely limited by low optical readout fidelity (photoluminescence rates are typically $\sim 10^5$ photons/second and spin state contrast is typically 20-30\%), thus necessitating large numbers of measurement shots per pixel, which lengthens measurement times. State-of-the-art single NV center probes typically require few-second-long acquisition times per pixel for imaging 10’s of nanotesla fields at optimal conditions (with NV depths $> 10\,\mathrm{nm}$). A potential solution to poor readout fidelity is single-shot readout at cryogenic temperatures via resonant red laser excitation, as demonstrated for bulk NV centers \cite{Robledo2011}, but charge state conversion under resonant excitation for shallow NV centers in fabricated structures makes resonant readout currently difficult for NV scanning probes.
NV-based imaging of CM systems promises many unique advantages over other state-of-the-art material characterization techniques, but several challenges limit widespread implementation of NV-based-sensing in the CM community. In addition to the above challenges, operation at low ($< \sim 1\,\mathrm{K}$) temperatures and in high ($\sim$1-10\,T) magnetic fields is often desirable for eliciting various CM phenomena. However, these environments present challenges for NV experiments due to laser and microwave induced heating, high microwave frequencies necessary for driving spin resonance in strong magnetic fields, and the need for precise alignment of the field to the NV axis to maintain high sensitivity. For certain CM targets, this field alignment requirement may conflict with a desired field direction with respect to the sample axes, such as in quantum Hall systems. NV probes of different crystallographic orientations, such as (111)- and (110)-faced tips \cite{Rohner2019, Welter2022}, are a potential solution in some cases, however, the production of such probes is still relatively rare due to diamond growth challenges. Moreover, it was recently shown that at low temperatures, NV center performance degrades due to spin mixing in the optical excited state [22].

\subsubsection*{\textbf{Advances in Science and Technology to Meet Challenges}}
\begin{figure}
    \centering
    \includegraphics[width = 0.75\textwidth]{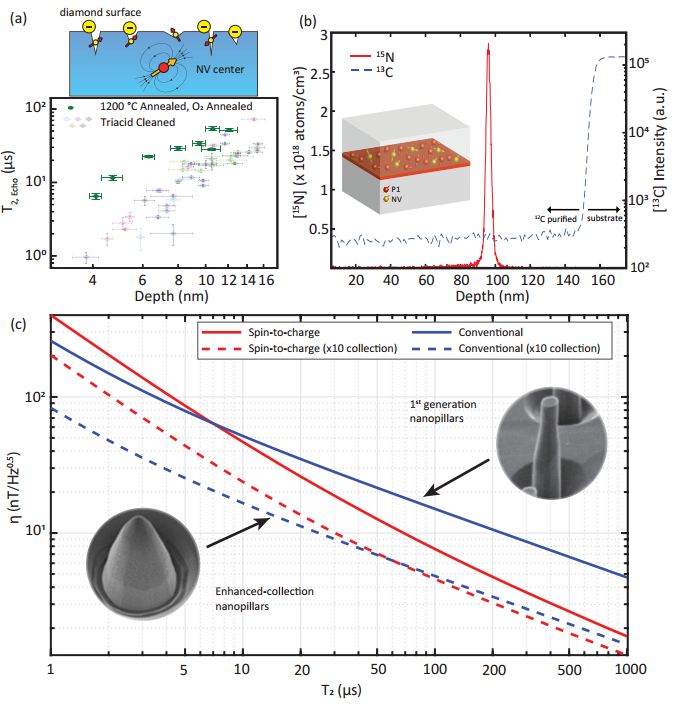}
    \caption{Scientific and technological improvements for NV SPM. (a) Coherence times of shallow NV centers (reproduced with permission from \cite{Welter2022}). The top panel illustrates how the surface can host defects that produce electric and magnetic field noise. The bottom panel shows Hahn echo coherence time $T_2$ as a function of the NV center’s depth, comparing a conventional triacid cleaning technique with high-temperature anneal + oxygen termination. (b) Secondary ion mass spectrometry measurement of delta-doped NV centers in an isotopically pure diamond. The red (blue) profiles show the $^{15}$N ($^{13}$C) abundance. (c) Projected magnetic sensitivity  for an NV center as a function of the NV center’s coherence time $T_2$. Solid (dashed) lines are for cylindrical (parabolic) shaped pillars. Blue (red) curves are for conventional (SCC) readout techniques. The 1st generation cylindrical nanopillar image was taken from \cite{Appel2016}, and the parabolically-shaped nanopillar image was taken from \cite{Wan2018}.}
    \label{Jayich:fig2}
\end{figure}

The role of the diamond surface - its proximity, morphology, and termination - is an active area of investigation, and advances in precise annealing, surface preparation, and termination show promise in mitigating the detrimental effects of the surface (Fig.\,\ref{Jayich:fig2}a) \cite{Sangtawesin2019}. Nevertheless, shallow NV centers still grossly underperform compared to their bulk counterparts, and hence the continued development and refinement of robust techniques for preparing the diamond surface prior to measurement, as well as for recovering a desirable surface during a scanning operation, is critical. Achieving these goals will require a deeper understanding of surface-mediated decoherence and surface-mediated charge instabilities (and the interplay between the two), as well as novel mitigation strategies such as surface passivation (e.g. encapsulation and termination) and Fermi level engineering approaches that, importantly, need to preserve nanometer-scale sensor-target separations. A complementary approach, which resigns itself to the presence of surface nonidealities, involves the in-situ manipulation of magnetic-noise sources and charge traps such as DC electric field biasing or surface spin driving or DC electric field biasing \cite{Zheng2022, Joos2022}.
Once surface effects are mitigated, or NV centers $> \sim 20\,\mathrm{nm}$ deep are employed, surface noise is no longer the limiting factor. For these NV depths, a promising approach to forming reproducible, stable, and coherent NV scanning probes involves nitrogen delta-doping (Fig.\,\ref{Jayich:fig2}b) during chemical vapor deposition (CVD) diamond growth \cite{Ohno2012} combined with 3D localization of NVs inside nanostructures by careful electron irradiation and annealing. This approach harbors several advantages: NV centers have consistently good coherence properties (even at high N-doping) due to the gentle nature of the nitrogen incorporation and low energy ($\sim$ 150 keV) electron irradiation, excellent depth localization down to 2 nm (in contrast to implanted NV centers), and ready use of isotopically purified $^{12}$C precursors for suppressing decoherence due to nuclear spins. Overall, this approach promises a high yield of high-performance scanning probes with consistent properties. Whether this approach, in conjunction with other surface improvements, can also lead to improved properties of shallow NV centers remains an open question. Forming NV scanning probes from delta-doped, CVD-grown, (110)- and (111)-oriented diamonds is another exciting frontier.
To address the fabrication challenges associated with making diamond NV scanning probes, more widespread availability of high-quality single diamond substrates, larger ($>4\times 4\,\mathrm{mm}^2$) substrates, and more reliable and rapid etching procedures would allow for more rapid development of optimized probe fabrication strategies \cite{Appel2016}. For forming thin ($\sim$micrometer scale) diamond slabs from which diamond cantilevers can be made, ion slicing is a promising method, akin to the Smart Cut$^\mathrm{TM}$ process for forming silicon on insulator (SOI). When combined with CVD diamond overgrowth, the membranes can host high quality NV centers. Compared to current approaches that involve starting with $> 100\,\upmu\mathrm{m}$ thick diamond pieces and removing most of the diamond via a combination of laser cutting, polishing, and etching, a time-consuming and wasteful process, this Smart Cut$^\mathrm{TM}$ approach is a promising avenue to a time efficient formation of diamond thin films while allowing for the reuse of diamond substrates.
To overcome low readout fidelities, hence reducing imaging times and enabling the use of advanced sensing techniques, there are several avenues of approach that should be pursued. Increased photon count rates can be achieved by careful geometric shaping of the NV-containing nanostructure, e.g. rounding the apex of a nanopillar to act as a parabolic mirror \cite{Wan2018}. For these pillar-based geometries, accurate 3D positioning of the NV center to a few tens of nanometers is critical and several approaches are being explored \cite{Wang2022a}. Higher collection efficiencies can facilitate the use of other advanced readout techniques such as spin-to-charge readout \cite{Shields2015}, resonant readout \cite{Robledo2011}, and nuclear-assisted repetitive readout schemes \cite{Haeberle2017}, which are especially useful for measurements with long sensing times, such as $T_1$-relaxometry imaging or AC measurements with long $T_2$ times, as shown in Fig.\,\ref{Jayich:fig2}c.

\subsubsection*{\textbf{Concluding Remarks}}
NV SPM harbors many advantages over other imaging techniques, with its combination of nanometer spatial resolution, nanotesla field sensitivity, noninvasiveness, quantitativeness, temperature versatility, and mulit-modal sensing capabilities. The full potential of the NV SPM is still far from being realized, however, due to the various reasons outlined above, but there are no known fundamental obstacles standing in the way. Hence, with coherent, charge-stable, near-surface NV centers that maintain sensitivity while scanning, reproducible high throughput production of scanning probes, and improvements in NV readout that approach single-shot readout, we expect widespread implementation of the NV SPM across condensed matter physics and eventually looking towards biological phenomena.

\subsubsection*{Acknowledgements}
The authors thank Sunghoon Kim and Simon Meynell for assistance with figures and comments on the text. P.L acknowledges support from the US Department of Energy (DOE) Office of Science, Basic Energy Sciences (BES) under Award No. DE-SC0019241, and A.C.B.J. acknowledges support from the Gordon and Betty Moore Foundation's EPiQS Initiative via Grant GBMF10279.

\newpage

\subsection{Quantum control for nanoscale NMR: status, challenges and frontiers}
\label{Ajoy}
Ashok Ajoy\\
Department of Chemistry, University of California, Berkeley CA 97420, USA \textit{and}\\
Chemical Sciences Division, Lawrence Berkeley National Laboratory, 1 Cyclotron Rd, Berkeley CA 94720, USA \textit{and}\\
Quantum Information Science Program, CIFAR, 661 University Ave., Toronto, ON M5G 1M1, Canada\\
Arjun Pillai\\
Department of Chemistry, University of California, Berkeley CA 97420, USA\\
ashokaj@berkeley.edu, apillai@berkeley.edu\\

\subsubsection*{\textbf{Status}}

A distinguishing feature of quantum sensor approaches for nanoscale NMR and MRI \cite{Degen2017} is how they leverage coherent dynamics of both analyte and sensor spins to provide the chemical specificity necessary for identifying individual molecules or mapping spatial distributions of nuclear species. The central challenge in the field is achieving high sensitivity, spatial resolution, and chemical spectral resolution simultaneously. Recent innovations in quantum control methods have been critical to making progress towards this. We discuss them here with a focus on NMR sensing with NV center electrons in diamond. 

\begin{figure}[h]
    \centering
    \includegraphics[width = 0.3\textwidth]{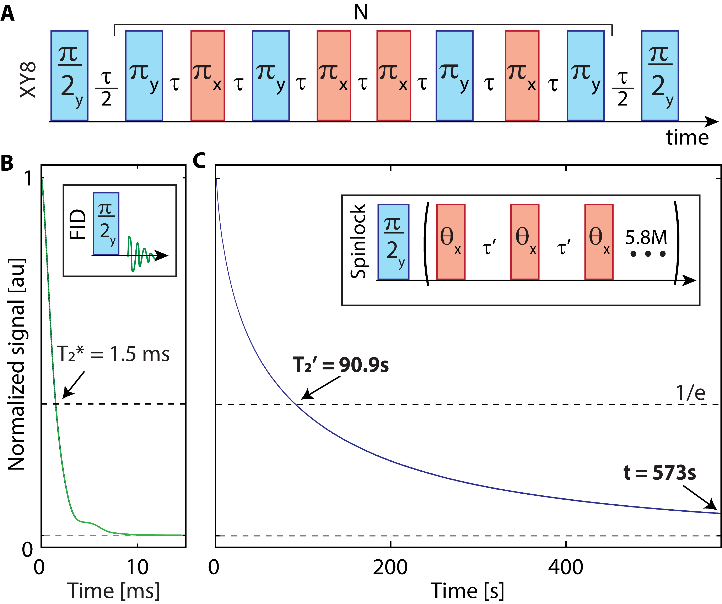}
    \caption{Quantum control schemes. (A) XY8 pulse sequence for sensing with NV centers. (B) Typical FID ($T_2^* = 1.5$ ms) of strongly coupled $^{13}$C nuclei in diamond. (C) Pulsed spinlocking (inset; here $\theta_x \neq \pi$ and lifetime depends parametrically on $\tau'$) allows for $>60000\times$ coherence time extension (Figure partially adapted from Ref.\,\cite{Takahashi2008}).}
    \label{Ajoy:fig1}
\end{figure}

One important quantum control strategy involves applying a train of pulses (e.g. CPMG/XY8) to the sensor spins with the interpulse spacing $\tau$ matched to one Larmor period $1/f_L$ of the analyte spins \cite{Staudacher2013, Schmitt2017, Glenn2018, Mamin2013, Konzelmann2018}. The sensor is prepared in a superposition state $+$, and the dipolar (hyperfine) interaction between the sensor and analyte spins induces its rotation towards an orthogonal state $-$, under the applied control. The extent of this rotation can be sensitively discerned and constitutes the signal from the selectively probed analyte species.

Achieving high sensitivity requires maintaining long sensor coherence times $T_2$ despite external perturbations or noise. In this regard, the sequence in Fig.\,\ref{Ajoy:fig1}A offers a natural solution by rendering sensor spins insensitive to fields that are mismatched in frequency from $f_L$. Related methods can also be used to ``spectrally decompose'' the noise fields so that they can be combated effectively \cite{Alvarez2011}. When this noise originates from other spins in the environment, such as those in the host lattice, control schemes decoupling them from sensor spins \cite{Bauch2018} can significantly improvement $T_2$ lifetimes without affecting sensor operation.  

There has also been considerable effort towards reducing sensor decoherence induced by imperfections in the quantum control itself, such as amplifier noise. The use of composite pulses, solid echo sequences \cite{Aiello2013}, and concatenated drives \cite{Cai2012} has been shown to provide significant protection against these errors. 

Furthermore, quantum control allows for the construction of effective (Floquet) Hamiltonians of the sensor-analyte interaction, which can aid in spectral simplification. For example, Ref. \cite{Ajoy2017} demonstrated an approach to ``interpolate'' the sensor dynamics to obtain quasi-continuous sampling of the delay interval $\tau$ despite instrumental limitations. Similarly, Ref.\cite{Casanova2015} exploited Hamiltonian engineering to suppress harmonics and aliasing artifacts, tune the spectral profile of the sensor bandwidth, and separate strongly and weakly coupled spins. The development of more sophisticated infrastructure (e.g. high-speed AWGs) will promote further such advances.

\subsubsection*{\textbf{Current and Future Challenges}}

\begin{figure}
    \centering
    \includegraphics{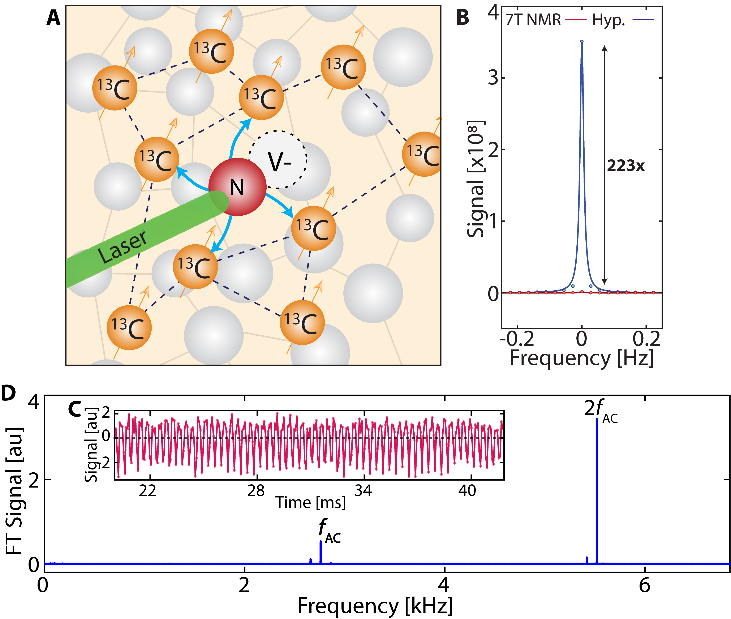}
    \caption{Quantum sensing with hyperpolarized $^{13}$C nuclei. (A) Diamond lattice schematic showing optical polarization of NV electrons and transfer to surrounding $^{13}$Cs via MWs. (B) Resulting signal is improved $>200\times$ compared to thermal NMR. (C) $^{13}$C NMR signal carries coherent oscillations when spins are subjected to AC field with frequency $f_\mathrm{AC}$. (D) FT of signal in (C) yields strong responses at $f_\mathrm{AC}$ and harmonics (Figure adapted from Refs. \cite{Gierth2020, Takahashi2008}).}
    \label{Ajoy:fig2}
\end{figure}

Several important experimental challenges remain towards achieving simultaneously high sensitivity, spatial resolution and chemical specificity. 
\begin{enumerate}
\item \textbf{High-field operation}: Currently, quantum sensing is predominantly carried out at low fields ($<0.1$\,T), where it is relatively easy to match the effective electron Rabi frequency to nuclear Larmor frequency at $B_0$ (typically $<1$\,MHz). However, operation at higher $B_0$ fields offers significant boosts in (i) the initial analyte polarization and (ii) chemical shift separation, both of which are proportional to $B_0$. Comparing sensing experiments at $0.1$\,T and $10$\,T, for example, the latter would allow for a $\sim 10^4$-fold increase in chemical resolving power from the increase in field alone. This increase in resolving power comes from an easier separation of the chemical shift peaks, and from them growing in intensity. 

However, accessing this regime is highly challenging due to a scarcity of sources and amplifiers in the ``THz-gap'' frequency range (e.g. electron Larmor frequency $\sim$280\,GHz at 10\,T). Matching electron Rabi frequencies to nuclear Larmor frequency in this regime is also difficult, as the timing condition in Fig.\,\ref{Ajoy:fig1}A would require trains of electron $\pi$-pulses separated by $<5$\,ns, which challenges the capabilities of available technology. As a result, it is not straightforward to directly utilize low-field CPMG/XY8 sequence variants for sensing at high-field.

\item \textbf{Low analyte spin polarization}: Quantum sensing involves detecting oscillating fields from Boltzmann-polarized nuclei, which constitute a small number at low-fields. Increasing $B_0$ can boost this signal linearly but doing so necessitates significant technical requirements. A promising approach is to transfer polarization from the optically polarized sensor electron to the nuclei to boost their polarization \cite{London2013} (Fig.\,\ref{Ajoy:fig2}A-B). Even a small amount of polarization transfer has a drastic effect: 1\% transfer from an NV center to analyte $^{13}$C nuclei can boost the signal $>10^4$-fold at 0.1\,T and room temperature. Producing such nuclear ``hyperpolarization'' remains an active area of research. Currently, it is possible to efficiently hyperpolarize nuclei within the diamond lattice, but transferring this polarization outside of diamond remains a challenge, likely due to fast-relaxing paramagnetic impurities on the diamond surface.

\item \textbf{Quantum sensing in the dense sensor limit}: Using an ensemble of $N$ sensors is an attractive way to increase sensitivity, with sensitivity theoretically expected to grow as $\sqrt{N}$. However, in practice, such gains are only feasible for well-separated (dilute) sensor spins. At higher concentrations, intersensor interactions lead to decreased sensor coherence time ($T_2^*$ in Fig.\,\ref{Ajoy:fig1}B), erasing the benefits of larger $N$. Overcoming this will require new sensing protocols that maintain sensor performance and coherence time $T_2$ even in the presence of intersensor couplings.

\end{enumerate}

\subsubsection*{\textbf{Advances in Science and Technology to Meet Challenges}}
We will now discuss recent developments in the field that address these challenges. 

\begin{enumerate}
    \item \textbf{High-field quantum control}: Accessing the sensitivity and resolution gains available at high-field requires significant effort, often involving the construction of an optical detection system within a superconducting magnet and the necessary infrastructure for high-frequency electron control. A noteworthy experiment by Ref.\cite{Aslam2017} involved operation at 3\,T, demonstrating chemical-shift resolved nanoscale NMR spectra of small molecules for the first time. Takahashi and co-workers have built an impressive apparatus for quantum sensing at 8.3\,T \cite{Fortman2021}, allowing spin manipulation directly at 230\,GHz using quasioptics and multiple frequency doubling chains. We anticipate further such developments, although the technical complexity required presents significant challenges. 

    \item \textbf{Nuclear spin quantum sensors}: An alternative approach to high-field quantum sensing is to use nuclear spins as sensors instead of electronic spins. Ref.\cite{Sahin2022} demonstrated this using diamond $^{13}$C nuclei optically hyperpolarized by NV centers (Fig.\,\ref{Ajoy:fig2}C-D). This approach overcomes the challenges of quantum control at high-field since the sensor Larmor frequency is only $\sim$100\,MHz even at 10\,T, making high-fidelity control easy. Moreover, these sensor spins have remarkably long $T_2$ lifetimes, with Ref.\cite{Beatrez2021} achieving $T_2$' exceeding 90\,s at room temperature (Fig.\,\ref{Ajoy:fig1}C). The extended coherence of nuclear spins offsets sensitivity losses from their lower gyromagnetic ratio. The sensing protocol here involves a train of spin-locking pulses with inductive interpulse sensor readout, similar to low-field strategies. Although preliminary sensitivity estimates ($\sim$410\,pT/Hz$^{1/2}$ at 7\,T) are lower than those of optically detected electrons, high-field operation reduces sensitivity requirements for NMR sensing. Lastly, the nuclear spins here operate in the high-density limit ($>10^4$ times denser than NV centers). In this regime, interactions between individual sensor spins serve as a means of ``stabilization'', increasing the robustness to control errors. 

    \item \textbf{Material science and quantum control approaches to slowing electron relaxation}: Several efforts to improve sensor coherence through materials advancements are underway, including controlled depletion methods for creating nuclear spin-free host lattices, plasma annealing for producing pristine diamond surfaces \cite{Sangtawesin2019}, and rapid high-temperature annealing to quench lattice paramagnetic impurities \cite{Gierth2020}. Additionally, there are promising prospects for employing sensing regimes where electronic decoherence is naturally suppressed. Of particular interest is the low-temperature high-field regime (e.g., 4\,K and 10\,T), where all paramagnetic spins are fully polarized, suppressing spin flip-flops with the sensor and significantly lengthening coherence \cite{Takahashi2008}. We anticipate more quantum sensing experiments leveraging this regime, particularly for nanoNMR applications. 

\end{enumerate}

\subsubsection*{\textbf{Concluding Remarks}}
In conclusion, we emphasize that ongoing developments in quantum control methods are crucial for quantum sensing and anticipate exciting prospects for the development of high specificity sub-micron-scale chemical sensors. These advancements portend several new applications in various fields. 

\subsubsection*{Acknowledgements}
We acknowledge funding from AFOSR YIP (FA9550-23-1-0106) and AFOSR DURIP (FA9550-22-1-0156).
\newpage

\subsection{Nano MRI using spin quantum sensors}
J\"org Wrachtrup\\
Physikalisches Institut, University of Stuttgart, Pfaffenwaldring 57, 70569 Stuttgart, Germany \textit{and}\\
Max Planck Institute for Solid State Research, Stuttgart, Germany\\
Fedor Jelezko\\
Institute of Quantum Optics, Ulm University, Ulm 89081, Germany\\
j.wrachtrup@pi3.uni-stuttgart.de, fedor.jelezko@uni-ulm.de

\subsubsection*{\textbf{Status}}
The use of spin ensembles or even single spins to measure various quantities, like magnetic- or electric field, temperature etc. has gone through spectacular developments in recent years \cite{Degen2017}. Since its initial demonstration in 2008 detection and imaging of magnetic fields from micro- and nanomagnetic structures, the imaging and detection of electric fields as well as temperature and temperature profiles, the detection of chemical compounds like triplet oxygen, or radicals have been achieved \cite{Degen2017}. All of the above is based on defects in diamond (or other materials like silicon carbide or, e.g., hexagonal boron nitride). Their electron spins Zeeman splitting and/or spin relaxation among them is used to measure those quantities. Optical detection of single spin sensors achieves such high sensitivity that, e.g., the magnetic field of single spins outside of the sensor can be measured. When incorporated into scanning probe devices, imaging with a spatial resolution only limited by the stand-off distance between sensor and sample is feasible. This has been used to image the magnetic field profile of micromagnetic structures with spatial resolution down to 20 nm \cite{Balasubramanian2008}. The sensitivity reached in these systems depends on the specific experimental modality and ranges from $\upmu$T/Hz$^{1/2}$ to a few nT/Hz$^{1/2}$. Imaging and detection of e.g. magnetic fields can also be done by wide field magnetometry \cite{Steinert2010, Ziem2013}. In this setting, a thin layer of spins is implanted in a diamond substrate and local magnetic fields are measured and imaged through detecting a spatially selected part of the diamond chip. Micromagnetic structures as well as dynamic magnetization has been measured this way. 
An ensemble of spins is also used for the so far most sensitive measurements of magnetic fields with solid state spins. Using $10^{14}$ spins in a volume of a few hundred $\upmu\mathrm{m}^3$ fields smaller than 1\,pT/Hz$^{1/2}$ were detected. Flux concentration further on reduced this sensitivity to below 900\,fT/Hz$^{1/2}$ \cite{Fescenko2020}.
Not only DC fields, but also AC fields with frequencies up to 120\,GHz can be detected \cite{Stepanov2015}. This allows for measuring NMR signals of small quantities of sample spins. First demonstrations comprise measurements of electron spins on diamond surfaces \cite{Mamin2012} as well as of spin labels on proteins. Later, this was extended to measuring nuclear spins \cite{Staudacher2013} down to the level of single proton - or nuclear spins on single proteins \cite{Lovchinsky2016}. Besides sensitivity in nanoscale NMR experiments, a central challenge is to achieve the required spectral resolution to resolve quantities like chemical shift or J-coupling. To this end, different methods have been developed. Synchronized readout uses the frequency stability of an external reference to achieve $\upmu$Hz resolution \cite{Degen2017}, while quantum memory-based techniques yield mHz spectral resolution \cite{Aslam2017}. Electron spin resonance as well as NMR detection was combined with spatial resolution, imaging for example a single electron spin \cite{Grinolds2013} or small amounts of nuclear spins with spatial resolution of around 20\,nm \cite{Haeberle2015}. For small standoff distances, the back action of quantum sensor on nuclear spins is not negligible and weak measurement based schemes can be employed to mitigate the effect of backaction \cite{Pfender2019, Cohen2020}.

\begin{figure}[h]
    \centering
    \includegraphics[width = 0.5\textwidth]{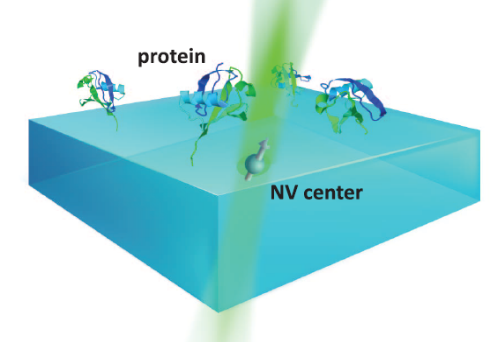}
    \caption{NMR signal of a single ubiquitin protein \cite{Lovchinsky2016} Let side: schematical representation of the sample geometry. Right side: $^{13}$C and $^2$H spectrum of a single ubiquitin protein. The protein was $^{13}$C and $^2$H labeled.}
    \label{Wrachtrup:fig1}
\end{figure}

\subsubsection*{\textbf{Current and Future Challenges}}
Combined with flux concentration, solid state spin sensors do have the potential to achieve sub 100\,fT/Hz$^{1/2}$ sensitivity while maintaining a rather large bandwidth. This requires material improvements, specifically achieving maximum signal contrast as well as an efficient conversion of AC- to DC-signals. The large bandwidth may then be used to design robust gradiometers. A further advance may come from oriented growth of defects which would improve sensitivity and facilitate operation of the magnetometer.
Imaging systems, like wide field - or scanning probe magnetometry would greatly benefit from an increase in sensitivity and spatial resolution. For wide field magnetometry the spatial resolution achieved so far is diffraction limited, while for scanning probe systems the sensitivity is limited because often single defect centers are used for detection. This limits their sensitivity, while the spatial resolution of the technique is still limited the probe to sample distance. 
The detection of NMR and EPR signals by spin quantum sensors relies on sensor to sample distance. Achieving closer proximity without degradation of the spin relaxation and dephasing times is a central challenge in the field. At the same time, high resolution NMR requires averaging of the nuclear dipolar interaction by fast molecular motion. This however leads to diffusion broadening in nanoscale NMR detection. On the other hand, spin ensemble-based NMR and EPR usually does not yield the expected increase in sensitivity as either the quality, i.e., coherence properties of the spin-sensor or the proximity to the sample is not ideal.  
Detection of nuclear spins in nanoscale NMR experiments often employs statistical polarization of nuclear spins. Although this method allows to detect nuclear spins in moderate or zero magnetic fields, the diffusion of spins leads to broadening of NMR lines, limiting spectral resolution \cite{Schwartz2019}. Confinement of nuclear spins in nanoscale structures adjusted to the electron spin sensor allows to reduce diffusion related broadening and opens a path to reach high spectral resolution \cite{Liu2022}. Another possibility to reduce diffusion broadening is related to use of the nuclear spin hyperpolarization combined with quantum sensing enabled by electron spins. External spins can be polarized by quantum sensor itself \cite{Shagieva2018, Healey2021} or using conventional hyperpolarization techniques \cite{Arunkumar2021}. Note that diffusion broadening, and line shape of NMR signals can provide important information about the motion of molecules near surfaces \cite{Cohen2020a}.

\subsubsection*{\textbf{Advances in Science and Technology to Meet Challenges}}
A distinct potential for further improvement comes from improved material quality. Nanoscale probes would greatly benefit from improved performance of spin centers close to surface, i.e., at distances less than 5\,nm. In combination with advanced tip geometries this might boost the current spatial resolution to below 10\,nm and at the same time improve the sensitivity of the method. Likewise, multi spin probes will enhance the sensitivity without degradation - they currently achieve spatial resolution of around 30\,nm. The sensitivity in ensemble magnetometry is also mostly determined by material parameter, like ensemble dephasing time, short $T_2^*$ and limited signal contrast. All of these parameters can potentially be improved by materials refinement. 
So far, experimental demonstrations of spin-based NMR quantum sensing were realized using bulk sensing material and shallow sensing spins. In the future, this approach can be extended to doped nano diamonds. Currently, diamond nanoparticles doped with NV centers were shown to be promising sensors for the detection external electron spins via relaxometry \cite{Mzyk2022}. Detection of nuclear spins using nanodiamonds remains challenging, and only a few demonstrations were reported so far \cite{Holzgrafe2020}. Optimization of nanodiamond growth including techniques allowing generation of nanoparticles with well-defined shape and dopant composition is essential for future development of quantum sensing. This includes self-nucleation-based growth of nanoparticles using high pressure high temperature approach and technique based on CVD growth.
Optimization of quantum materials hosting quantum sensors needs to be combined with the development of quantum sensing protocols, allowing to reach long coherence time and improve sensitivity. Promising techniques include optimized dynamical decoupling, control tools tailored to specific noise environment including pulsed and continuous approaches \cite{Degen2017}. Dynamical decoupling suitable for cancelling low frequency noise can be combined with quantum error correction capable to protect sensing qubits from arbitrary noise spectrum of low amplitude.

\subsubsection*{\textbf{Concluding Remarks}}
Spin based quantum sensing provides a versatile tool allowing to improve sensitivity of nuclear magnetic resonance and reach the limit of detection of single molecules. Future development of this technique will be directed to application in reconstruction of the structure and dynamics of single molecule and unraveling chemical reactions at nanoscale.  Future development will require significant advances in material and quantum control approaches.

\subsubsection*{Acknowledgements}
JW acknowledges funding by the BMBF via the cluster4future Qsens, DiaQNOS and NeuroQ the BW foundation via project SPOC, the EU via projects AMADEUS and C-QuENS as well as the Max Planck Society. FJ acknowledges support by the BMBF, the DFG and the ERC.

\newpage

\section{STM-ESR}
\label{STM}

\subsection{An atomic spin sensor on surfaces}
Yujeong Bae and Andreas J. Heinrich\\
Center for Quantum Nanoscience, Institute for Basic Science, Seoul 03760, South Korea \textit{and}\\
Department of Physics, Ewha Womans University, Seoul 03760, South Korea\\
bae.yujeong@qns.science, heinrich.andreas@qns.science\\

\subsubsection*{\textbf{Status}}
\begin{figure}[h]
    \centering
    \includegraphics{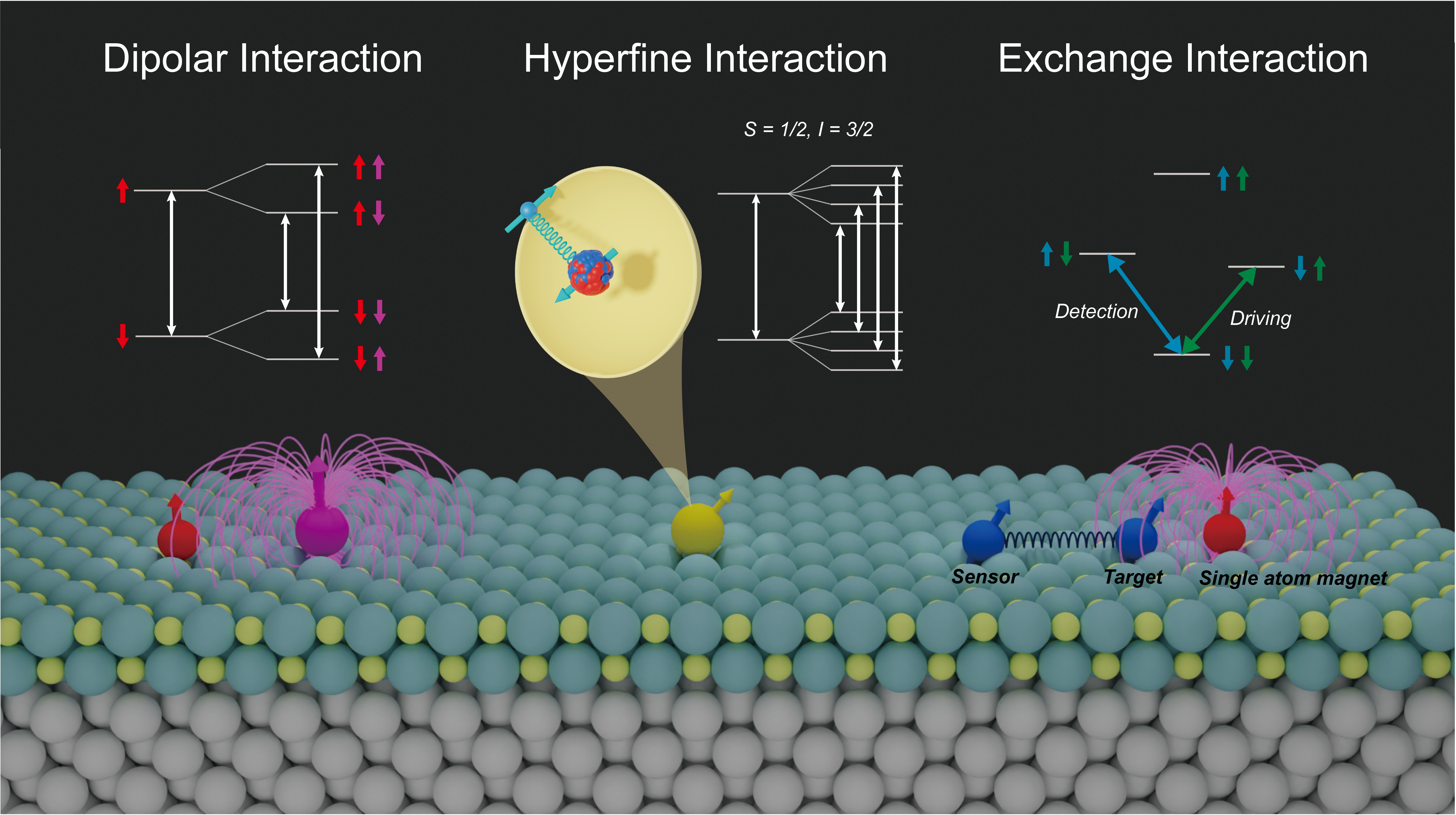}
    \caption{Magnetic interactions used for indirect sensing in ESR-STM and observable ESR transitions. Left: Dipolar interaction between a sensor atom (red) and a target atom (magenta), where the energy levels of the sensor atom are split by the dipole fields from the target atom. Middle: Hyperfine interaction between electron and nuclear spins with ESR transitions at different nuclear spin states. Right: Exchange interaction between two electron spins (one from sensor and one from target), where the coherent manipulation and detection of the target spin are implemented via a single atom magnet and the sensor spin, respectively.}
    \label{Heinrich:fig1}
\end{figure}

Integrating two experimental approaches is often vital in innovation of research, allowing more complex problems to be tackled and more sophisticated technological advances to be achieved. Electron spin resonance (ESR) combined with scanning tunneling microscopy (STM) is one of those examples, which enables us to build artificial quantum structures atom by atom on a surface and characterize their quantum states with tens of nano-eV energy resolution at the atomic scale \cite{Chen2022}. The capability to structure matter at the nanoscale and probe weak electrical signals at the single atomic level renders ESR-STM an appealing technique for exploring quantum phenomena in engineered nanostructures. 
Since its first demonstration \cite{Baumann2015}, ESR-STM has been intensively used to explore different atomic/molecular species on surfaces \cite{Yang2017, Bae2018, Yang2018, Kovarik2022, Zhang2021a, Kawaguchi2022}. Beside control and detection of single electron spins, advanced applications of ESR-STM can be seen in coupled spin systems designed for ``indirect'' sensing of target spins. Indirect sensing is appropriate i) to probe spins with no detectable spin resonance transitions \cite{Natterer2017, Singha2021} or ii) to coherently control multiple spins, including one spin at the STM junction and ``remote'' spins coupled to and indirectly read out through the spin at the STM junction \cite{Wang2023}. 
In indirect sensing, a ``sensor'' spin is located at the STM junction and experiences resonant excitations through direct coupling with the magnetic tip, which has been described by either a piezo electric coupling \cite{Seifert2020} or a tunnel barrier modulation \cite{Delgado2021, ReinaGalvez2023}. Its resonance signal appears as a variation of tunnel current due to the DC and AC tunneling magnetoresistance effects \cite{Bae2018, Seifert2020}. The target spin is magnetically coupled to the sensor spin, where the coupling strength is precisely controlled by adjusting the atomic separations \cite{Choi2017} or the chemical environment \cite{Willke2018} using atom manipulation in STM. As shown in Fig.\,\ref{Heinrich:fig1}, three different couplings have been employed for indirect sensing: i) the dipole-dipole coupling \cite{Choi2017} to determine the magnetic moments and spin lifetime of target spins that have no spectral features in spin resonance or inelastic tunneling spectroscopies (e.g. Ho \cite{Natterer2017} and Dy \cite{Singha2021}), ii) the hyperfine interaction to probe the nuclear spin state and the chemical environment \cite{Yang2018, Willke2018, Farinacci2022, Kim2022}, and iii) the exchange interaction between two electron spins largely detuned in energy for remote driving and detection of target spins \cite{Wang2023}. 
Advancing these indirect sensing methods would impact both the metrological practice and the applications of ESR-STM: it would allow us to improve the quantum coherent properties of target spins, as well as to explore different material systems with fewer prerequisites of target spins for ESR-STM measurement. These impacts will facilitate the generalized use of ESR-STM to explore a variety of material systems for quantum applications.

\subsubsection*{\textbf{Current and Future Challenges}}
\begin{figure}[h]
    \centering
    \includegraphics{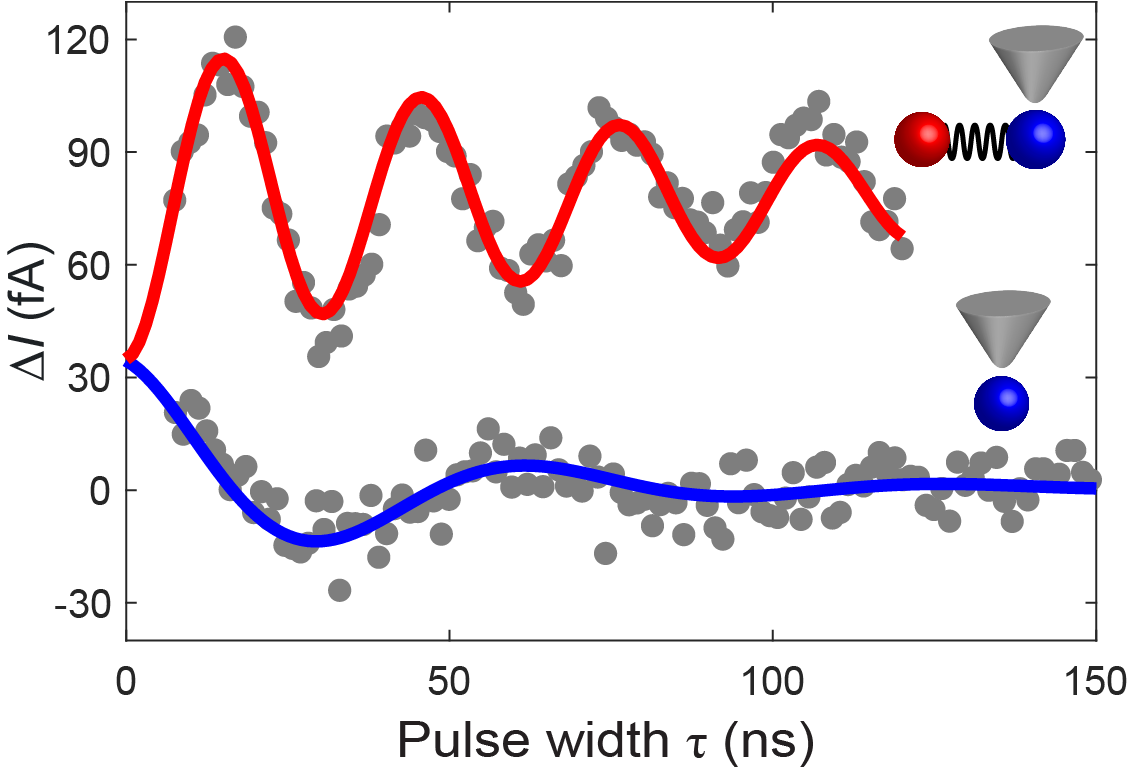}
    \caption{Rabi oscillations of a remote spin (red) through indirect sensing and a spin at the STM junction (blue), measured using ESR-STM. The remote spin (red), as measured using double electron-electron resonance scheme, shows Rabi oscillations with better signal-to-noise ratio and longer spin coherence time, compared to the spin at the STM junction (blue) (11). Measurement conditions: $B = 0.656$ T, $T = 0.4$ K, (red) $V_\mathrm{DC} = 50$ mV, $I_\mathrm{set} = 10$ pA, $V_\mathrm{RF} = 60$ mV \& 120 mV, (blue) $V_\mathrm{DC} = 50$ mV, $I_\mathrm{set} = 5$ pA, $V_\mathrm{RF} = 120$ mV.}
    \label{Heinrich:fig2}
\end{figure}

One of the main challenges of ESR-STM is to improve its sensitivity and energy resolution. ESR signals in STM are based on changes in tunnel current depending on the relative alignment of two spins at the STM junction: one at the STM tip apex and one on a surface. Using a typical magnetic tip, the ESR signals appear as about 1-5\% changes in tunnel current (about 0.2-1\,pA changes at 20\,pA set-point current). In addition, the best energy resolution achieved with ESR-STM is several tens of nano-eV (several MHz) with no significant improvements over the last 7 years. At the current sensitivity and energy resolution, spins with small magnetic moments, such as a spin-1/2 system, are detectable only if the sensor spin is located within 1.5\,nm of the target spin in the dipole-field sensing. In addition, while the hyperfine interaction has been measured only for atoms which carry both electron and nuclear spins \cite{Yang2018, Willke2018, Farinacci2022, Kim2022}, the super-hyperfine interaction can be of great interest for molecules with radicals, which requires both higher sensitivity and better energy resolution. 
Improving the energy resolution is closely related to enhancing the relaxation and coherence times of spins on surfaces. Previous studies found that the tunnel current, coupling with substrates, and magnetic field fluctuations caused by the magnetic tip are the most dominant sources of decoherence \cite{Willke2018a, Yang2019}. In indirect sensing, where the target spin is remotely controlled with assistance of a single atom magnet and monitored through a sensor spin, the target spin is not directly exposed to those decoherence sources occurring at the STM junction \cite{Wang2023}. As shown in Fig.\,\ref{Heinrich:fig2}, the target spin in indirect sensing shows significantly enhanced signal-to-noise ratios and spin coherence. Further improvement on the coherent properties of target spins might be achieved by increasing the spin relaxation time of the sensor spin and by reducing the coupling strength with substrate electrons. 
Finally, one interesting challenge is to have a sensor spin in diverse environments, e.g., at the tip apex. At the moment, the only material systems used as a sensor are transition metals (Fe and hydrogenated-Ti) on two monolayers of MgO/Ag(100). However, there are nearly infinite combinations of atoms or molecules on different surfaces that will show different spin states. Furthermore, many interesting molecules, e.g., carbon-based molecules, barely adsorb on insulating surfaces such as MgO and NaCl. To apply this ESR-STM technique to diverse material systems, it is therefore necessary to develop a robust sensor spin that works well in different environments.

\subsubsection*{\textbf{Advances in Science and Technology to Meet Challenges}}
While using a spin-polarized tip is essential to drive and detect ESR in STM, the magnetic states of the tip are not well-known, barely controllable, and described in a classical picture. Understanding the role and effects of magnetic tips on spins on surfaces for ESR would help to improve the sensitivity and energy resolution of ESR-STM. While most magnetic tips for ESR-STM have been prepared by attaching several magnetic atoms to the tip apex, tips functionalized by better characterized spin-carrying objects might be beneficial for ESR measurement in STM, e.g., spin states of a molecule or spin-polarized Yu-Shiba-Rusinov states. Having the quantized states at the tip might provide even more sensitive probes to detect and control spins on a surface. 
Aside from increasing sensitivity and energy resolutions, having quantized spin systems at the tip apex enables scanning magnetometry at the atomic scale. In case of using a sensor spin on a surface to characterize target spins, the spatial resolution is determined by the surface lattice on which the target and sensor spins can be located. In addition, the spin states of the sensor would change on different surfaces. In contrast, having a robust quantum sensor spin at the tip apex as a mobile sensor provides a universal solution in characterizing different spin systems and enables 3D mapping of the target spin with atomic resolution. 
These advances in the ESR-STM technique would also be beneficial for mitigating the impact of decoherence sources. Improving sensitivity and energy resolution facilitates adjusting the measurement conditions, such as the set-point tunnel current and the tip-sample distances, in a way to minimize sources of decoherence. Additionally, even though we protect spins by remotely driving and detecting their states, the coherence time of the remote spin is still limited by coupling with substrates. Decoupling the spin from the substrate by increasing thickness of insulating layers or using spins carried by a molecule with tailored spacing ligands would enhance the relaxation and coherence times of the spins on surfaces. In addition, well-protected spin states like nuclear spins or 4f-electrons have been barely explored for their coherent properties using ESR-STM. Identifying different spin systems that can be probed with ESR-STM is a necessary step for making this technique more versatile for generalized use.

\subsubsection*{\textbf{Concluding Remarks}}
Recent progress on indirect sensing in ESR-STM has led to the first observation of the hyperfine interaction at the single atom level, as well as the dipolar fields from single atom magnets on a surface. Taking advantage of precise atom manipulation, indirect sensing of spin states becomes more beneficial for STM-based studies. The newly proposed approach to control and sense remote spins, which are not located directly in the STM junction, opens an avenue to the simultaneous but independent control of multiple spins in nanostructures designed with atomic precision. Considerable improvements in the energy resolution of ESR-STM and the spin relaxation and coherence times of both a sensor and a target are still necessary to achieve more precise control and detection of quantum spin architectures built on a surface. In addition, functionalizing the STM tip using molecules with well-isolated spin states would advance ESR-STM for achieving the quantum sensing with atomic resolutions. Since the ESR-STM technique is still relatively new and less explored, applying this technique to different spin systems would certainly enlarge its importance in the fields of quantum technology and quantum materials.

\subsubsection*{Acknowledgements}
We acknowledge Kyungju Noh and Yu Wang for their fruitful discussions. This work has been supported by the Institute for Basic Science, Korea (Grant No. IBS-R027-D1) and the Asian Office of Aerospace Research and Development (Grant No. FA2386-20-1-4052).

\newpage
\subsection{Sensing the electron spin resonance at the atomic scale}
Christian R.\,Ast\\
Max Planck Institute for Solid State Research, Heisenbergstraße
1, 70569 Stuttgart, Germany\\
c.ast@fkf.mpg.de

\subsubsection*{\textbf{Status}}
\begin{figure}
    \centering
    \includegraphics[width = 0.5\textwidth]{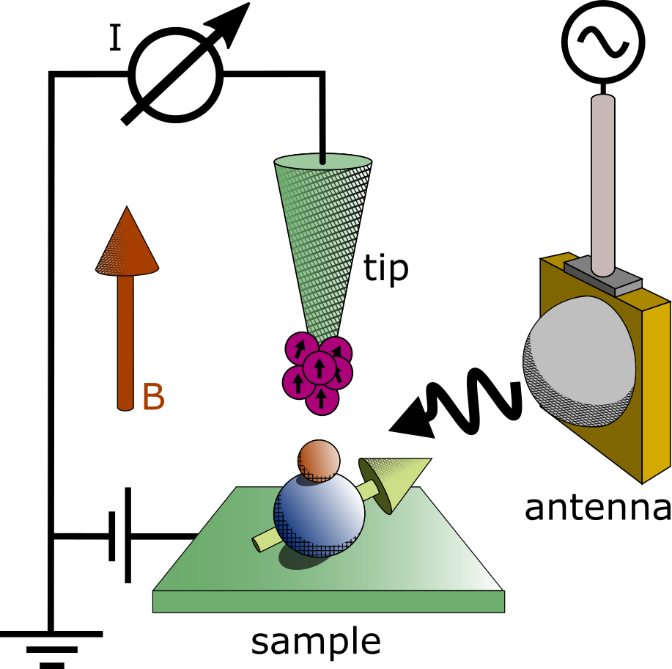}
    \caption{Principles of STM-based ESR. The spin system (here a TiH molecule) lies on an insulating MgO layer on Ag(100) (sample), which feels a Zeeman splitting from an external magnetic field. The tip, which is spin-polarized by a cluster of Fe atoms (purple), measures the tunneling current through the spin system. The microwave excites the spin system (here from an antenna).}
    \label{Ast:fig1}
\end{figure}

The idea to measure the electron spin resonance (ESR) signal from a single atom or molecule is probably as old as the technique itself. Combining the ESR concept with an atomically resolving technique, such as scanning tunneling microscopy (STM), only seems natural and has been attempted already more than twenty years ago \cite{Balatsky2002, Balatsky2012}. However, it was not until a few years ago, when the breakthrough was demonstrated that pushed the ESR-STM technique over the top \cite{Baumann2015}. Since then, the ESR-STM technique has steadily developed and provided new insight by combining the capabilities of STM with the possibilities of ESR \cite{Kovarik2022, Zhang2021a}. Many of the concepts that have been successfully applied in ``conventional'' ESR experiments, such as magnetic resonance imaging (MRI) \cite{Willke2019}, pulsed-ESR \cite{Yang2019}, and nuclear magnetic resonance (NMR) measurements \cite{Willke2018}, just to name a few, have been successfully applied to single atoms and molecules in the STM.
The locally confined and atomically resolved detection of the ESR signal on a single atom or molecule is achieved through the local detection of the tunneling current through the atomically sharp tip \cite{Paul2016}.
This is probably the most decisive difference to conventional ESR experiments. The excitation of the spin system by the microwave is similarly confined as the current measurement. The strong field enhancement in the vicinity of the tip apex confines the area of excitation to the region under the tip. The ESR signal is detected through the change in the spin-polarization of the tunneling current on resonance vs. off resonance. This is achieved by employing an appropriately spin-polarized tip. The experimental setup is schematically shown in Fig\,\ref{Ast:fig1}.
It was quickly realized that the interaction of the tunneling current with the spin system is a significant source of perturbation and decoherence for the spin system. Still, the interaction is necessary for the spin state to be detected. Even at small tunneling currents in the low pA range, the tunneling current remains the biggest source of decoherence \cite{Willke2018a}. This detection mode is referred to as ``direct sensing'', where the tunneling current passes through the spin system under investigation. In order to circumvent this problem, an ``indirect sensing'' technique is being developed, where the spin polarized tunneling current interacts with a sensor spin, which in turn is magnetically coupled to the sample spin under investigation \cite{Choi2017}. In this way, the sample spin system is not perturbed by the tunneling current and it is easier to probe the intrinsic properties of the spin system. The concept of ``indirect sensing'' will be discussed in the next section.
The ESR-STM technique is still young and many aspects not yet fully understood. For example, the coupling mechanism how the microwave excites the spin system is still under debate \cite{Baumann2015, Seifert2020, ReinaGalvez2019}. Also, the overall number of spin systems and substrates that have been investigated are rather limited. The set of atoms and molecules that can be studied, as well as the different substrates on which the ESR measurement works gradually increases. As such, ESR-STM will greatly benefit from reaching a sophistication that allows to reliably address arbitrary and unknown spin systems on an atomic scale.

\subsubsection*{\textbf{Current and Future Challenges}}
The challenges faced by the ESR-STM technique can be divided into two categories, the challenges that are inherent to the ESR technique, which are common to all implementations of sensing an ESR signal as well as the challenges that are specific to ESR-STM. The common challenges can be summarized as improving the measurement sensitivity and enhancing the coherence time. The measurements' sensitivity has already reached its limit for ESR-STM in the sense that a single spin system can be detected. However, as the tunneling current goes through the spin system under investigation in the direct sensing mode, the tunneling current is a major source of decoherence \cite{Willke2018a}, such that reducing the tunneling current to a minimum while still being able to detect an ESR signal is the sensitivity that has to be optimized. The challenge here is that only a small fraction of the total current that is being measured carries the ESR signal. To filter out the ESR current signal, the microwave signal is chopped with a certain frequency (usually around 100\,Hz) and then the ESR current signal is detected with a lock-in amplifier at that frequency, which is essentially equivalent to measuring the current difference between microwave on and off \cite{Paul2016}. The resulting ESR signal current is usually two to three orders of magnitude smaller than the total tunneling current, but all of the current contributes to the decoherence. Therefore, the ESR signal is likely lost before the decoherence from the tunneling current is significantly suppressed. Alternative detection schemes (discussed below) and indirect sensing (discussed in the next section) can provide improvements to decoherence during the measurement.
On a more ESR-STM specific challenge, the spin system under investigation is typically rather strongly bound to its substrate. This is another, more intrinsic source of decoherence, which can be similar in magnitude as the tunneling current. The challenge here is to reduce the coupling to the substrate as much as possible, but keeping a measurable tunneling current. This can be implemented by increasing the thickness of the insulating layer. The coupling to the substrate does not only influence the coherence times, it also changes the properties of the spin system through a crystal field splitting as well as spin-orbit coupling. Therefore, a measurement of the g-factor, for example, has to take the local environment into account and cannot be simply attributed to the spin system under investigation \cite{Steinbrecher2021, Kot2023}. With an improved understanding of the ESR-STM mechanism, which is still under debate, the influence of the substrate and the local environment can be accounted for more clearly.
The challenges the ESR-STM technique is currently facing can be in large part attributed to the youth of this technique. Many of these challenges are currently being addressed and will be addressed in the future.

\subsubsection*{\textbf{Advances in Science and Technology to Meet Challenges}}
The tunneling current as a source of decoherence is likely the foremost challenge in improving the ESR-STM technique. Staying with the direct sensing concept, where the tunneling current flows through the spin system of interest, probably the best way to reduce the tunneling current to a minimum is to employ homodyne detection of the ESR signal at zero bias voltage \cite{Bae2018}. The only prerequisite for measuring a sizeable homodyne signal is a magnetization of the spin-polarized tip, which has a sizeable component perpendicular to the magnetization of the spin system. This has, in principle, been demonstrated experimentally already, but it has not been implemented and tested as a routine measurement, yet. Going beyond direct sensing to indirect sensing is discussed in the next section. This would eliminate the decoherence through the tunneling current, but not the decoherence through the coupling to the substrate. In order to decrease the decoherence through the substrate, the thickness of the insulating layer has to increase while still being able to measure a tunneling current as well as an ESR signal. The material that is mostly used here is MgO. Different layer thicknesses have been investigated already with increasing lifetimes of the spin systems as the thickness increases \cite{Paul2016a}. More recently, NaCl has been explored as an insulating layer and other materials will follow \cite{Kawaguchi2022}. The challenge here is that it is still unclear, which part the substrate plays in the ESR mechanism.
Overall, the ESR-STM technique would greatly benefit from a better theoretical understanding of the ESR mechanism in the STM. Several aspects are still under debate, such as the excitation of the spin system by the microwave field. While there is a strong enhancement of the electric field under the tip apex, it is not quite clear how the electric field couples to the spin system. One possibility is through a piezoelectric-like motion of the spin system moving in the magnetic field gradient of the spin-polarized tip \cite{Baumann2015}. With a better theoretical understanding, technological improvements to the technique could be more targeted and effective.

\subsubsection*{\textbf{Concluding Remarks}}
The ESR-STM technique is still young compared to the ESR technique as a whole. The advancements in the past few years have been tremendous, but there are still many aspects to be understood and explored. The full potential of the technique is by no means exhausted or even realized. In this sense, the ESR-STM technique is still in an adolescent ``assessment stage'' rather than in a more mature ``refinement stage''. Therefore, only a few major construction sites could be identified within this roadmap. It will be exciting to see the next few years and witness how this technique evolves towards maturity.

\subsubsection*{Acknowledgements}
We gratefully acknowledge Piotr Kot, Maneesha Ismail, and Janis Siebrecht for fruitful discussions. This work has been funded in part by the ERC Consolidator Grant AbsoluteSpin (Grant No. 681164) and by the Center for Integrated Quantum Science and Technology (IQ$^\mathrm{ST}$).

\newpage
\section{Superconducting Resonators}
\label{SC}

\subsection{Nanoscale magnetic resonance with superconducting circuits}
Patrice Bertet\\
Université Paris-Saclay, CEA, CNRS, SPEC, 91191 Gif-sur-Yvette, France\\
patrice.bertet@cea.fr

\subsubsection*{\textbf{Status}}
In conventional Electron Spin Resonance (ESR) spectroscopy, paramagnetic impurities are detected by their interaction with the magnetic component $B_1$  of the microwave field inside a resonator of frequency $\omega_0$. This approach applies to all types of spin systems and samples, but it usually lacks sensitivity to detect small numbers of spins, preventing its use for small samples or for spin-based quantum computing where single-spin-qubit addressing is needed. 
In the quest to improve resonator-based ESR sensitivity, low-mode-volume metallic-thin-film-based planar microwave resonators such as micro-coils were developed, enabling sensitivities in the $10^6$-$10^8$ spin/Hz$^{1/2}$ \cite{Narkowicz2005, Dayan2018}. Maximizing sensitivity however requires not only a small resonator mode volume, but also a high quality factor, bringing strong motivation for patterning the planar resonator with superconducting thin-films \cite{Wallace1991}. One difficulty is that a relatively large magnetic field  is often needed to tune the spin Larmor frequency at $\omega_0$, which may induce losses and hysteresis in a superconducting resonator. By applying  $B_0$ in the plane of the superconducting thin-film \cite{Wallace1991, Sigillito2014}, carefully designing the resonator, and choosing a superconductor with high critical field such as Nb, NbN, or NbTiN, fields of order 1\,T were reached, sufficient for X-band EPR \cite{Graaf2012, Zollitsch2019}. A major impact of coupling spins to a high-quality-factor low-mode-volume resonator is that the rate of microwave photon spontaneous emission becomes enhanced via the Purcell effect, to the point where microwave radiation can become the dominant spin energy relaxation mechanism \cite{Bienfait2016, Eichler2017}.
Further sensitivity gains are obtained using nearly-noiseless Josephson Parametric Amplifiers (JPAs), developed in the context of quantum computing with superconducting qubits \cite{Macklin2015}, to amplify the spin signal. Pulsed ESR measurements with superconducting micro-resonators and JPAs at millikelvin temperatures demonstrated a dramatically enhanced sensitivity, in the $10$-$10^4$ spin/Hz$^{1/2}$ range \cite{Eichler2017, Bienfait2015, Probst2017, Ranjan2020}. In these experiments, the spin echo signal is detected by linear amplification and demodulation, and is therefore sensitive to the quantum fluctuations of the $B_1$ field, which impose a physical limit to the achievable sensitivity. It is possible to reduce the quantum fluctuations below this limit on the quadrature on which the spin-echo is emitted, at the expense of increased noise on the orthogonal quadrature to satisfy Heisenberg uncertainty relations; this quantum squeezing approach demonstrated however modest sensitivity gain \cite{Bienfait2017}. A more promising approach is to detect the microwave photons emitted by the spins upon radiative relaxation with an energy detector such as a superconducting-qubit-based Single Microwave Photon Detector (SMPD) \cite{Lescanne2020}, which is then insensitive to the quantum fluctuations of $B_1$ \cite{Albertinale2021}. With this approach, resonator-based single-spin ESR spectroscopy was recently reported, with a spin sensitivity reaching 0.6 spin/Hz$^{1/2}$ \cite{Wang2023a}.

\begin{figure}
    \centering
    \includegraphics[width = 0.5\textwidth]{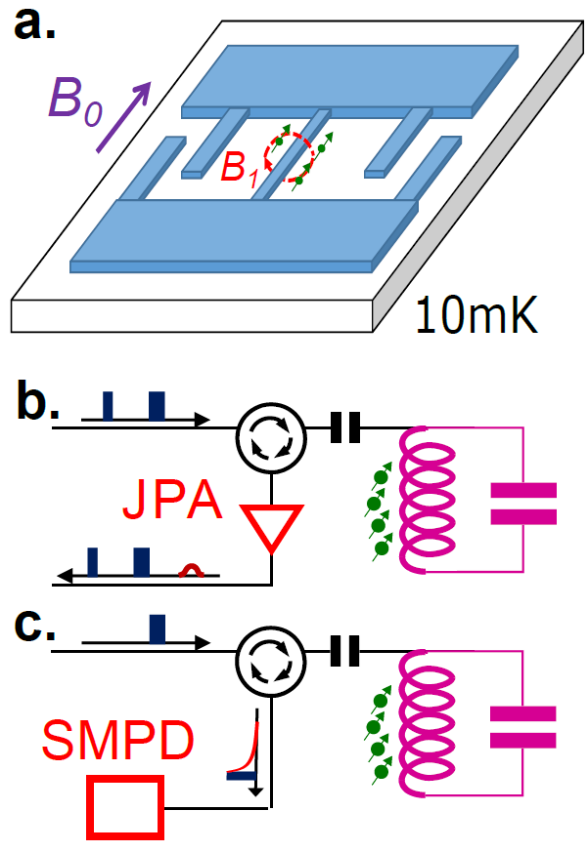}
    \caption{(a) Schematics of the experiments. A planar LC resonator is patterned out of a superconducting thin film, on top of a crystal host containing the spins (green arrows). The spins couple to the resonator by dipolar coupling with the magnetic field $B_1$ (red dashed arrow), which is strongest near the inductive wire. The sample is cooled to 10\,mK in a dilution refrigerator, and a dc magnetic field $B_0$ is applied parallel to the resonator to tune the spin frequency in resonance. (b) Inductive Detection. The spins are probed by a Hahn echo sequence $(\pi/2)-\tau-\pi-\tau-\mathrm{echo}$, and the echo emission is amplified by a Josephson Parametric Amplifier (JPA) and detected in quadrature. (c) Fluorescence Detection. The spins are excited by a $\pi$ pulse, and the incoherent microwave spontaneous emission is detected by a Single Microwave Photon Detector (SMPD).}
    \label{Bertet:fig1}
\end{figure}

\subsubsection*{\textbf{Current and Future Challenges}}
\begin{figure}[h]
    \centering
    \includegraphics[width = 0.6\textwidth]{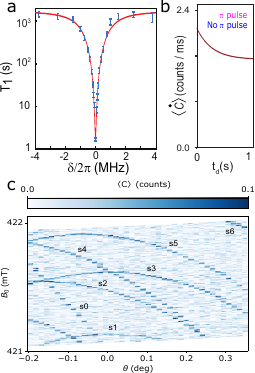}
    \caption{(a) Observation of the Purcell effect for spins. The relaxation time $T_1$ of donors in silicon (blue dots) is plotted as a function of the detuning $\delta$ with the resonator. Blue dots are data, and red solid line is a fit to $1/\left( \Gamma_R(\delta) + \Gamma_{NR}\right)$. Reproduced with permission from Ref.\,\cite{Bienfait2016}. (b) Observation of the spin microwave fluorescence by microwave photon counting. The average count rate $\langle C \rangle$ is shown as a function of the time $t_d$ from an excitation $\pi$ pulse. Blue bars are without pulse, and magenta bars are with pulse applied. Solid purple line is an exponential fit yielding the radiative decay rate $\Gamma_R^{-1}=0.3$\,s. Reproduced with permission from Ref.\,\cite{Albertinale2021} (c) Rotation pattern showing the resonance signal from seven individual electron spins labelled from s0 to s6, detected by the fluorescence method. The number of counts following an excitation pulse integrated for 2ms is plotted as a function of the magnitude and angle of the applied magnetic field. Reproduced with permission from Ref.\,\cite{Wang2023a}.}
    \label{Bertet:fig2}
\end{figure}

\textit{Inductively-detected pulsed EPR at the quantum limit.} 
We now describe the experiments in more details. The resonator consists of a finger capacitor shunted by a micron-wide inductive wire, patterned out of a planar thin superconducting film (typically 50\,nm-thick niobium) directly on top of the substrate containing the electron spins (see Fig.\,\ref{Bertet:fig1}a). Each spin couples to the resonator via the magnetic dipole interaction, with a strength characterized by the coupling constant $g_0 = \gamma B_1\cdot \gamma \cdot \langle \downarrow | S | \uparrow \rangle$. In this equation, $\delta B_1(r)$ is the root-mean-square vacuum fluctuation of $B_1$ at the spin location $r$, $\gamma$ is the gyromagnetic tensor, and $\langle \downarrow | S | \uparrow \rangle$ is the spin operator matrix element between the ground state $\downarrow$ and the excited state $\uparrow$. 
The resonator is moreover coupled to a measurement line, through which spin driving pulses are sent and which also collects the spin signal, with a microwave energy decay rate $\kappa = \omega_0/Q$, $Q$ being the quality factor. The spin-resonator coupling enhances the spin radiative relaxation rate $\Gamma_R = \kappa g_0^2/\left( \delta^2 + \kappa^2/4\right)$, $\delta = \omega_s(B_0)-\omega_0$ being the spin-resonator detuning; in particular, at resonance ($\delta=0$), $\Gamma_R = 4g_0^2/\kappa$. Signatures of the Purcell regime include the dependence of the spin relaxation time $T_1$ on the detuning \cite{Bienfait2016, Eichler2017} (see Fig.\,\ref{Bertet:fig2}a) and the thermalization of the spin ensemble to the temperature of the intra-resonator microwave field instead of the lattice \cite{Albanese2020}. 
In conventional, inductively-detected pulsed EPR spectroscopy, the ensemble of $N$ spins is probed by the Hahn echo pulse sequence $(\pi/2)-\tau-\pi-\tau-\mathrm{echo}$ (see Fig.\,\ref{Bertet:fig1}b). The echo occurs because of the transient rephasing of the spin dipoles, during the Free-Induction-Decay time $\Gamma_2^{-1}$. It leads to the emission of a microwave pulse, detected in quadrature by amplification and demodulation. The resulting sensitivity $\sim \sqrt{\Gamma_2}/\left(\sqrt{\eta}\Gamma_R\right)$ spin/Hz$^{1/2}$ (taking into account a finite detection efficiency $\eta$) is limited by the vacuum fluctuations of the $B_1$ field, even using an ideal noiseless JPA. A large coupling constant $g_0$ is therefore key to a high sensitivity, as was confirmed in a series of experiments \cite{Bienfait2015, Probst2017, Ranjan2020} where the resonator mode volume was reduced from picoliter to femtoliter, leading to a coupling constant $g_0/2\pi$ increase from 60\,Hz to 2.7\,kHz, and a corresponding sensitivity increase from $1.4\cdot 10^3$ to $12$ spin/Hz$^{1/2}$. 
With such high sensitivity, pulsed EPR measurements of samples with very low concentrations of paramagnetic impurities (at the sub-ppb level or lower) become possible. This gives access to long spin coherence times, as demonstrated with Er$^{3+}$:CaWO$_4$ \cite{LeDantec2021}. Another application is the spatially-resolved spectroscopy of a small ensemble of strain-shifted donors in silicon, with sub-$\upmu\mathrm{m}$ resolution \cite{Ranjan2021}.

\textit{Microwave fluorescence detected EPR spectroscopy}
Another promising approach consists in detecting the microwave photons spontaneously emitted by the spins during their return to equilibrium after excitation by a $\pi$ pulse (the microwave fluorescence), as shown in Fig.\,\ref{Bertet:fig1}c. The radiofrequency spontaneous emission from an ensemble of nuclear spins was detected in 1985 \cite{Sleator1985}, for the purpose of fundamental studies of light-matter interaction. Only recently has it been recognized as a potential sensitive spin detection method. Indeed, if the spins are in the Purcell regime, they will emit $N$ photons following the excitation pulse, and an ideal photon counter will detect them all, noiselessly. In an actual experiment, finite detection efficiency $\eta$ and dark count rate $\alpha$ yield an expected sensitivity $\sim\sqrt{\alpha}/\left(\eta \Gamma_R\right)$ (in spin/Hz$^{1/2}$). Therefore, the sensitivity is only limited by SMPD imperfections, and can be higher than in Inductive Detection as long as $\alpha/\eta<\Gamma_2$.
Single Microwave Photon Detectors (SMPDs) are not available commercially, but first prototype devices were recently developed. They work by mapping the presence (or absence) of a photon at the SMPD input onto the state of a superconducting qubit, and by reading it out subsequently. By repeating this detection cycle, time traces of clicks are obtained, analogous to optical photon detectors \cite{Lescanne2020}. Thanks to the high level of control and high-fidelity readout achieved in the Circuit Quantum Electrodynamics architecture, devices with high efficiency ($\sim 0.5$) and low dark count rates (down to $\alpha\sim 100\,\mathrm{s}^{-1}$) have been reported \cite{Wang2023a}. 
Spin detection by microwave photon counting was demonstrated in 2021 \cite{Albertinale2021}, with an ensemble of donors in silicon. Figure \ref{Bertet:fig2}b shows the SMPD average count rate $\langle C \rangle$ following an excitation $\pi$ pulse. The spin fluorescence appears as an excess of counts that decays exponentially at the radiative relaxation rate, $\Gamma_R=3\,\mathrm{s}^{-1}$ in this experiment. A signal-to-noise analysis showed a sensitivity of $10^3$ spin/Hz$^{1/2}$. More recently, the results were considerably improved by increasing $\eta$ and $\Gamma_R$ and using an improved SMPD with lower dark count rate $\alpha$, reaching a sensitivity of 0.6 spin/Hz$^{1/2}$ \cite{Wang2023a}. With such signal-to-noise, resonator-based single spin detection and control was demonstrated, opening new perspectives for nanoscale EPR spectroscopy as evidenced by the rotation pattern of a small ensemble of seven individual erbium ion spins in a scheelite crystal shown in Fig.\,\ref{Bertet:fig2}c. Even higher sensitivites can be expected by improving the SMPD devices (aiming at higher efficiency and lower dark count rate), and also enhancing further the spin-resonator coupling and therefore the radiative relaxation rate, with optimized resonator designs for instance.


\subsubsection*{\textbf{Concluding Remarks}}
Superconducting quantum technology has enabled the development of new tools for the control and detection of microwave fields at the single photon level. These devices find a direct application in EPR spectroscopy. Planar microwave resonators with low-mode-volume and high-quality-factor enhance the spin-resonator coupling and the spin radiative relaxation rate. Superconducting parametric amplifiers and microwave photon counters enable low-noise detection of the spin signal. Combining these tools, a dramatic enhancement of EPR spectroscopy sensitivity was demonstrated, to the point that the detection of individual paramagnetic impurities has become possible. This opens exciting new perspectives for operational single-spin EPR spectroscopy at millikelvin temperatures. 

\newpage
\section{Quantum control schemes for enhanced spin detection}
\label{SEQ}

\subsection{Composite pulse design for ESR/NMR}
\label{Cappellaro}
Paola Cappellaro\\
Massachusetts Institute of Technology (77 Massachusetts Ave, Cambridge, MA 02139, USA)\\
pcappell@mit.edu

\subsubsection*{\textbf{Status}}
Magnetic resonance has a long history of developing control techniques to manipulate the dynamics of quantum spins, and efficiently extract information about the spin system and their surroundings. Many techniques used in NMR (nuclear magnetic resonance), ESR (electron spin resonance) and MRI (Magnetic Resonance Imaging) have been evolved to be used in quantum technology applications ranging from quantum metrology to quantum computation. Starting from the simple pulsed manipulation of single spins by resonant radiofrequency (rf) and microwave ($\upmu\mathrm{w}$) fields and its mathematical description in terms of a rotating frame, magnetic resonance control has introduced many tools in spin control encompassing hardware, protocols, theoretical analysis and computational design. A few examples stand out for their impact. 
In the quest to measure relaxation times of NMR experiments, E. Hahn developed the first ``quantum error correction'' method, the spin echo \cite{Hahn1950}. By introducing pulsed control (short rf bursts) he was able to measure not only the first FID (free induction decay) but also its revival due to a second pulse. The second pulse creates an effective time reversal, refocusing the relative phases accumulated by the spins during the first evolution period. While the first echo experiment was done with two $\pi/2$ pulses, soon it was discovered that a $\pi$ pulse was more effective, and a sequence of them even more so, when in the presence of slowly varying field inhomogeneities. This train of $\pi$-pulses (Carr-Purcell-Meiboom-Gill or CPMG sequence \cite{Meiboom1958, Carr1954}) has been very effectively adopted in quantum technology and extended to many dynamical decoupling (DD) sequences that protect a quantum system against dephasing (or even more complex noise sources) \cite{Viola1999}. DD sequences have found a particularly fruitful application in quantum sensing, where they are used as band-pass filters to detect AC signals \cite{Taylor2008, Biercuk2011, Zhao2012, Bylander2011, Alvarez2011, Kotler2011, Young2012, Yuge2011} (such as arising from nuclear spins or other magnetic sources at the nanoscale) while cancelling noise contributions at other frequencies.     
The need to robustly apply the desired excitation to a broad (or narrow) sets of spins has later led NMR practitioners to develop a wide set of shaped and composite pulses. Replacing ``rectangular'' pulses with amplitude and phase modulated excitation can correct for practical pulse imperfections (such as distortions or ringdowns) or achieve broadband excitation \cite{Baum1985}.
Finally, to unravel the complex information stored in the measured spectra, NMR researchers developed an array of powerful 2D spectroscopy methods that use correlations to uncover spatial and chemical couplings \cite{Bax1986}. By varying the control sequences, different part of the spin evolution pathways could be singled out, thus providing selective information on the underlying Hamiltonian.   
As this rich toolset of control methods is revisited in the context of quantum technology, and in particular in nano-MRI applications, there is a need to adapt the control sequences to practical and fundamental differences that arise from addressing a small ensemble of, or even single, spins, as well as to novel applications.

\subsubsection*{\textbf{Current and Future Challenges}}
With the advent of several techniques for the measurement of magnetic fields at the nanoscale, including MRFM, single-spin ODMR, and STM-ESR, traditional magnetic resonance control techniques have been embraced to meet new challenges. The methods that have been introduced to achieve MRI at the nanoscale have quickly demonstrated their ability to achieve unprecedented sensitivity, shortly reaching the level needed to measure single electronic and nuclear spins (an 8-order of magnitude improvement in sensitivity with respect to conventional induction methods). 
Soon, it was however realized that to enable imaging and spectroscopy, more complex control sequences were required. For example, DD sequences allow to isolate magnetic fields oscillating at the frequency matching the $\pi$-pulse spacing. While these sequences have been routinely used in NMR, applying them at the nanoscale brought new challenges. For example, combining DD pulses with the presence of high magnetic field gradients as required for MRFM is challenging, due to the degradation in the pulse fidelity over the sample inhomogeneities in either the resonance or driving frequency. Hardware \cite{Eberhardt2008} or optimal control \cite{Rose2018} solutions have been developed to tackle such challenge. 
As the target of NanoMRI is often other spins, their quantum nature might bring additional challenges when using techniques developed in other contexts. For example, evolution due to the pulses can lead not only to the appearance of spurious harmonics \cite{Loretz2015}, leading to faulty identification of nuclear spin, but also to coherent polarization exchange. The exquisite frequency resolution provided by the long coherence times of single spins might exceed the hardware timing resolution capabilities. To address these challenges, precise understanding of the quantum system dynamics and the development of robust pulse control (including phase modulation \cite{Shu2017, Cai2012, Aiello2013, Hirose2012, Souza2011}, aperiodic pulse sequences \cite{Casanova2015, Zhao2011}, pulse shaping \cite{Zopes2017}, and quantum interpolation \cite{Ajoy2017}) need to be further developed. Another challenge that emerges when bringing magnetic resonance techniques to the nanoscale is to precisely take into account the quantum nature of the system. For example, while in traditional MRI experiments the measurement can be treated classically, detection of a single spin is a strong, projective measurement, which would typically forbid stroboscopic measurements. To overcome this issue, it is imperative to minimize the measurement back-action by using weakly coupled nuclear spins and optimized sensing protocols \cite{Pfender2019, Cujia2019}.

\subsubsection*{\textbf{Advances in Science and Technology to Meet Challenges}}
The challenges and opportunities presented by quantum sensing with individual spins have been met with the development of novel techniques that build on magnetic resonance tools, but also often introduce innovative strategies. 
As mentioned, composite pulses have been adopted to tackle new challenges such as large gradients in MRFM and the presence of driving inhomogeneities in spin defect control. Going beyond robust control \cite{Dong2021, Dong2022}, drive modulation has been further extended to achieve new tasks in sensing, such as vectorial \cite{Wang2021} and broadband sensing \cite{Wang2022a}. An exemplary extension of DD to novel goals has been the successful application of DD to sense – and control – nuclear spins \cite{Zhao2012, Taminiau2012, Kolkowitz2012}. It was noticed that DD could not only act as filter for classical baths – thus aiding in the reconstruction of their noise spectrum – but also to detect quantum spins in the environment. Soon, it was realized that these sensing protocols could be interpreted as controlling the nuclear spins. In this picture, the central electronic spin becomes a quantum actuator, inducing controlled evolution of the nearby spin qubits. This method has been applied to control tens of qubits \cite{Abobeih2019}, and can be further used to detect and manipulate spin-spin interactions for NanoMRI \cite{Ajoy2019}. 
In addition to pursuing novel applications of traditional magnetic resonance techniques, the novel experimental platforms call for developing alternative modalities for control, as well as integrated hardware for the miniaturization of the overall device. Recent advances in the first direction include the mechanical control of spin qubits \cite{Lee2016, Maity2020, MacQuarrie2013, Chen2020} and even of nearby nuclear spins \cite{Maity2022}, electrical control \cite{Forneris2017, Wang2020a, Gulka2021}, and all-optical control \cite{Yale2013}. These control modalities not only could provide practical advantages in developing compact and robust devices, but they also allow directly manipulating transitions that magnetic resonance cannot drive, such as the double-quantum transition from the NV $m_s=-1$ to $+1$ levels. Conversely, these techniques (e.g., mechanical drive) are enabled by the nano- or micro-scopic size of the material platforms. 
In order to fully take advantage of the quantum NanoMRI system size, it is also often necessary to miniaturize the overall apparatus. A few first steps have been taken in this direction, with the control hardware for NV-based sensing, including a microwave generator, optical filter and photodetector, integrated in a compact, $(200\,\upmu\mathrm{m})^2$ CMOS device \cite{Kim2019}. Other compact devices have also been presented \cite{Stuerner2021, Patel2020, Du2021, Wang2022b, Ibrahim2021, Misonou2020}, and further steps towards integration of photonics \cite{Wan2020} and microwave delivery \cite{Omirzakhov2022, Shalaginov2020} would be highly desirable, while at the same time pushing for commercialization.

\begin{figure}
    \centering
    \includegraphics{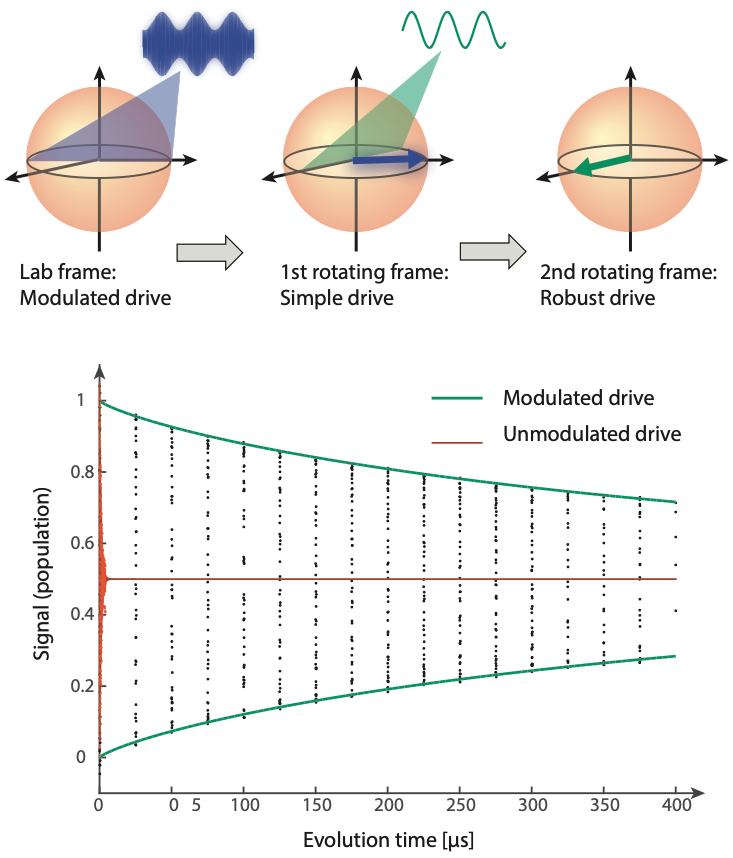}
    \caption{Example of robust control of spin defects with composite pulses, here a modulated driving scheme. Top: Concatenated Driving Decoupling allows to achieve robust driving in a second rotate frame, by correcting inhomogeneities and fluctuations of the main drive. Bottom: Example of implementation of the robust driving scheme in a large ensemble of Nitrogen-Vacancy electronic spins. The coherence time of the unmodulated (Rabi) driving (red) is increased by two orders of magnitude \cite{Wang2020a}.}
    \label{Cappellaro:fig1}
\end{figure}

\subsubsection*{\textbf{Concluding Remarks}}
The rich toolset developed by magnetic resonance and quantum control is being brought to bear to enable and improve novel techniques developed to reach NanoMRI. Thanks to the opportunities and challenges brought forward by operating with systems at the nanoscale, a broad range of novel techniques and hardware have started to emerge and will ensure reaching the ultimate limit of sensitivity and spatial resolution.  

\subsubsection*{Acknowledgements}
This work was in part supported by HRI-001835, NSF PHY1734011, DARPA DRINQS. D18AC00024) and Q-Diamond W911NF13D0001.

\newpage

\subsection{Statistical learning for nanoscale magnetic resonance}
\label{Bonato}
Cristian Bonato and Erik Gauger\\
SUPA, Institute of Photonics and Quantum Sciences, School of Engineering and Physical Sciences, Heriot-Watt University, Edinburgh EH14 4AS, UK\\

Yoann Altmann\\
Institute of Signals, Sensors and Systems, School of Engineering and Physical Sciences, Heriot-Watt University, Edinburgh EH14 4AS, UK \\
c.bonato@hw.ac.uk, y.altmann@hw.ac.uk, e.gauger@hw.ac.uk

\subsubsection*{\textbf{Status}}
\begin{figure}
    \centering
    \includegraphics[width = 0.8 \textwidth]{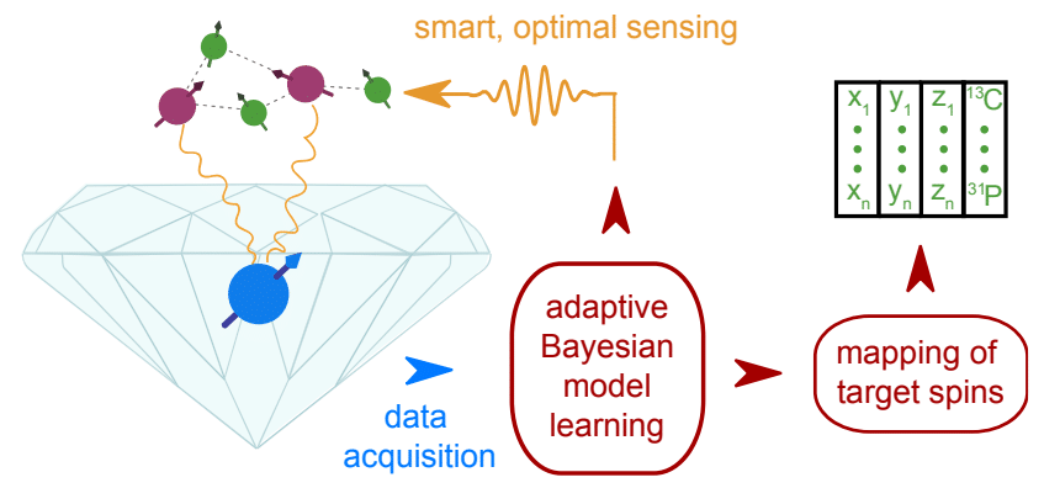}
    \caption{Schematic of a nano MRI sensing approach with an NV-center quantum spin sensor probing a target molecule via spin-spin interactions. Data acquisition proceeds over many repetitions with adaptively optimized control sequences. The right branch shows the associated task of spatially mapping the positions and type of spins that make up the target.}
    \label{Bonato:fig1}
\end{figure}

The extension of magnetic resonance imaging to the nanoscale regime could unlock unprecedented opportunities in chemistry, biology, and nanoscience by enabling direct three-dimensional reconstruction of molecular structures with chemical specificity. State-of-the-art nano-MRI experiments \cite{Taminiau2012, Bradley2019, Cujia2019, Abobeih2019, Cujia2022} with a single electron spin sensor detect the positions of individual nearby nuclear spins by their impact on the precession of the electron spin due through the hyperfine interaction. These measurements are still very time consuming, since achieving high spectral selectivity requires long pulse sequences. Whilst this issue can be partially mitigated by further improvements to the readout process, another–potentially even more powerful –possibility is to optimize how data is acquired \cite{Gebhart2023}. The standard measurement approach consists of sweeping parameters, such as inter-pulse delays and the phases of the applied pulses, across pre-determined ranges, chosen by the researcher based on educated guesses. Smarter ‘adaptive’ approaches that do not just sweep parameters but decide adaptively which sensing parameter values provide the most useful information, could significantly accelerate data acquisition by optimizing the information extracted by each measurement. This is typically based on statistical modelling and in particular Bayesian inference \cite{2007}, exploiting Bayes’ rule to update the probability distributions of parameters of interest (e.g. the hyperfine values $\left\{ \vec{A}_i  \right\}$) after each measurement outcome $m_n$ as:
$$
P\left(\left\{ \vec{A}_i | m_n,\ldots ,m_1 \right\} \right) \propto P\left(m_n | \left\{ \vec{A}_i  \right\} \right)\times P\left(\left\{ \vec{A}_i  \right\} | m_{n-1}, \ldots , m_1\right)
$$
The function $P\left(m_n | \left\{ \vec{A}_i  \right\} \right)$, known as the likelihood, represents the statistical model describing the system, and indicates how likely the outcome $m_n$ is, given a certain set of hyperfine couplings $\left\{ \vec{A}_i  \right\}$. The recursive formula above is then used, in combination with tools from information theory and/or heuristic mechanisms, to choose future parameter settings for optimizing a loss function, e.g., by minimizing the uncertainty on the estimated hyperfine interactions (see Fig.\,\ref{Bonato:fig1}). Pioneering experiments have shown that adaptive techniques can improve relatively simple measurements, such as the estimation of decoherence timescales \cite{Arshad2022, CaouetteMansour2022} (Fig.\,\ref{Bonato:fig2}) and static magnetic fields \cite{Bonato2015, Santagati2019, Joas2021} by quantum sensors.
An additional challenge is the post-processing of the acquired data, to interpret the results in terms of nuclear spin species and their three-dimensional positions. This requires fitting a large and typically unknown number of parameters, a non-trivial and time-consuming step that could potentially be optimized automatically. Seminal experimental breakthroughs in nano-MRI have been supported by different types of data analysis, starting from ``manual'' curve fitting \cite{Taminiau2012} to more sophisticated approaches based on maximum likelihood estimation \cite{Cujia2022}. Deep learning approaches have also been employed to fit data from dynamical decoupling spectroscopy \cite{Jung2021}, employing a suite of neural networks to de-noise the data and identify the features expected from theory.

\subsubsection*{\textbf{Current and Future Challenges}}
There are several challenges associated with online optimization of a spin quantum sensor to detect multiple nuclear spins through their hyperfine interaction. A major issue is the extension to the estimation of multiple parameters, e.g., the hyperfine components of a few dozen nuclear spins. First, estimating large numbers of parameters online is challenging due to the limited amount of information in each measurement and the increasing uncertainties as the number of model parameters grows. This requires the use of scalable algorithms to reliably quantify uncertainties based on non-standard multivariate distributions. Additionally, while simple adaptive rules can be found, often analytically, for single-parameter estimation, the extension to multi-parameter cases tends to be highly non-trivial, typically resulting in intractable objective functions. Moreover, adaptation of the experimental settings can also target the optimization of multiple objective functions, such as multiple types of measurements, overall acquisition time, parameter uncertainty. Prioritizing objective functions automatically is difficult, and so is deriving corresponding adaptive heuristics. Intuitive heuristics have been proposed for frequency estimation in the context of Hamiltonian learning (e.g. the particle guess heuristic \cite{Santagati2019, Wiebe2014}) but their scaling beyond the few-parameters case is still unclear. A possible solution to address this issue is to use data-driven (e.g. neural-network based) approaches trained on simulated data to learn smarter adaptation rules \cite{Fiderer2021}. This strategy could be particularly interesting when coupled with transfer learning approaches, whereby data-driven methods could be pre-trained on realistic (but potentially oversimplified) simulated data, and then refined on actual measurements \cite{Han2021}.
Another important challenge is that the number of parameters is typically unknown \textit{a-priori} (model-order selection problems), as the number of nuclear spins is \textit{a-priori} unknown. This can be addressed by running different models in parallel and performing model selection or model averaging using Bayes’ factors or similar criteria \cite{2007}. This might, however, not be practical when comparing hundreds or thousands of models. Another approach, closer to reversible-jump MCMC (RJ-MCMC) strategies, consists of attempting jumps across multiple models to identify the most likely models. In such cases, designing efficient jumping strategies will be crucial, and it would be interesting to assess whether data-driven strategies can help to tailor those jumps.
A further possibility consists of using artificial intelligence, for example reinforcement learning methods \cite{Baum2021, Foesel2018}, to invent novel estimation sequences, instead of optimizing existing ones. One of the advantages of this idea is that, while standard estimation techniques have been developed aiming for broad generality and insensitivity to noise, one could possibly design sequences specifically tailored for a given experimental setup and application.

\subsubsection*{\textbf{Advances in Science and Technology to Meet Challenges}}
The detection of NanoMRI signals could be improved by ad-hoc adaptive algorithms capable of estimating an unknown number of parameters, with fast-converging performance and the possibility to quantify uncertainties. This goal is still work in progress in the signal processing and machine learning communities. Whether model-based or data-driven approaches are adopted, methods with Bayesian interpretations are likely to be the most promising to handle uncertainties efficiently. Model-based approaches are easier to assess, but as experiments depart from simplistic scenarios, the distributions become non-standard and exact Bayesian inference might no longer be possible. Sampling methods such as Markov chain Monte Carlo (MCMC) methods \cite{2007} can handle complex distributions, but are computationally expensive. More general theoretical results are still required to facilitate the deployment of such methods to arbitrary distributions. It should also be noted that (approximate) Bayesian inference without analytical forward model has also progressed significantly \cite{Green2015, Chopin2015}. In particular, likelihood-free methods can be derived when simulating data quickly is possible. Such methods are particularly well suited for low-dimensional problems, but their scalability needs to be improved. Sequential Monte Carlo algorithms, i.e., particle filters are particularly well suited for online estimation but do not scale well yet with large numbers of unknown parameters, especially for highly multimodal distributions. In such cases, it might be beneficial to combine particle filters with MCMC updates to locally optimize the particles.
One important point to consider is that the computation time required to adaptively choose optimal parameters for each measurement adds an overhead to the estimation, and must be kept much smaller than the probing time. Thus, important efforts should be made to design algorithms that simultaneously provide estimation performance guarantees whilst being compatible with real-time and low SWaP constraints. Again, neural-network based generative models could be used to approximate, at a lower cost, exact models that are expensive to compute or simulate. Regarding efficient implementation, it has already been shown that, for simple measurements, the processing time can be limited to few tens of microseconds with microcontrollers \cite{2007, CaouetteMansour2022}, and could be faster with FPGAs. The availability of fast low-latency interfacing between the computation of optimal parameters and generation of the corresponding waveform (by an arbitrary waveform generator) is also crucial.
The construction of a physical model through Bayesian inference requires the capability to compute the system dynamics. While simple analytical models exist in the case of diluted clusters, e.g.in the detection of $^{13}$C spins surrounding the nitrogen-vacancy (NV) center in the diamond lattice, the computational burden becomes exponentially more complicated when scaling up to larger clusters of close-by strongly interacting spins. In this case, related to the more general topic of Hamiltonian learning \cite{Gentile2021}, the future availability of small-scale quantum simulators could be crucial.

\begin{figure}
    \centering
    \includegraphics[width = 0.8\textwidth]{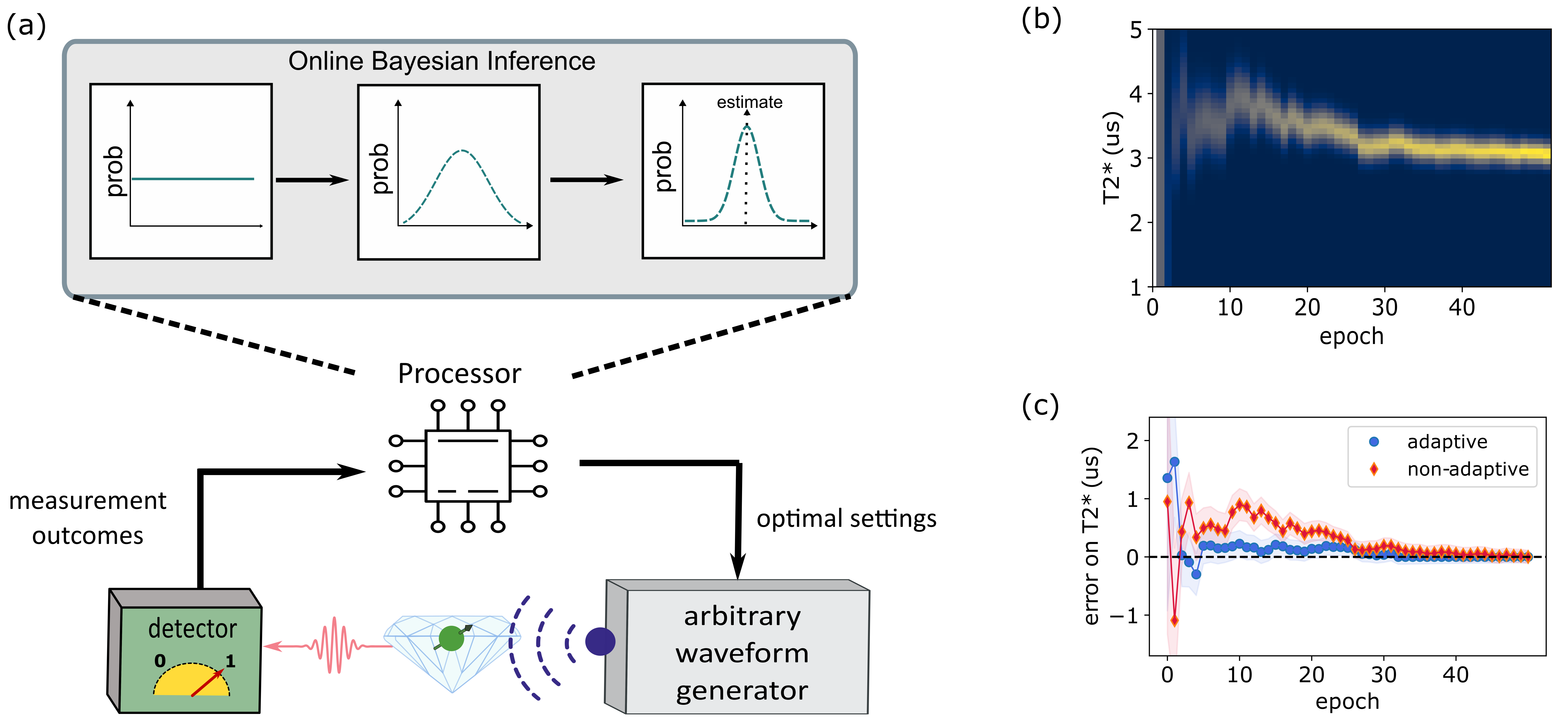}
    \caption{(a) online adaptive learning configuration. A processor updates the probability distribution for the parameters of interest (e.g.the hyperfine values for several nuclear spins) using Bayes’ rule and computes optimal settings for the next measurement. The optimal setting values are then passed to the arbitrary waveform generator, which constructs control sequences accordingly to drive the quantum sensor. Measurement outcomes, in the form of emitted photons in the case of the NV center in diamond, are then processed for the next learning step. (b) Example of Bayesian estimation [8], for the estimation of the dephasing time $T_2^*$ of a single spin \cite{2007} The probability for each possible value of $T_2^*$, $p(T_2^*)$, is plotted as a function of the measurement number (epoch). The probability distribution for $T_2^*$ is initially uniform, as no information is available, and become more and more peaked around the true value as more measurement outcomes are processed. (c) Comparison between experiments using adaptive and non-adaptive setting choice in the estimation of $T_2^*$ \cite{Arshad2022}, showing how adaptive protocols can reduce errors faster than non-adaptive ones.}
    \label{Bonato:fig2}
\end{figure}

\subsubsection*{\textbf{Concluding Remarks}}
The development and adoption of advanced signal processing and machine learning methods holds the promise to improve and speed-up data acquisition and model building for nanoscale magnetic resonance experiments. Online optimization of control pulses and pulse sequences is expected to boost sensitivity and reduce the number of measurements required to acquire a magnetic resonance spectrum. Further work on model learning, potentially empowered by data-driven methods, can simplify and automate the association of a spectrum to a nuclear spin configuration.In combination with further advanced in physics and quantum technology, advanced signal processing will enable imaging larger and larger nuclear spins cluster, down to demonstrating magnetic resonance imaging of single molecules. While in this contribution we have mostly focused on the case of detecting multiple individual nuclear spins with a quantum sensor based on a single electron spin, the algorithms described here can be readily applied to any other detection techniques.

\subsubsection*{Acknowledgements}
We thank Muhammad Junaid Arshad and Ben Haylock for the data in Fig.\,\ref{Bonato:fig2}(b)/(c). This work is funded by the Engineering and Physical Sciences Research Council (EP/S000550/1 and EP/V053779/1), the Leverhulme Trust (RPG-2019-388), the Royal Academy of Engineering under the Research Fellowship scheme RF201617/16/31 and the European Commission (QuanTELCO, grant agreement No.\,862721).

\section*{References}

\bibliographystyle{iopart-num}
\bibliography{main.bib}

\providecommand{\newblock}{}
\begin{thebibliography}{100}
\expandafter\ifx\csname url\endcsname\relax
  \def\url#1{{\tt #1}}\fi
\expandafter\ifx\csname urlprefix\endcsname\relax\def\urlprefix{URL }\fi
\providecommand{\eprint}[2][]{\url{#2}}

\bibitem{Sidles1991}
Sidles J~A 1991 {\em Appl. Phys. Lett.\/} {\bf 58} 2854--2856

\bibitem{Rugar1992}
Rugar D, Yannoni C~S and Sidles J~A 1992 {\em Nature\/} {\bf 360} 563--566

\bibitem{Poggio2018}
Poggio M and Herzog B~E 2018 Force‐detected nuclear magnetic resonance

\bibitem{Sidles1992}
Sidles J~A 1992 {\em Phys. Rev. Lett.\/} {\bf 68} 1124--1127

\bibitem{Braakman2019}
Braakman F~R and Poggio M 2019 {\em Nanotechnology\/} {\bf 30} 332001

\bibitem{Mamin2001}
Mamin H~J and Rugar D 2001 {\em Appl. Phys. Lett.\/} {\bf 79} 3358--3360

\bibitem{Chui}
Chui B, Hishinuma Y, Budakian R, Mamin H, Kenny T and Rugar D 2003 Mass-loaded cantilevers with suppressed higher-order modes for magnetic resonance force microscopy {\em TRANSDUCERS ’03. 12th International Conference on Solid-State Sensors, Actuators and Microsystems. Digest of Technical Papers (Cat. No.03TH8664)\/} SENSOR-03 (IEEE)

\bibitem{Heritier2018}
Héritier M, Eichler A, Pan Y, Grob U, Shorubalko I, Krass M~D, Tao Y and Degen C~L 2018 {\em Nano Lett.\/} {\bf 18} 1814--1818

\bibitem{Moser2014}
Moser J, Eichler A, Güttinger J, Dykman M~I and Bachtold A 2014 {\em Nat. Nanotechnol.\/} {\bf 9} 1007--1011

\bibitem{Nichol2012}
Nichol J~M, Hemesath E~R, Lauhon L~J and Budakian R 2012 {\em Phys. Rev. B\/} {\bf 85} 054414

\bibitem{Degen2009}
Degen C~L, Poggio M, Mamin H~J, Rettner C~T and Rugar D 2009 {\em Proc. Natl. Acad. Sci.\/} {\bf 106} 1313--1317

\bibitem{Nichol2013}
Nichol J~M, Naibert T~R, Hemesath E~R, Lauhon L~J and Budakian R 2013 {\em Phys. Rev. X\/} {\bf 3} 031016

\bibitem{Gloppe2014}
Gloppe A, Verlot P, Dupont-Ferrier E, Siria A, Poncharal P, Bachelier G, Vincent P and Arcizet O 2014 {\em Nat. Nanotechnol.\/} {\bf 9} 920--926

\bibitem{Rossi2019}
Rossi N, Gross B, Dirnberger F, Bougeard D and Poggio M 2019 {\em Nano Lett.\/} {\bf 19} 930--936

\bibitem{Tavernarakis2018}
Tavernarakis A, Stavrinadis A, Nowak A, Tsioutsios I, Bachtold A and Verlot P 2018 {\em Nat. Commun.\/} {\bf 9} 662

\bibitem{Tao2015}
Tao Y, Navaretti P, Hauert R, Grob U, Poggio M and Degen C~L 2015 {\em Nanotechnology\/} {\bf 26} 465501

\bibitem{Rugar2004}
Rugar D, Budakian R, Mamin H~J and Chui B~W 2004 {\em Nature\/} {\bf 430} 329--332

\bibitem{Rossi2016}
Rossi N, Braakman F~R, Cadeddu D, Vasyukov D, Tütüncüoglu G, Fontcuberta~i Morral A and Poggio M 2016 {\em Nat. Nanotechnol.\/} {\bf 12} 150--155

\bibitem{Lepinay2016}
de~Lépinay L~M, Pigeau B, Besga B, Vincent P, Poncharal P and Arcizet O 2016 {\em Nat. Nanotechnol.\/} {\bf 12} 156--162

\bibitem{Siria2017}
Siria A and Niguès A 2017 {\em Sci. Rep.\/} {\bf 7} 11595

\bibitem{Reiche2015}
Reiche C~F, Körner J, Büchner B and Mühl T 2015 {\em Nanotechnology\/} {\bf 26} 335501

\bibitem{Mattiat2020}
Mattiat H, Rossi N, Gross B, Pablo-Navarro J, Magén C, Badea R, Berezovsky J, De~Teresa J~M and Poggio M 2020 {\em Phys. Rev. Applied\/} {\bf 13} 044043

\bibitem{Kisiel2011}
Kisiel M, Gnecco E, Gysin U, Marot L, Rast S and Meyer E 2011 {\em Nat. Mater.\/} {\bf 10} 119--122

\bibitem{Kisiel2015}
Kisiel M, Pellegrini F, Santoro G, Samadashvili M, Pawlak R, Benassi A, Gysin U, Buzio R, Gerbi A, Meyer E and Tosatti E 2015 {\em Phys. Rev. Lett.\/} {\bf 115} 046101

\bibitem{Marchiori2021}
Marchiori E, Ceccarelli L, Rossi N, Lorenzelli L, Degen C~L and Poggio M 2021 {\em Nat. Rev. Phys.\/} {\bf 4} 49--60

\bibitem{Unterreithmeier2010}
Unterreithmeier Q~P, Faust T and Kotthaus J~P 2010 {\em Phys. Rev. Lett.\/} {\bf 105} 027205

\bibitem{Schmid2011}
Schmid S, Jensen K~D, Nielsen K~H and Boisen A 2011 {\em Phys. Rev. B\/} {\bf 84} 165307

\bibitem{Tsaturyan2017}
Tsaturyan Y, Barg A, Polzik E~S and Schliesser A 2017 {\em Nat. Nanotechnol.\/} {\bf 12} 776--783

\bibitem{Ghadimi2018}
Ghadimi A~H, Fedorov S~A, Engelsen N~J, Bereyhi M~J, Schilling R, Wilson D~J and Kippenberg T~J 2018 {\em Science\/} {\bf 360} 764--768

\bibitem{Gisler2022}
Gisler T, Helal M, Sabonis D, Grob U, Héritier M, Degen C~L, Ghadimi A~H and Eichler A 2022 {\em Phys. Rev. Lett.\/} {\bf 129} 104301

\bibitem{Bereyhi2022}
Bereyhi M~J, Arabmoheghi A, Beccari A, Fedorov S~A, Huang G, Kippenberg T~J and Engelsen N~J 2022 {\em Phys. Rev. X\/} {\bf 12} 021036

\bibitem{Eichler2022}
Eichler A 2022  (\textit{Preprint} \eprint{2209.05183})

\bibitem{Heritier2021}
Héritier M, Pachlatko R, Tao Y, Abendroth J~M, Degen C~L and Eichler A 2021 {\em Phys. Rev. Lett.\/} {\bf 127} 216101

\bibitem{Krass2022}
Krass M~D, Prumbaum N, Pachlatko R, Grob U, Takahashi H, Yamauchi Y, Degen C~L and Eichler A 2022 {\em Phys. Rev. Applied\/} {\bf 18} 034052

\bibitem{Aspelmeyer2014}
Aspelmeyer M, Kippenberg T~J and Marquardt F 2014 {\em Rev. Mod. Phys.\/} {\bf 86} 1391--1452

\bibitem{Poggio2010}
Poggio M and Degen C~L 2010 {\em Nanotechnology\/} {\bf 21} 342001

\bibitem{Haelg2021}
Hälg D, Gisler T, Tsaturyan Y, Catalini L, Grob U, Krass M~D, Héritier M, Mattiat H, Thamm A~K, Schirhagl R, Langman E~C, Schliesser A, Degen C~L and Eichler A 2021 {\em Phys. Rev. Applied\/} {\bf 15} l021001

\bibitem{Scozzaro2016}
Scozzaro N, Ruchotzke W, Belding A, Cardellino J, Blomberg E, McCullian B, Bhallamudi V, Pelekhov D and Hammel P 2016 {\em J. Magn. Reson.\/} {\bf 271} 15--20

\bibitem{Fischer2019}
Fischer R, McNally D~P, Reetz C, Assumpção G~G~T, Knief T, Lin Y and Regal C~A 2019 {\em New J. Phys.\/} {\bf 21} 043049

\bibitem{Gavartin2012}
Gavartin E, Verlot P and Kippenberg T~J 2012 {\em Nat. Nanotechnol.\/} {\bf 7} 509--514

\bibitem{Dougherty1996}
Dougherty W~M, Bruland K~J, Garbini J~L and Sidles J~A 1996 {\em Meas. Sci. Technol.\/} {\bf 7} 1733--1739

\bibitem{Kosata2020}
Košata J, Zilberberg O, Degen C~L, Chitra R and Eichler A 2020 {\em Phys. Rev. Applied\/} {\bf 14} 014042

\bibitem{Sidles1992a}
Sidles J~A, Garbini J~L and Drobny G~P 1992 {\em Rev. Sci. Instrum.\/} {\bf 63} 3881--3899

\bibitem{Sidles1995}
Sidles J~A, Garbini J~L, Bruland K~J, Rugar D, Züger O, Hoen S and Yannoni C~S 1995 {\em Rev. Mod. Phys.\/} {\bf 67} 249--265

\bibitem{Haas2022}
Haas H, Tabatabaei S, Rose W, Sahafi P, Piscitelli M, Jordan A, Priyadarsi P, Singh N, Yager B, Poole P~J, Dalacu D and Budakian R 2022 {\em Proc. Natl. Acad. Sci.\/} {\bf 119}

\bibitem{Rose2018}
Rose W, Haas H, Chen A~Q, Jeon N, Lauhon L~J, Cory D~G and Budakian R 2018 {\em Phys. Rev. X\/} {\bf 8} 011030

\bibitem{Eberhardt2007}
Eberhardt K~W, Degen C~L and Meier B~H 2007 {\em Phys. Rev. B\/} {\bf 76} 180405

\bibitem{Eberhardt2007a}
Eberhardt K~W, Lin Q, Meier U, Hunkeler A and Meier B~H 2007 {\em Phys. Rev. B\/} {\bf 75} 184430

\bibitem{Issac2016}
Issac C~E, Gleave C~M, Nasr P~T, Nguyen H~L, Curley E~A, Yoder J~L, Moore E~W, Chen L and Marohn J~A 2016 {\em Phys. Chem. Chem. Phys.\/} {\bf 18} 8806--8819

\bibitem{Tao2016}
Tao Y, Eichler A, Holzherr T and Degen C~L 2016 {\em Nat. Commun.\/} {\bf 7} 12714

\bibitem{Sahafi2019}
Sahafi P, Rose W, Jordan A, Yager B, Piscitelli M and Budakian R 2019 {\em Nano Lett.\/} {\bf 20} 218--223

\bibitem{Moser2013}
Moser J, Güttinger J, Eichler A, Esplandiu M~J, Liu D~E, Dykman M~I and Bachtold A 2013 {\em Nat. Nanotechnol.\/} {\bf 8} 493--496

\bibitem{Stipe2001}
Stipe B~C, Mamin H~J, Stowe T~D, Kenny T~W and Rugar D 2001 {\em Phys. Rev. Lett.\/} {\bf 87} 096801

\bibitem{Tao2015a}
Tao Y and Degen C~L 2015 {\em Nano Lett.\/} {\bf 15} 7893--7897

\bibitem{Haas2019}
Haas H, Puzzuoli D, Zhang F and Cory D~G 2019 {\em New J. Phys.\/} {\bf 21} 103011

\bibitem{Tabatabaei2021}
Tabatabaei S, Haas H, Rose W, Yager B, Piscitelli M, Sahafi P, Jordan A, Poole P~J, Dalacu D and Budakian R 2021 {\em Phys. Rev. Applied\/} {\bf 15} 044043

\bibitem{Zhang1996}
Zhang Z, Hammel P~C and Wigen P~E 1996 {\em Appl. Phys. Lett.\/} {\bf 68} 2005--2007

\bibitem{Wago1998}
Wago K, Botkin D, Yannoni C~S and Rugar D 1998 {\em Appl. Phys. Lett.\/} {\bf 72} 2757--2759

\bibitem{Urban2006}
Urban R, Putilin A, Wigen P~E, Liou S~H, Cross M~C, Hammel P~C and Roukes M~L 2006 {\em Phys. Rev. B\/} {\bf 73} 212410

\bibitem{Mewes2006}
Mewes T, Kim J, Pelekhov D~V, Kakazei G~N, Wigen P~E, Batra S and Hammel P~C 2006 {\em Phys. Rev. B\/} {\bf 74} 144424

\bibitem{Loubens2007}
de~Loubens G, Naletov V~V, Klein O, Youssef J~B, Boust F and Vukadinovic N 2007 {\em Phys. Rev. Lett.\/} {\bf 98} 127601

\bibitem{Obukhov2008}
Obukhov Y, Pelekhov D~V, Kim J, Banerjee P, Martin I, Nazaretski E, Movshovich R, An S, Gramila T~J, Batra S and Hammel P~C 2008 {\em Phys. Rev. Lett.\/} {\bf 100} 197601

\bibitem{Klein2008}
Klein O, de~Loubens G, Naletov V~V, Boust F, Guillet T, Hurdequint H, Leksikov A, Slavin A~N, Tiberkevich V~S and Vukadinovic N 2008 {\em Phys. Rev. B\/} {\bf 78} 144410

\bibitem{Lee2010}
Lee I, Obukhov Y, Xiang G, Hauser A, Yang F, Banerjee P, Pelekhov D~V and Hammel P~C 2010 {\em Nature\/} {\bf 466} 845--848

\bibitem{Lee2011}
Lee I, Obukhov Y, Hauser A~J, Yang F~Y, Pelekhov D~V and Hammel P~C 2011 {\em J. Appl. Phys.\/} {\bf 109} 07D313

\bibitem{Du2015}
Du C, Lee I, Adur R, Obukhov Y, Hamann C, Buchner B, McCord J, Pelekhov D~V and Hammel P~C 2015 {\em Phys. Rev. B\/} {\bf 92} 214413

\bibitem{Wu2020}
Wu G, White S~P, Ruane W~T, Brangham J~T, Pelekhov D~V, Yang F and Hammel P~C 2020 {\em Phys. Rev. B\/} {\bf 101} 184409

\bibitem{Wu2022}
Wu G, Wang D, Verma N, Rao R, Cheng Y, Guo S, Cao G, Watanabe K, Taniguchi T, Lau C~N, Yang F, Randeria M, Bockrath M and Hammel P~C 2022 {\em Nano Lett.\/} {\bf 22} 1115--1121

\bibitem{Adur2014}
Adur R, Du C, Wang H, Manuilov S~A, Bhallamudi V~P, Zhang C, Pelekhov D~V, Yang F and Hammel P~C 2014 {\em Phys. Rev. Lett.\/} {\bf 113} 176601

\bibitem{Banerjee2010}
Banerjee P, Wolny F, Pelekhov D~V, Herman M~R, Fong K~C, Weissker U, Mühl T, Obukhov Y, Leonhardt A, Büchner B and Hammel P~C 2010 {\em Appl. Phys. Lett.\/} {\bf 96} 252505

\bibitem{Wolny2011}
Wolny F, Obukhov Y, Mühl T, Weißker U, Philippi S, Leonhardt A, Banerjee P, Reed A, Xiang G, Adur R, Lee I, Hauser A, Yang F, Pelekhov D, Büchner B and Hammel P 2011 {\em Ultramicroscopy\/} {\bf 111} 1360--1365

\bibitem{Hickman2010}
Hickman S~A, Moore E~W, Lee S, Longenecker J~G, Wright S~J, Harrell L~E and Marohn J~A 2010 {\em ACS Nano\/} {\bf 4} 7141--7150

\bibitem{Chia2012}
Chia H~J, Guo F, Belova L~M and McMichael R~D 2012 {\em Phys. Rev. Lett.\/} {\bf 108} 087206

\bibitem{Zhang2017}
Zhang C, Pu Y, Manuilov S~A, White S~P, Page M~R, Blomberg E~C, Pelekhov D~V and Hammel P~C 2017 {\em Phys. Rev. Applied\/} {\bf 7} 054019

\bibitem{Zhang2021}
Zhang C, Lee I, Pu Y, Manuilov S~A, Pelekhov D~V and Hammel P~C 2021 {\em Nano Lett.\/} {\bf 21} 10208--10214

\bibitem{Julsgaard2001}
Julsgaard B, Kozhekin A and Polzik E~S 2001 {\em Nature\/} {\bf 413} 400--403

\bibitem{Polzik2014}
Polzik E~S and Hammerer K 2015 {\em Ann. Phys.\/} {\bf 527} A15–A20

\bibitem{Hammerer2009}
Hammerer K, Aspelmeyer M, Polzik E~S and Zoller P 2009 {\em Phys. Rev. Lett.\/} {\bf 102} 020501

\bibitem{Tsang2012}
Tsang M and Caves C~M 2012 {\em Phys. Rev. X\/} {\bf 2} 031016

\bibitem{Woolley2013}
Woolley M~J and Clerk A~A 2013 {\em Phys. Rev. A\/} {\bf 87} 063846

\bibitem{Zhang2013}
Zhang K, Meystre P and Zhang W 2013 {\em Phys. Rev. A\/} {\bf 88} 043632

\bibitem{OckeloenKorppi2018}
Ockeloen-Korppi C~F, Damskägg E, Pirkkalainen J~M, Asjad M, Clerk A~A, Massel F, Woolley M~J and Sillanpää M~A 2018 {\em Nature\/} {\bf 556} 478--482

\bibitem{Riedinger2018}
Riedinger R, Wallucks A, Marinković I, Löschnauer C, Aspelmeyer M, Hong S and Gröblacher S 2018 {\em Nature\/} {\bf 556} 473--477

\bibitem{Moeller2017}
Møller C~B, Thomas R~A, Vasilakis G, Zeuthen E, Tsaturyan Y, Balabas M, Jensen K, Schliesser A, Hammerer K and Polzik E~S 2017 {\em Nature\/} {\bf 547} 191--195

\bibitem{Thomas2020}
Thomas R~A, Parniak M, Østfeldt C, Møller C~B, Bærentsen C, Tsaturyan Y, Schliesser A, Appel J, Zeuthen E and Polzik E~S 2020 {\em Nat. Phys.\/} {\bf 17} 228--233

\bibitem{Karg2020}
Karg T~M, Gouraud B, Ngai C~T, Schmid G~L, Hammerer K and Treutlein P 2020 {\em Science\/} {\bf 369} 174--179

\bibitem{Wasilewski2010}
Wasilewski W, Jensen K, Krauter H, Renema J~J, Balabas M~V and Polzik E~S 2010 {\em Phys. Rev. Lett.\/} {\bf 104} 133601

\bibitem{Khalili2018}
Khalili F and Polzik E 2018 {\em Phys. Rev. Lett.\/} {\bf 121} 031101

\bibitem{Zeuthen2019}
Zeuthen E, Polzik E~S and Khalili F~Y 2019 {\em Phys. Rev. D\/} {\bf 100} 062004

\bibitem{Zeuthen2022}
Zeuthen E, Polzik E~S and Khalili F~Y 2022 {\em PRX Quantum\/} {\bf 3} 020362

\bibitem{Rossi2018}
Rossi M, Mason D, Chen J, Tsaturyan Y and Schliesser A 2018 {\em Nature\/} {\bf 563} 53--58

\bibitem{Shao2019}
Shao L, Maity S, Zheng L, Wu L, Shams-Ansari A, Sohn Y~I, Puma E, Gadalla M, Zhang M, Wang C, Hu E, Lai K and Lončar M 2019 {\em Phys. Rev. Applied\/} {\bf 12} 014022

\bibitem{MacCabe2020}
MacCabe G~S, Ren H, Luo J, Cohen J~D, Zhou H, Sipahigil A, Mirhosseini M and Painter O 2020 {\em Science\/} {\bf 370} 840--843

\bibitem{Wood2022}
Wood B~D, Stimpson G~A, March J~E, Lekhai Y~N~D, Stephen C~J, Green B~L, Frangeskou A~C, Ginés L, Mandal S, Williams O~A and Morley G~W 2022 {\em Phys. Rev. B\/} {\bf 105} 205401

\bibitem{Hahn2021}
Hahn W and Dobrovitski V~V 2021 {\em New J. Phys.\/} {\bf 23} 073029

\bibitem{Ajoy2015}
Ajoy A, Bissbort U, Lukin M, Walsworth R and Cappellaro P 2015 {\em Phys. Rev. X\/} {\bf 5} 011001

\bibitem{Kost2015}
Kost M, Cai J and Plenio M~B 2015 {\em Sci. Rep.\/} {\bf 5} 11007

\bibitem{Wang2016}
Wang Z~Y, Haase J~F, Casanova J and Plenio M~B 2016 {\em Phys. Rev. B\/} {\bf 93} 174104

\bibitem{Perunicic2016}
Perunicic V~S, Hill C~D, Hall L~T and Hollenberg L 2016 {\em Nat. Commun.\/} {\bf 7} 12667

\bibitem{Perunicic2021}
Perunicic V~S, Usman M, Hill C~D and Hollenberg L~C~L 2021  (\textit{Preprint} \eprint{2112.03623})

\bibitem{Shi2015}
Shi F, Zhang Q, Wang P, Sun H, Wang J, Rong X, Chen M, Ju C, Reinhard F, Chen H, Wrachtrup J, Wang J and Du J 2015 {\em Science\/} {\bf 347} 1135--1138

\bibitem{Lovchinsky2016}
Lovchinsky I, Sushkov A~O, Urbach E, de~Leon N~P, Choi S, De~Greve K, Evans R, Gertner R, Bersin E, Müller C, McGuinness L, Jelezko F, Walsworth R~L, Park H and Lukin M~D 2016 {\em Science\/} {\bf 351} 836--841

\bibitem{Abobeih2019}
Abobeih M~H, Randall J, Bradley C~E, Bartling H~P, Bakker M~A, Degen M~J, Markham M, Twitchen D~J and Taminiau T~H 2019 {\em Nature\/} {\bf 576} 411--415

\bibitem{Yudilevich2022}
Yudilevich D, Stöhr R, Denisenko A and Finkler A 2022 {\em Phys. Rev. Applied\/} {\bf 18} 054016

\bibitem{Cujia2022}
Cujia K~S, Herb K, Zopes J, Abendroth J~M and Degen C~L 2022 {\em Nat. Commun.\/} {\bf 13} 1260

\bibitem{Cai2013}
Cai J, Retzker A, Jelezko F and Plenio M~B 2013 {\em Nat. Phys.\/} {\bf 9} 168--173

\bibitem{Randall2021}
Randall J, Bradley C~E, van~der Gronden F~V, Galicia A, Abobeih M~H, Markham M, Twitchen D~J, Machado F, Yao N~Y and Taminiau T~H 2021 {\em Science\/} {\bf 374} 1474--1478

\bibitem{Janitz2022}
Janitz E, Herb K, Völker L~A, Huxter W~S, Degen C~L and Abendroth J~M 2022 {\em J. Mater. Chem. C\/} {\bf 10} 13533--13569

\bibitem{Abendroth2022}
Abendroth J~M, Herb K, Janitz E, Zhu T, Völker L~A and Degen C~L 2022 {\em Nano Lett.\/} {\bf 22} 7294--7303

\bibitem{Bayliss2020}
Bayliss S~L, Laorenza D~W, Mintun P~J, Kovos B~D, Freedman D~E and Awschalom D~D 2020 {\em Science\/} {\bf 370} 1309--1312

\bibitem{Xie2022}
Xie M, Yu X, Rodgers L~V~H, Xu D, Chi-Durán I, Toros A, Quack N, de~Leon N~P and Maurer P~C 2022 {\em Proc. Natl. Acad. Sci.\/} {\bf 119} e2114186119

\bibitem{Wang2022}
Wang G, Liu Y~X, Schloss J~M, Alsid S~T, Braje D~A and Cappellaro P 2022 {\em Phys. Rev. X\/} {\bf 12} 021061

\bibitem{Luehmann2019}
Lühmann T, John R, Wunderlich R, Meijer J and Pezzagna S 2019 {\em Nat. Commun.\/} {\bf 10} 4956

\bibitem{Nguyen2019}
Nguyen C~T, Sukachev D~D, Bhaskar M~K, Machielse B, Levonian D~S, Knall E~N, Stroganov P, Chia C, Burek M~J, Riedinger R, Park H, Lončar M and Lukin M~D 2019 {\em Phys. Rev. B\/} {\bf 100} 165428

\bibitem{Glenn2018}
Glenn D~R, Bucher D~B, Lee J, Lukin M~D, Park H and Walsworth R~L 2018 {\em Nature\/} {\bf 555} 351--354

\bibitem{Smits2019}
Smits J, Damron J~T, Kehayias P, McDowell A~F, Mosavian N, Fescenko I, Ristoff N, Laraoui A, Jarmola A and Acosta V~M 2019 {\em Sci. Adv.\/} {\bf 5} aaw7895

\bibitem{Bucher2020}
Bucher D~B, Glenn D~R, Park H, Lukin M~D and Walsworth R~L 2020 {\em Phys. Rev. X\/} {\bf 10} 021053

\bibitem{Arunkumar2021}
Arunkumar N, Bucher D~B, Turner M~J, TomHon P, Glenn D, Lehmkuhl S, Lukin M~D, Park H, Rosen M~S, Theis T and Walsworth R~L 2021 {\em PRX Quantum\/} {\bf 2} 010305

\bibitem{Aslam2017}
Aslam N, Pfender M, Neumann P, Reuter R, Zappe A, Fávaro~de Oliveira F, Denisenko A, Sumiya H, Onoda S, Isoya J and Wrachtrup J 2017 {\em Science\/} {\bf 357} 67--71

\bibitem{Fortman2021}
Fortman B, Mugica-Sanchez L, Tischler N, Selco C, Hang Y, Holczer K and Takahashi S 2021 {\em J. Appl. Phys.\/} {\bf 130} 083901

\bibitem{Ren2023}
Ren Y, Selco C, Kawashiri D, Coumans M, Fortman B, Bouchard L~S, Holczer K and Takahashi S 2023 {\em Phys. Rev. B\/} {\bf 108} 045421

\bibitem{Kruger2022}
Kruger D, Zhang A, Hinton H, Arnal V~M, Song Y~Q, Tang Y, Lei K~M, Anders J and Ham D 2022 A portable cmos-based mri system with $67 \times 67 \times 83 \mathrm{\mu m}^3$ image resolution {\em ESSCIRC 2022- IEEE 48th European Solid State Circuits Conference (ESSCIRC)\/} (IEEE)

\bibitem{Zhou2020}
Zhou H, Choi J, Choi S, Landig R, Douglas A~M, Isoya J, Jelezko F, Onoda S, Sumiya H, Cappellaro P, Knowles H~S, Park H and Lukin M~D 2020 {\em Phys. Rev. X\/} {\bf 10} 031003

\bibitem{Choi2020}
Choi J, Zhou H, Knowles H~S, Landig R, Choi S and Lukin M~D 2020 {\em Phys. Rev. X\/} {\bf 10} 031002

\bibitem{Arunkumar2023}
Arunkumar N, Olsson K~S, Oon J~T, Hart C~A, Bucher D~B, Glenn D~R, Lukin M~D, Park H, Ham D and Walsworth R~L 2023 {\em Phys. Rev. Lett.\/} {\bf 131} 100801

\bibitem{Balasubramanian2008}
Balasubramanian G, Chan I~Y, Kolesov R, Al-Hmoud M, Tisler J, Shin C, Kim C, Wojcik A, Hemmer P~R, Krueger A, Hanke T, Leitenstorfer A, Bratschitsch R, Jelezko F and Wrachtrup J 2008 {\em Nature\/} {\bf 455} 648--651

\bibitem{Maletinsky2012}
Maletinsky P, Hong S, Grinolds M~S, Hausmann B, Lukin M~D, Walsworth R~L, Loncar M and Yacoby A 2012 {\em Nat. Nanotechnol.\/} {\bf 7} 320--324

\bibitem{Grinolds2014}
Grinolds M~S, Warner M, De~Greve K, Dovzhenko Y, Thiel L, Walsworth R~L, Hong S, Maletinsky P and Yacoby A 2014 {\em Nat. Nanotechnol.\/} {\bf 9} 279--284

\bibitem{Bian2021}
Bian K, Zheng W, Zeng X, Chen X, Stöhr R, Denisenko A, Yang S, Wrachtrup J and Jiang Y 2021 {\em Nat. Commun.\/} {\bf 12} 2457

\bibitem{Arai2015}
Arai K, Belthangady C, Zhang H, Bar-Gill N, DeVience S~J, Cappellaro P, Yacoby A and Walsworth R~L 2015 {\em Nat. Nanotechnol.\/} {\bf 10} 859--864

\bibitem{Grinolds2013}
Grinolds M~S, Hong S, Maletinsky P, Luan L, Lukin M~D, Walsworth R~L and Yacoby A 2013 {\em Nat. Phys.\/} {\bf 9} 215--219

\bibitem{Tetienne2014}
Tetienne J~P, Hingant T, Kim J~V, Diez L~H, Adam J~P, Garcia K, Roch J~F, Rohart S, Thiaville A, Ravelosona D and Jacques V 2014 {\em Science\/} {\bf 344} 1366--1369

\bibitem{Thiel2019}
Thiel L, Wang Z, Tschudin M~A, Rohner D, Gutiérrez-Lezama I, Ubrig N, Gibertini M, Giannini E, Morpurgo A~F and Maletinsky P 2019 {\em Science\/} {\bf 364} 973--976

\bibitem{Sun2021}
Sun Q~C, Song T, Anderson E, Brunner A, Förster J, Shalomayeva T, Taniguchi T, Watanabe K, Gräfe J, Stöhr R, Xu X and Wrachtrup J 2021 {\em Nat. Commun.\/} {\bf 12} 1989

\bibitem{Pelliccione2016}
Pelliccione M, Jenkins A, Ovartchaiyapong P, Reetz C, Emmanouilidou E, Ni N and Bleszynski~Jayich A~C 2016 {\em Nat. Nanotechnol.\/} {\bf 11} 700--705

\bibitem{Thiel2016}
Thiel L, Rohner D, Ganzhorn M, Appel P, Neu E, Müller B, Kleiner R, Koelle D and Maletinsky P 2016 {\em Nat. Nanotechnol.\/} {\bf 11} 677--681

\bibitem{Jenkins2022}
Jenkins A, Baumann S, Zhou H, Meynell S~A, Daipeng Y, Watanabe K, Taniguchi T, Lucas A, Young A~F and Bleszynski~Jayich A~C 2022 {\em Phys. Rev. Lett.\/} {\bf 129} 087701

\bibitem{Casola2018}
Casola F, van~der Sar T and Yacoby A 2018 {\em Nat. Rev. Mater.\/} {\bf 3} 17088

\bibitem{Wang2019}
Wang P, Chen S, Guo M, Peng S, Wang M, Chen M, Ma W, Zhang R, Su J, Rong X, Shi F, Xu T and Du J 2019 {\em Sci. Adv.\/} {\bf 5} aau8038

\bibitem{Ariyaratne2018}
Ariyaratne A, Bluvstein D, Myers B~A and Jayich A~C~B 2018 {\em Nat. Commun.\/} {\bf 9} 2406

\bibitem{Bluvstein2019}
Bluvstein D, Zhang Z and Jayich A~C~B 2019 {\em Phys. Rev. Lett.\/} {\bf 122} 076101

\bibitem{Myers2014}
Myers B, Das A, Dartiailh M, Ohno K, Awschalom D and Bleszynski~Jayich A 2014 {\em Phys. Rev. Lett.\/} {\bf 113} 027602

\bibitem{Chen2019}
Chen Y~C, Griffiths B, Weng L, Nicley S~S, Ishmael S~N, Lekhai Y, Johnson S, Stephen C~J, Green B~L, Morley G~W, Newton M~E, Booth M~J, Salter P~S and Smith J~M 2019 {\em Optica\/} {\bf 6} 662--667

\bibitem{Robledo2011}
Robledo L, Childress L, Bernien H, Hensen B, Alkemade P~F~A and Hanson R 2011 {\em Nature\/} {\bf 477} 574--578

\bibitem{Rohner2019}
Rohner D, Happacher J, Reiser P, Tschudin M~A, Tallaire A, Achard J, Shields B~J and Maletinsky P 2019 {\em Appl. Phys. Lett.\/} {\bf 115} 192401

\bibitem{Welter2022}
Welter P, Rhensius J, Morales A, Wörnle M~S, Lambert C~H, Puebla-Hellmann G, Gambardella P and Degen C~L 2022 {\em Appl. Phys. Lett.\/} {\bf 120} 074003

\bibitem{Sangtawesin2019}
Sangtawesin S, Dwyer B~L, Srinivasan S, Allred J~J, Rodgers L~V, De~Greve K, Stacey A, Dontschuk N, O’Donnell K~M, Hu D, Evans D~A, Jaye C, Fischer D~A, Markham M~L, Twitchen D~J, Park H, Lukin M~D and de~Leon N~P 2019 {\em Phys. Rev. X\/} {\bf 9} 031052

\bibitem{Zheng2022}
Zheng W, Bian K, Chen X, Shen Y, Zhang S, Stöhr R, Denisenko A, Wrachtrup J, Yang S and Jiang Y 2022 {\em Nat. Phys.\/} {\bf 18} 1317--1323

\bibitem{Joos2022}
Joos M, Bluvstein D, Lyu Y, Weld D and Bleszynski~Jayich A 2022 {\em npj Quantum Inf.\/} {\bf 8} 47

\bibitem{Ohno2012}
Ohno K, Joseph~Heremans F, Bassett L~C, Myers B~A, Toyli D~M, Bleszynski~Jayich A~C, Palmstrøm C~J and Awschalom D~D 2012 {\em Appl. Phys. Lett.\/} {\bf 101} 082413

\bibitem{Appel2016}
Appel P, Neu E, Ganzhorn M, Barfuss A, Batzer M, Gratz M, Tschöpe A and Maletinsky P 2016 {\em Rev. Sci. Instrum.\/} {\bf 87} 063703

\bibitem{Wan2018}
Wan N~H, Shields B~J, Kim D, Mouradian S, Lienhard B, Walsh M, Bakhru H, Schröder T and Englund D 2018 {\em Nano Lett.\/} {\bf 18} 2787--2793

\bibitem{Wang2022a}
Wang M, Sun H, Ye X, Yu P, Liu H, Zhou J, Wang P, Shi F, Wang Y and Du J 2022 {\em Sci. Adv.\/} {\bf 8} abn9573

\bibitem{Shields2015}
Shields B, Unterreithmeier Q, de~Leon N, Park H and Lukin M 2015 {\em Phys. Rev. Lett.\/} {\bf 114} 136402

\bibitem{Haeberle2017}
H\"aberle T, Oeckinghaus T, Schmid-Lorch D, Pfender M, de~Oliveira F~F, Momenzadeh S~A, Finkler A and Wrachtrup J 2017 {\em Rev. Sci. Instrum.\/} {\bf 88} 013702

\bibitem{Degen2017}
Degen C, Reinhard F and Cappellaro P 2017 {\em Rev. Mod. Phys.\/} {\bf 89} 035002

\bibitem{Takahashi2008}
Takahashi S, Hanson R, van Tol J, Sherwin M~S and Awschalom D~D 2008 {\em Phys. Rev. Lett.\/} {\bf 101} 047601

\bibitem{Staudacher2013}
Staudacher T, Shi F, Pezzagna S, Meijer J, Du J, Meriles C~A, Reinhard F and Wrachtrup J 2013 {\em Science\/} {\bf 339} 561--563

\bibitem{Schmitt2017}
Schmitt S, Gefen T, Stürner F~M, Unden T, Wolff G, Müller C, Scheuer J, Naydenov B, Markham M, Pezzagna S, Meijer J, Schwarz I, Plenio M, Retzker A, McGuinness L~P and Jelezko F 2017 {\em Science\/} {\bf 356} 832--837

\bibitem{Mamin2013}
Mamin H~J, Kim M, Sherwood M~H, Rettner C~T, Ohno K, Awschalom D~D and Rugar D 2013 {\em Science\/} {\bf 339} 557--560

\bibitem{Konzelmann2018}
Konzelmann P, Rendler T, Bergholm V, Zappe A, Pfannenstill V, Garsi M, Ziem F, Niethammer M, Widmann M, Lee S~Y, Neumann P and Wrachtrup J 2018 {\em New J. Phys.\/} {\bf 20} 123013

\bibitem{Alvarez2011}
Álvarez G~A and Suter D 2011 {\em Phys. Rev. Lett.\/} {\bf 107} 230501

\bibitem{Bauch2018}
Bauch E, Hart C~A, Schloss J~M, Turner M~J, Barry J~F, Kehayias P, Singh S and Walsworth R~L 2018 {\em Phys. Rev. X\/} {\bf 8} 031025

\bibitem{Aiello2013}
Aiello C~D, Hirose M and Cappellaro P 2013 {\em Nat. Commun.\/} {\bf 4} 1419

\bibitem{Cai2012}
Cai J~M, Naydenov B, Pfeiffer R, McGuinness L~P, Jahnke K~D, Jelezko F, Plenio M~B and Retzker A 2012 {\em New J. Phys.\/} {\bf 14} 113023

\bibitem{Ajoy2017}
Ajoy A, Liu Y~X, Saha K, Marseglia L, Jaskula J~C, Bissbort U and Cappellaro P 2017 {\em Proc. Natl. Acad. Sci.\/} {\bf 114} 2149--2153

\bibitem{Casanova2015}
Casanova J, Wang Z~Y, Haase J~F and Plenio M~B 2015 {\em Phys. Rev. A\/} {\bf 92} 042304

\bibitem{London2013}
London P, Scheuer J, Cai J~M, Schwarz I, Retzker A, Plenio M~B, Katagiri M, Teraji T, Koizumi S, Isoya J, Fischer R, McGuinness L~P, Naydenov B and Jelezko F 2013 {\em Phys. Rev. Lett.\/} {\bf 111} 067601

\bibitem{Gierth2020}
Gierth M, Krespach V, Shames A~I, Raghavan P, Druga E, Nunn N, Torelli M, Nirodi R, Le S, Zhao R, Aguilar A, Lv X, Shen M, Meriles C~A, Reimer J~A, Zaitsev A, Pines A, Shenderova O and Ajoy A 2020 {\em Adv. Quantum Technol.\/} {\bf 3} 2000050

\bibitem{Sahin2022}
Sahin O, de~Leon~Sanchez E, Conti S, Akkiraju A, Reshetikhin P, Druga E, Aggarwal A, Gilbert B, Bhave S and Ajoy A 2022 {\em Nat. Commun.\/} {\bf 13} 5486

\bibitem{Beatrez2021}
Beatrez W, Janes O, Akkiraju A, Pillai A, Oddo A, Reshetikhin P, Druga E, McAllister M, Elo M, Gilbert B, Suter D and Ajoy A 2021 {\em Phys. Rev. Lett.\/} {\bf 127} 170603

\bibitem{Steinert2010}
Steinert S, Dolde F, Neumann P, Aird A, Naydenov B, Balasubramanian G, Jelezko F and Wrachtrup J 2010 {\em Rev. Sci. Instrum.\/} {\bf 81} 043705

\bibitem{Ziem2013}
Ziem F~C, Götz N~S, Zappe A, Steinert S and Wrachtrup J 2013 {\em Nano Lett.\/} {\bf 13} 4093--4098

\bibitem{Fescenko2020}
Fescenko I, Jarmola A, Savukov I, Kehayias P, Smits J, Damron J, Ristoff N, Mosavian N and Acosta V~M 2020 {\em Phys. Rev. Research\/} {\bf 2} 023394

\bibitem{Stepanov2015}
Stepanov V, Cho F~H, Abeywardana C and Takahashi S 2015 {\em Appl. Phys. Lett.\/} {\bf 106} 063111

\bibitem{Mamin2012}
Mamin H~J, Sherwood M~H and Rugar D 2012 {\em Phys. Rev. B\/} {\bf 86} 195422

\bibitem{Haeberle2015}
H\"aberle T, Schmid-Lorch D, Reinhard F and Wrachtrup J 2015 {\em Nat Nanotechnol.\/} {\bf 10} 125--128

\bibitem{Pfender2019}
Pfender M, Wang P, Sumiya H, Onoda S, Yang W, Dasari D~B~R, Neumann P, Pan X~Y, Isoya J, Liu R~B and Wrachtrup J 2019 {\em Nat. Commun.\/} {\bf 10} 594

\bibitem{Cohen2020}
Cohen D, Gefen T, Ortiz L and Retzker A 2020 {\em npj Quantum Inf.\/} {\bf 6} 83

\bibitem{Schwartz2019}
Schwartz I, Rosskopf J, Schmitt S, Tratzmiller B, Chen Q, McGuinness L~P, Jelezko F and Plenio M~B 2019 {\em Sci. Rep.\/} {\bf 9} 6938

\bibitem{Liu2022}
Liu K~S, Ma X, Rizzato R, Semrau A~L, Henning A, Sharp I~D, Fischer R~A and Bucher D~B 2022 {\em Nano Lett.\/} {\bf 22} 9876--9882

\bibitem{Shagieva2018}
Shagieva F, Zaiser S, Neumann P, Dasari D~B~R, St\"ohr R, Denisenko A, Reuter R, Meriles C~A and Wrachtrup J 2018 {\em Nano Lett.\/} {\bf 18} 3731--3737

\bibitem{Healey2021}
Healey A, Hall L, White G, Teraji T, Sani M~A, Separovic F, Tetienne J~P and Hollenberg L 2021 {\em Phys. Rev. Applied\/} {\bf 15} 054052

\bibitem{Cohen2020a}
Cohen D, Nigmatullin R, Kenneth O, Jelezko F, Khodas M and Retzker A 2020 {\em Sci. Rep.\/} {\bf 10} 5298

\bibitem{Mzyk2022}
Mzyk A, Sigaeva A and Schirhagl R 2022 {\em Acc. Chem. Res.\/} {\bf 55} 3572--3580

\bibitem{Holzgrafe2020}
Holzgrafe J, Gu Q, Beitner J, Kara D~M, Knowles H~S and Atatüre M 2020 {\em Phys. Rev. Applied\/} {\bf 13} 044004

\bibitem{Chen2022}
Chen Y, Bae Y and Heinrich A~J 2022 {\em Adv. Mater.\/} {\bf 35} 2107534

\bibitem{Baumann2015}
Baumann S, Paul W, Choi T, Lutz C~P, Ardavan A and Heinrich A~J 2015 {\em Science\/} {\bf 350} 417--420

\bibitem{Yang2017}
Yang K, Bae Y, Paul W, Natterer F~D, Willke P, Lado J~L, Ferrón A, Choi T, Fernández-Rossier J, Heinrich A~J and Lutz C~P 2017 {\em Phys. Rev. Lett.\/} {\bf 119} 227206

\bibitem{Bae2018}
Bae Y, Yang K, Willke P, Choi T, Heinrich A~J and Lutz C~P 2018 {\em Sci. Adv.\/} {\bf 4} aau4159

\bibitem{Yang2018}
Yang K, Willke P, Bae Y, Ferrón A, Lado J~L, Ardavan A, Fernández-Rossier J, Heinrich A~J and Lutz C~P 2018 {\em Nat. Nanotechnol.\/} {\bf 13} 1120--1125

\bibitem{Kovarik2022}
Kovarik S, Robles R, Schlitz R, Seifert T~S, Lorente N, Gambardella P and Stepanow S 2022 {\em Nano Lett.\/} {\bf 22} 4176--4181

\bibitem{Zhang2021a}
Zhang X, Wolf C, Wang Y, Aubin H, Bilgeri T, Willke P, Heinrich A~J and Choi T 2021 {\em Nat. Chem.\/} {\bf 14} 59--65

\bibitem{Kawaguchi2022}
Kawaguchi R, Hashimoto K, Kakudate T, Katoh K, Yamashita M and Komeda T 2022 {\em Nano Lett.\/} {\bf 23} 213--219

\bibitem{Natterer2017}
Natterer F~D, Yang K, Paul W, Willke P, Choi T, Greber T, Heinrich A~J and Lutz C~P 2017 {\em Nature\/} {\bf 543} 226--228

\bibitem{Singha2021}
Singha A, Willke P, Bilgeri T, Zhang X, Brune H, Donati F, Heinrich A~J and Choi T 2021 {\em Nat. Commun.\/} {\bf 12} 4179

\bibitem{Wang2023}
Wang Y, Chen Y, Bui H~T, Wolf C, Haze M, Mier C, Kim J, Choi D~J, Lutz C~P, Bae Y, Phark S~h and Heinrich A~J 2023 {\em Science\/} {\bf 382} 87--92

\bibitem{Seifert2020}
Seifert T~S, Kovarik S, Juraschek D~M, Spaldin N~A, Gambardella P and Stepanow S 2020 {\em Sci. Adv.\/} {\bf 6} abc5511

\bibitem{Delgado2021}
Delgado F and Lorente N 2021 {\em Prog. Surf. Sci.\/} {\bf 96} 100625

\bibitem{ReinaGalvez2023}
Reina-Gálvez J, Wolf C and Lorente N 2023 {\em Phys. Rev. B\/} {\bf 107} 235404

\bibitem{Choi2017}
Choi T, Paul W, Rolf-Pissarczyk S, Macdonald A~J, Natterer F~D, Yang K, Willke P, Lutz C~P and Heinrich A~J 2017 {\em Nat. Nanotechnol.\/} {\bf 12} 420--424

\bibitem{Willke2018}
Willke P, Bae Y, Yang K, Lado J~L, Ferrón A, Choi T, Ardavan A, Fernández-Rossier J, Heinrich A~J and Lutz C~P 2018 {\em Science\/} {\bf 362} 336--339

\bibitem{Farinacci2022}
Farinacci L, Veldman L~M, Willke P and Otte S 2022 {\em Nano Lett.\/} {\bf 22} 8470--8474

\bibitem{Kim2022}
Kim J, Noh K, Chen Y, Donati F, Heinrich A~J, Wolf C and Bae Y 2022 {\em Nano Lett.\/} {\bf 22} 9766--9772

\bibitem{Willke2018a}
Willke P, Paul W, Natterer F~D, Yang K, Bae Y, Choi T, Fernández-Rossier J, Heinrich A~J and Lutz C~P 2018 {\em Sci. Adv.\/} {\bf 4} aaq1543

\bibitem{Yang2019}
Yang K, Paul W, Phark S~H, Willke P, Bae Y, Choi T, Esat T, Ardavan A, Heinrich A~J and Lutz C~P 2019 {\em Science\/} {\bf 366} 509--512

\bibitem{Balatsky2002}
Balatsky A~V, Manassen Y and Salem R 2002 {\em Phys. Rev. B\/} {\bf 66} 195416

\bibitem{Balatsky2012}
Balatsky A~V, Nishijima M and Manassen Y 2012 {\em Adv. Phys.\/} {\bf 61} 117--152

\bibitem{Willke2019}
Willke P, Yang K, Bae Y, Heinrich A~J and Lutz C~P 2019 {\em Nat. Phys.\/} {\bf 15} 1005--1010

\bibitem{Paul2016}
Paul W, Baumann S, Lutz C~P and Heinrich A~J 2016 {\em Rev. Sci. Instrum.\/} {\bf 87} 074703

\bibitem{ReinaGalvez2019}
Reina~Gálvez J, Wolf C, Delgado F and Lorente N 2019 {\em Phys. Rev. B\/} {\bf 100} 035411

\bibitem{Steinbrecher2021}
Steinbrecher M, van Weerdenburg W~M~J, Walraven E~F, van Mullekom N~P~E, Gerritsen J~W, Natterer F~D, Badrtdinov D~I, Rudenko A~N, Mazurenko V~V, Katsnelson M~I, van~der Avoird A, Groenenboom G~C and Khajetoorians A~A 2021 {\em Phys. Rev. B\/} {\bf 103} 155405

\bibitem{Kot2023}
Kot P, Ismail M, Drost R, Siebrecht J, Huang H and Ast C~R 2023 {\em Nat. Commun.\/} {\bf 14} 6612

\bibitem{Paul2016a}
Paul W, Yang K, Baumann S, Romming N, Choi T, Lutz C and Heinrich A 2016 {\em Nat. Phys.\/} {\bf 13} 403--407

\bibitem{Narkowicz2005}
Narkowicz R, Suter D and Stonies R 2005 {\em J. Magn. Reson.\/} {\bf 175} 275--284

\bibitem{Dayan2018}
Dayan N, Ishay Y, Artzi Y, Cristea D, Reijerse E, Kuppusamy P and Blank A 2018 {\em Rev. Sci. Instrum.\/} {\bf 89} 124707

\bibitem{Wallace1991}
Wallace W~J and Silsbee R~H 1991 {\em Rev. Sci. Instrum.\/} {\bf 62} 1754--1766

\bibitem{Sigillito2014}
Sigillito A~J, Malissa H, Tyryshkin A~M, Riemann H, Abrosimov N~V, Becker P, Pohl H~J, Thewalt M~L~W, Itoh K~M, Morton J~J~L, Houck A~A, Schuster D~I and Lyon S~A 2014 {\em Appl. Phys. Lett.\/} {\bf 104} 222407

\bibitem{Graaf2012}
Graaf S~E~d, Danilov A~V, Adamyan A, Bauch T and Kubatkin S~E 2012 {\em J. Appl. Phys.\/} {\bf 112} 123905

\bibitem{Zollitsch2019}
Zollitsch C~W, O’Sullivan J, Kennedy O, Dold G and Morton J~J~L 2019 {\em AIP Adv.\/} {\bf 9} 125225

\bibitem{Bienfait2016}
Bienfait A, Pla J~J, Kubo Y, Zhou X, Stern M, Lo C~C, Weis C~D, Schenkel T, Vion D, Esteve D, Morton J~J~L and Bertet P 2016 {\em Nature\/} {\bf 531} 74--77

\bibitem{Eichler2017}
Eichler C, Sigillito A, Lyon S and Petta J 2017 {\em Phys. Rev. Lett.\/} {\bf 118} 037701

\bibitem{Macklin2015}
Macklin C, O’Brien K, Hover D, Schwartz M~E, Bolkhovsky V, Zhang X, Oliver W~D and Siddiqi I 2015 {\em Science\/} {\bf 350} 307--310

\bibitem{Bienfait2015}
Bienfait A, Pla J~J, Kubo Y, Stern M, Zhou X, Lo C~C, Weis C~D, Schenkel T, Thewalt M~L~W, Vion D, Esteve D, Julsgaard B, Mølmer K, Morton J~J~L and Bertet P 2015 {\em Nat. Nanotechnol.\/} {\bf 11} 253--257

\bibitem{Probst2017}
Probst S, Bienfait A, Campagne-Ibarcq P, Pla J~J, Albanese B, Da~Silva~Barbosa J~F, Schenkel T, Vion D, Esteve D, Mølmer K, Morton J~J~L, Heeres R and Bertet P 2017 {\em Appl. Phys. Lett.\/} {\bf 111} 202604

\bibitem{Ranjan2020}
Ranjan V, Probst S, Albanese B, Schenkel T, Vion D, Esteve D, Morton J~J~L and Bertet P 2020 {\em Appl. Phys. Lett.\/} {\bf 116} 184002

\bibitem{Bienfait2017}
Bienfait A, Campagne-Ibarcq P, Kiilerich A, Zhou X, Probst S, Pla J, Schenkel T, Vion D, Esteve D, Morton J, Moelmer K and Bertet P 2017 {\em Phys. Rev. X\/} {\bf 7} 041011

\bibitem{Lescanne2020}
Lescanne R, Deléglise S, Albertinale E, Réglade U, Capelle T, Ivanov E, Jacqmin T, Leghtas Z and Flurin E 2020 {\em Phys. Rev. X\/} {\bf 10} 021038

\bibitem{Albertinale2021}
Albertinale E, Balembois L, Billaud E, Ranjan V, Flanigan D, Schenkel T, Estève D, Vion D, Bertet P and Flurin E 2021 {\em Nature\/} {\bf 600} 434--438

\bibitem{Wang2023a}
Wang Z, Balembois L, Rančić M, Billaud E, Le~Dantec M, Ferrier A, Goldner P, Bertaina S, Chanelière T, Esteve D, Vion D, Bertet P and Flurin E 2023 {\em Nature\/} {\bf 619} 276--281

\bibitem{Albanese2020}
Albanese B, Probst S, Ranjan V, Zollitsch C~W, Pechal M, Wallraff A, Morton J~J~L, Vion D, Esteve D, Flurin E and Bertet P 2020 {\em Nat. Phys.\/} {\bf 16} 751--755

\bibitem{LeDantec2021}
Le~Dantec M, Rančić M, Lin S, Billaud E, Ranjan V, Flanigan D, Bertaina S, Chanelière T, Goldner P, Erb A, Liu R~B, Estève D, Vion D, Flurin E and Bertet P 2021 {\em Sci. Adv.\/} {\bf 7} abj9786

\bibitem{Ranjan2021}
Ranjan V, Albanese B, Albertinale E, Billaud E, Flanigan D, Pla J, Schenkel T, Vion D, Esteve D, Flurin E, Morton J, Niquet Y and Bertet P 2021 {\em Phys. Rev. X\/} {\bf 11} 031036

\bibitem{Sleator1985}
Sleator T, Hahn E~L, Hilbert C and Clarke J 1985 {\em Phys. Rev. Lett.\/} {\bf 55} 1742--1745

\bibitem{Hahn1950}
Hahn E~L 1950 {\em Phys. Rev.\/} {\bf 80} 580--594

\bibitem{Meiboom1958}
Meiboom S and Gill D 1958 {\em Rev. Sci. Instrum.\/} {\bf 29} 688--691

\bibitem{Carr1954}
Carr H~Y and Purcell E~M 1954 {\em Phys. Rev.\/} {\bf 94} 630--638

\bibitem{Viola1999}
Viola L, Knill E and Lloyd S 1999 {\em Phys. Rev. Lett.\/} {\bf 82} 2417--2421

\bibitem{Taylor2008}
Taylor J~M, Cappellaro P, Childress L, Jiang L, Budker D, Hemmer P~R, Yacoby A, Walsworth R and Lukin M~D 2008 {\em Nat. Phys.\/} {\bf 4} 810--816

\bibitem{Biercuk2011}
Biercuk M~J, Doherty A~C and Uys H 2011 {\em J. Phys. B: At. Mol. Opt. Phys.\/} {\bf 44} 154002

\bibitem{Zhao2012}
Zhao N, Honert J, Schmid B, Klas M, Isoya J, Markham M, Twitchen D, Jelezko F, Liu R~B, Fedder H and Wrachtrup J 2012 {\em Nat. Nanotechnol.\/} {\bf 7} 657--662

\bibitem{Bylander2011}
Bylander J, Gustavsson S, Yan F, Yoshihara F, Harrabi K, Fitch G, Cory D~G, Nakamura Y, Tsai J~S and Oliver W~D 2011 {\em Nat. Phys.\/} {\bf 7} 565--570

\bibitem{Kotler2011}
Kotler S, Akerman N, Glickman Y, Keselman A and Ozeri R 2011 {\em Nature\/} {\bf 473} 61--65

\bibitem{Young2012}
Young K~C and Whaley K~B 2012 {\em Phys. Rev. A\/} {\bf 86} 012314

\bibitem{Yuge2011}
Yuge T, Sasaki S and Hirayama Y 2011 {\em Phys. Rev. Lett.\/} {\bf 107} 170504

\bibitem{Baum1985}
Baum J, Tycko R and Pines A 1985 {\em Phys. Rev. A\/} {\bf 32} 3435--3447

\bibitem{Bax1986}
Bax A and Lerner L 1986 {\em Science\/} {\bf 232} 960--967

\bibitem{Eberhardt2008}
Eberhardt K, Degen C, Hunkeler A and Meier B 2008 {\em Angew. Chem. - Int. Ed.\/} {\bf 47} 8961--8963

\bibitem{Loretz2015}
Loretz M, Boss J, Rosskopf T, Mamin H, Rugar D and Degen C 2015 {\em Phys. Rev. X\/} {\bf 5} 021009

\bibitem{Shu2017}
Shu Z, Zhang Z, Cao Q, Yang P, Plenio M~B, Müller C, Lang J, Tomek N, Naydenov B, McGuinness L~P, Jelezko F and Cai J 2017 {\em Phys. Rev. A\/} {\bf 96} 051402

\bibitem{Hirose2012}
Hirose M, Aiello C~D and Cappellaro P 2012 {\em Phys. Rev. A\/} {\bf 86} 062320

\bibitem{Souza2011}
Souza A~M, Álvarez G~A and Suter D 2011 {\em Phys. Rev. Lett.\/} {\bf 106} 240501

\bibitem{Zhao2011}
Zhao N, Hu J~L, Ho S~W, Wan J~T~K and Liu R~B 2011 {\em Nat. Nanotechnol.\/} {\bf 6} 242--246

\bibitem{Zopes2017}
Zopes J, Sasaki K, Cujia K, Boss J, Chang K, Segawa T, Itoh K and Degen C 2017 {\em Phys. Rev. Lett.\/} {\bf 119} 260501

\bibitem{Cujia2019}
Cujia K~S, Boss J~M, Herb K, Zopes J and Degen C~L 2019 {\em Nature\/} {\bf 571} 230--233

\bibitem{Dong2021}
Dong Y, Xu J~Y, Zhang S~C, Zheng Y, Chen X~D, Zhu W, Wang G~Z, Guo G~C and Sun F~W 2021 {\em Funct. Diam.\/} {\bf 1} 125--134

\bibitem{Dong2022}
Dong Y, Gao X~D, Yu C, Feng Z~H, Lin H~B, Chen X~D, Zhu W and Sun F~W 2022 {\em Appl. Phys. Lett.\/} {\bf 120} 194001

\bibitem{Wang2021}
Wang G, Liu Y~X, Zhu Y and Cappellaro P 2021 {\em Nano Lett.\/} {\bf 21} 5143--5150

\bibitem{Taminiau2012}
Taminiau T~H, Wagenaar J~J~T, van~der Sar T, Jelezko F, Dobrovitski V~V and Hanson R 2012 {\em Phys. Rev. Lett.\/} {\bf 109} 137602

\bibitem{Kolkowitz2012}
Kolkowitz S, Unterreithmeier Q~P, Bennett S~D and Lukin M~D 2012 {\em Phys. Rev. Lett.\/} {\bf 109} 137601

\bibitem{Ajoy2019}
Ajoy A, Bissbort U, Poletti D and Cappellaro P 2019 {\em Phys. Rev. Lett.\/} {\bf 122} 013205

\bibitem{Lee2016}
Lee K~W, Lee D, Ovartchaiyapong P, Minguzzi J, Maze J~R and Bleszynski~Jayich A~C 2016 {\em Phys. Rev. Applied\/} {\bf 6} 034005

\bibitem{Maity2020}
Maity S, Shao L, Bogdanović S, Meesala S, Sohn Y~I, Sinclair N, Pingault B, Chalupnik M, Chia C, Zheng L, Lai K and Lončar M 2020 {\em Nat. Commun.\/} {\bf 11} 193

\bibitem{MacQuarrie2013}
MacQuarrie E~R, Gosavi T~A, Jungwirth N~R, Bhave S~A and Fuchs G~D 2013 {\em Phys. Rev. Lett.\/} {\bf 111} 227602

\bibitem{Chen2020}
Chen H~Y, Bhave S~A and Fuchs G~D 2020 {\em Phys. Rev. Applied\/} {\bf 13} 054068

\bibitem{Maity2022}
Maity S, Pingault B, Joe G, Chalupnik M, Assumpção D, Cornell E, Shao L and Lončar M 2022 {\em Phys. Rev. X\/} {\bf 12} 011056

\bibitem{Forneris2017}
Forneris J, Ditalia~Tchernij S, Tengattini A, Enrico E, Grilj V, Skukan N, Amato G, Boarino L, Jakšić M and Olivero P 2017 {\em Carbon\/} {\bf 113} 76--86

\bibitem{Wang2020a}
Wang X, Xiao Y, Liu C, Lee-Wong E, McLaughlin N~J, Wang H, Wu M, Wang H, Fullerton E~E and Du C~R 2020 {\em npj Quantum Inf.\/} {\bf 6} 78

\bibitem{Gulka2021}
Gulka M, Wirtitsch D, Ivády V, Vodnik J, Hruby J, Magchiels G, Bourgeois E, Gali A, Trupke M and Nesladek M 2021 {\em Nat. Commun.\/} {\bf 12} 4421

\bibitem{Yale2013}
Yale C~G, Buckley B~B, Christle D~J, Burkard G, Heremans F~J, Bassett L~C and Awschalom D~D 2013 {\em Proc. Natl. Acad. Sci.\/} {\bf 110} 7595--7600

\bibitem{Kim2019}
Kim D, Ibrahim M~I, Foy C, Trusheim M~E, Han R and Englund D~R 2019 {\em Nat. Electron.\/} {\bf 2} 284--289

\bibitem{Stuerner2021}
Stürner F~M, Brenneis A, Buck T, Kassel J, Rölver R, Fuchs T, Savitsky A, Suter D, Grimmel J, Hengesbach S, Förtsch M, Nakamura K, Sumiya H, Onoda S, Isoya J and Jelezko F 2021 {\em Adv. Quantum Technol.\/} {\bf 4} 2000111

\bibitem{Patel2020}
Patel R, Zhou L, Frangeskou A, Stimpson G, Breeze B, Nikitin A, Dale M, Nichols E, Thornley W, Green B, Newton M, Edmonds A, Markham M, Twitchen D and Morley G 2020 {\em Phys. Rev. Applied\/} {\bf 14} 044058

\bibitem{Du2021}
Du B, Huang K, Nie Y, Zhang Z, Xu R, Cui J and Li J 2021 {\em IEEE Sensors J.\/} {\bf 21} 24665--24671

\bibitem{Wang2022b}
Wang X, Zheng D, Wang X, Liu X, Wang Q, Zhao J, Guo H, Qin L, Tang J, Ma Z and Liu J 2022 {\em IEEE Sensors J.\/} {\bf 22} 5580--5587

\bibitem{Ibrahim2021}
Ibrahim M~I, Foy C, Englund D~R and Han R 2021 {\em IEEE J. Solid-State Circuits\/} {\bf 56} 1001--1014

\bibitem{Misonou2020}
Misonou D, Sasaki K, Ishizu S, Monnai Y, Itoh K~M and Abe E 2020 {\em AIP Adv.\/} {\bf 10} 025206

\bibitem{Wan2020}
Wan N~H, Lu T~J, Chen K~C, Walsh M~P, Trusheim M~E, De~Santis L, Bersin E~A, Harris I~B, Mouradian S~L, Christen I~R, Bielejec E~S and Englund D 2020 {\em Nature\/} {\bf 583} 226--231

\bibitem{Omirzakhov2022}
Omirzakhov K, Idjadi M~H, Huang T~Y, Breitweiser S~A, Hopper D~A, Bassett L~C and Aflatouni F 2022 An integrated quantum spin control system in 180nm cmos {\em 2022 IEEE Radio Frequency Integrated Circuits Symposium (RFIC)\/}

\bibitem{Shalaginov2020}
Shalaginov M~Y, Bogdanov S~I, Lagutchev A~S, Kildishev A~V, Boltasseva A and Shalaev V~M 2020 {\em ACS Photonics\/} {\bf 7} 2018--2026

\bibitem{Bradley2019}
Bradley C, Randall J, Abobeih M, Berrevoets R, Degen M, Bakker M, Markham M, Twitchen D and Taminiau T 2019 {\em Phys. Rev. X\/} {\bf 9} 031045

\bibitem{Gebhart2023}
Gebhart V, Santagati R, Gentile A~A, Gauger E~M, Craig D, Ares N, Banchi L, Marquardt F, Pezzè L and Bonato C 2023 {\em Nat. Rev. Phys.\/} {\bf 5} 141--156

\bibitem{2007}
Robert C~P 2007 {\em The Bayesian Choice\/} (Springer New York) ISBN 9780387715988

\bibitem{Arshad2022}
Arshad M~J, Bekker C, Haylock B, Skrzypczak K, White D, Griffiths B, Gore J, Morley G~W, Salter P, Smith J, Zohar I, Finkler A, Altmann Y, Gauger E~M and Bonato C 2022  (\textit{Preprint} \eprint{2210.06103})

\bibitem{CaouetteMansour2022}
Caouette-Mansour M, Solyom A, Ruffolo B, McMichael R~D, Sankey J and Childress L 2022 {\em Phys. Rev. Applied\/} {\bf 17} 064031

\bibitem{Bonato2015}
Bonato C, Blok M~S, Dinani H~T, Berry D~W, Markham M~L, Twitchen D~J and Hanson R 2015 {\em Nat. Nanotechnol.\/} {\bf 11} 247--252

\bibitem{Santagati2019}
Santagati R, Gentile A, Knauer S, Schmitt S, Paesani S, Granade C, Wiebe N, Osterkamp C, McGuinness L, Wang J, Thompson M, Rarity J, Jelezko F and Laing A 2019 {\em Phys. Rev. X\/} {\bf 9} 021019

\bibitem{Joas2021}
Joas T, Schmitt S, Santagati R, Gentile A~A, Bonato C, Laing A, McGuinness L~P and Jelezko F 2021 {\em npj Quantum Inf.\/} {\bf 7} 56

\bibitem{Jung2021}
Jung K, Abobeih M~H, Yun J, Kim G, Oh H, Henry A, Taminiau T~H and Kim D 2021 {\em npj Quantum Inf.\/} {\bf 7} 41

\bibitem{Wiebe2014}
Wiebe N, Granade C, Ferrie C and Cory D 2014 {\em Phys. Rev. Lett.\/} {\bf 112} 190501

\bibitem{Fiderer2021}
Fiderer L~J, Schuff J and Braun D 2021 {\em PRX Quantum\/} {\bf 2} 020303

\bibitem{Han2021}
Han H and Choi S 2021 {\em J. Phys. Chem. Lett.\/} {\bf 12} 3662--3668

\bibitem{Baum2021}
Baum Y, Amico M, Howell S, Hush M, Liuzzi M, Mundada P, Merkh T, Carvalho A~R and Biercuk M~J 2021 {\em PRX Quantum\/} {\bf 2} 040324

\bibitem{Foesel2018}
Fösel T, Tighineanu P, Weiss T and Marquardt F 2018 {\em Phys. Rev. X\/} {\bf 8} 031084

\bibitem{Green2015}
Green P~J, Łatuszyński K, Pereyra M and Robert C~P 2015 {\em Stat. Comput.\/} {\bf 25} 835--862

\bibitem{Chopin2015}
Chopin N, Gadat S, Guedj B, Guyader A and Vernet E 2015 {\em ESAIM Proc. Surveys\/} {\bf 51} 293--319

\bibitem{Gentile2021}
Gentile A~A, Flynn B, Knauer S, Wiebe N, Paesani S, Granade C~E, Rarity J~G, Santagati R and Laing A 2021 {\em Nat. Phys.\/} {\bf 17} 837--843

\end{thebibliography}
\end{document}